\documentclass{aa}  

\usepackage{graphicx}
\usepackage{txfonts}
\usepackage{comment}
\usepackage{adjustbox}
\usepackage{tabu}
\DeclareUnicodeCharacter{2212}{-}
\usepackage{placeins}

\usepackage{xspace}
\newcommand{\kms}{km~s$^{-1}$\xspace}

\usepackage[breaklinks=true, colorlinks=true, citecolor=blue, linkcolor=blue,urlcolor=blue]{hyperref}

\usepackage{threeparttable}

\begin{document}

\title{The JWST/NIRSpec view of the nuclear region in the prototypical merging galaxy NGC 6240}

\author{M. Ceci \inst{1,2}
\and G. Cresci \inst{2}
\and S. Arribas \inst{3}
\and T. B{\"o}ker\inst{4}
\and A. J. Bunker\inst{5}
\and S. Charlot \inst{6}
\and K. Fahrion \inst{7}
\and I. Lamperti \inst{1,2}
\and A.~Marconi \inst{1,2}
\and M. Perna \inst{3}
\and G. Tozzi \inst{8}
\and L. Ulivi \inst{1,2,9}
}
\institute{Università di Firenze, Dipartimento di Fisica e Astronomia, via G. Sansone 1, 50019 Sesto F.no, Firenze, Italy\\\email{matteo.ceci@unifi.it}
\and 
INAF - Osservatorio Astrofisico di Arcetri, Largo E. Fermi 5, I-50125 Firenze, Italy
\and 
Centro de Astrobiología (CAB), CSIC–INTA,  Ctra. de Ajalvir Km. 4, 28850 – Torrejón de Ardoz, Madrid, Spain
\and 
Space Telescope Science Institute, 3700 San Martin Drive, Baltimore, MD 21218, USA
\and 
Department of Physics, University of Oxford, Denys Wilkinson Building, Keble Road, Oxford OX1 3RH, UK
\and 
Sorbonne Universit\'e, CNRS, UMR 7095, Institut d'Astrophysique de Paris, 98 bis bd Arago, 75014 Paris, France
\and 
Department of Astrophysics, University of Vienna, T\"{u}rkenschanzstra{\ss}e 17, 1180 Wien, Austria
\and 
Max-Planck-Institut für Extraterrestrische Physik (MPE), Giessenbachstraße 1, D-85748 Garching, Germany
\and 
University of Trento, Via Sommarive 14, I-38123 Trento, Italy
}
    \titlerunning{NGC 6240}
    \authorrunning{Ceci, M., et al.}

    \date{Received September 11, 2024; accepted January 03, 2025}

 
  \abstract
   {
    Merger events are thought to be an important phase in the assembly of massive galaxies. At the same time, active galactic nuclei (AGN) play a fundamental role in the evolution of their star formation histories. Both phenomena can be observed at work in NGC~6240, a local prototypical merger classified as an ultraluminous infrared galaxy (ULIRG) thanks to its  elevated infrared luminosity. Interestingly,  NGC\,6240  hosts two AGN separated by 1.5\arcsec($\sim$ 735 pc), detected in both X-ray and radio band.
    
    Taking advantage of the unprecedented sensitivity and wavelength coverage provided by the integral field unit (IFU) of the NIRSpec instrument on board JWST, we observed the nuclear region of NGC\,6240 in a field of view of 3.7\arcsec~$\times$~3.7\arcsec($1.9 \, \times \, 1.9$ kpc$^2$) in order to investigate gas kinematics and interstellar medium (ISM) properties with a high spatial resolution of $\sim$ 0.1\arcsec (or $\sim$ 50 pc).
    
    We characterized the 2D stellar kinematics, separated the different gas kinematic components through multi-Gaussian fitting, and studied the excitation properties of the ISM from the near-infrared diagnostic diagram based on the H$_2$ 1-0 S(1)/Br${\gamma}$ and [Fe II]$\lambda$1.257$\mu$m/Pa$\beta$ lines ratios. We isolated the ionization cones of the two nuclei and detected coronal line emission from both of them. Using H$_2$ line ratios, we found that the molecular hydrogen gas is excited mostly by thermal processes. We computed a hot molecular gas mass of $1.3 \times 10^5$ M$_{\odot}$ and an ionized gas mass in the range of $10^5$ - $10^7$ M$_{\odot}$, depending on assumptions. We studied with unprecedented spatial resolution and sensitivity the kinematics of the molecular and ionized gas phases, and we revealed the complex structure of the molecular gas and found a blueshifted outflow near the southern nucleus, together with filaments connecting a highly redshifted H$_2$ cloud with the two nuclei. We speculate on the possible nature of this H$_2$ cloud and propose two possible scenarios: outflowing gas or a tidal cloud falling onto the nuclei.
   }



    \maketitle
%

\section{Introduction}
\label{sect: intro}
Ultraluminous infrared galaxies (ULIRG) are among the most luminous objects in the local Universe, and they represent a fundamental phase in the evolution of active galactic nuclei (AGN). This extremely luminous episode, due to a combination of a starburst and/or AGN activity in a dust-enshrouded environment, occurs probably during a major merger of gas-rich galaxies (\citealt{sanders_ultraluminous_1988, sanders_luminous_1996,lipari_extreme_2003, hopkins_unified_2006}), when gas and dust are both channeled into the nuclear regions of the progenitors (\citealt{u_spectral_2012, lonsdale_ultraluminous_2006}). In this scenario, ULIRGs are seen as a rapid growth phase when the merger initiates an intense period of star formation, fueling the growth of a supermassive black hole (SMBH) and the eventual AGN activity. Throughout the merger process, powerful outflows generated by both starburst and AGN activity shape the galactic surrounding, influencing subsequent star formation and black hole accretion through feedback mechanisms (e.g., \citealt{silk_unleashing_2013}, \citealt{alexander_what_2012}).

In this paper, we focus on NGC 6240. Despite having an IR luminosity of 6.3~$\times$~10$^{11}$ L$_\odot$, which is just below the canonical ULIRG luminosity boundary of 10$^{12}$ L$_\odot$, it is frequently associated with the ULIRG category and stands out as one of the closest and most extensively examined objects within this category (e.g., \citealt{genzel_what_1998}). At z $\sim$ 0.02, NGC 6240 is considered the prototypical merger galaxy in the local Universe. This means it is observed during the short-lived and therefore rarely observed transient phase between the first encounter and the final coalescence of two gas-rich spiral galaxies (\citealt{tacconi_gas_1999, engel_ngc6240_2010}). It hosts two X-ray confirmed AGN that are positioned approximately 735 pc apart (\citealt{komossa_discovery_2003,max_locating_2007}) and driving kiloparsec-scale superwinds detected across various wavelengths: radio (\citealt{baan_h_2007}), far-IR (\citealt{veilleux_fast_2013}), near-infrared (NIR; \citealt{van_der_werf_near-infrared_1993}, \citealt{max_core_2005}), optical (\citealt{heckman_nature_1990, veilleux_search_2003}), and X-ray (\citealt{nardini_exceptional_2013, wang_fast_2014}). Despite both being active, the two nuclei might not exert enough influence or may be too obscured to ionize the interstellar medium (ISM) significantly. Indeed, star formation likely contributes 76\%-80\% of the IR luminosity (\citealt{armus_detection_2006, nardini_exploring_2009}).

NGC 6240 exhibits one of the most intense NIR H$_2$ emission observed in any galaxy, with a luminosity (L$_{\rm H_2}$) of approximately 10$^8$ L$_\odot$ (\citealt{joseph_detection_1984}). \cite{cicone_alma_2018} found a molecular outflow extending more than 10 kpc along the E-W direction and peaking between the two AGN rather than on either of the two. Numerous studies have charted the H$_2$ line emission within the inner ($\sim$1 kpc) region (e.g.,  \citealt{tecza_stellar_2000, max_core_2005, ilha_origin_2016}), indicating that shock excitation is the predominant mechanism.  \cite{treister_molecular_2020} and \cite{fyhrie_molecular_2021} argue that their observations are consistent with a tidal bridge between the two nuclei and an observed high velocity redshifted gas. Gas that is subject to gravitational torques, such as that in a tidal bridge in the nuclear region, will likely fall into the nuclear regions by the time of final coalescence (\citealt{souchay_tides_2012}). However, it is also possible that the structure will dissipate before this streaming occurs, given the high velocity dispersion relative to the vertical extent of the gas.

One of the most peculiar features of NGC 6240 concerns the positions of the two galaxy nuclei. The distance between them appears consistent across different wavelengths: the two AGN are separated by approximately 1.5\arcsec ($\sim$ 700 pc) in X-ray \citep{komossa_discovery_2003}, radio \citep{gallimore_parsec-scale_2004}, and at 3–5 $\mu$m \citep{risaliti_double_2006}. The first high-resolution view of the NIR nuclei by \citet{max_core_2005} revealed the impact of dust on the shapes and positions of the nuclei from optical to 2.2 $\mu$m, with \citet{max_locating_2007}, demonstrating the consistency between the radio/X-ray positions and those at high resolution in the NIR (separation $\sim$1.5"). However, the bright spots seen in the NIR and optical, separated by about 1.8\arcsec ($\sim$ 900 pc; e.g., \citealt{gerssen_hubble_2004}), are significantly affected by dust extinction. As a result, these spots do not correspond to the true positions of the nuclei, which only become visible at wavelengths longer than 2–3 $\mu$m, where the dust obscuration is less pronounced.
\cite{kollatschny_ngc6240_2020} propose the presence of a third SMBH close to the southern one based on narrow field mode (NFM) MUSE data. \cite{fabbiano_revisiting_2020} investigated the existence of this putative third nucleus through a reanalysis of the cumulative ACIS Chandra data but did not find any strong evidence for it, so the question is still under discussion.

This paper presents new observations of the nuclear region of NGC 6240 obtained with the integral field unit (IFU) of the Near Infrared Spectrograph (NIRSpec) instrument on board the James Webb Space Telescope (JWST; \citealt{jakobsen_near-infrared_2022, boker_near-infrared_2022}) and collected as part of the JWST/NIRSpec IFS Guaranteed Time Observations (GTO) survey Resolved structure and kinematics of the nuclear regions of nearby galaxies (program lead: Torsten B{\"o}ker). Other galaxies studied in this program include NGC 4654 (\citealp{fahrion_growing_2024}), Arp 220 (\citealp{perna_no_2024, ulivi_arp220_2024}), and Mrk 231 (Ulivi et al. in prep.). In collaboration with the Mid-Infrared Instrument (MIRI) GTO team, this program offers a complete NIR and mid-IR perspective on the nuclear region of NGC 6240 (see the MIRI NGC 6240 paper by \citealp{MIRI_6240}).

This manuscript is organized as follows.
Section \ref{sec:Obs} presents the NIRSpec observations and data reduction. Section \ref{sec:Analysis} presents the spectral analysis using a multi-Gaussian fitting procedure. In Section \ref{sec:Results}, we discuss our results through the characterization of the ISM properties and the spatially resolved kinematics of the ionized and molecular gas in the central region of NGC 6240. Finally, Section \ref{sec:Conclusions} summarizes our conclusions. Throughout this work, we assume $\Omega_m=0.286$ and $H_0=69.9$ \kms Mpc$^{-1}$ (\citealt{bennett_1_2014}).

\section{Observation and data reduction}
\label{sec:Obs}

\subsection{JWST/NIRSpec data}
NGC 6240 was observed with the JWST/NIRSpec instrument in IFU mode as part of the GTO programme 1265 (PI: A. Alonso-Herrero), on 17th July 2023. Centered on an intermediate position between the two nuclei of NGC 6240 (coordinates from far-IR continuum emission; see \citealt{scoville_alma_2015}), the dataset utilizes the following high-resolution (R $\sim$ 2700) configurations, with their respective spectral ranges: [0.97 $\mu$m, 1.89 $\mu$m] for G140H/F100LP (cube1 hereafter), [1.66 $\mu$m, 3.17 $\mu$m] for G235H/F170LP (cube2) and [2.87 $\mu$m, 5.27 $\mu$m] for G395H/F290LP (cube3). The observations were taken with an NRSIRS2RAPID readout pattern with 15 groups per integration, and one integration per exposure, using a four-point medium-cycling dithering. Every configuration has a total integration time of 1867 s. However, this work focuses solely on the first two configurations, as our aim is to study the kinematics of the emitting gas, with its strongest emission lines below 2.4 $\mu$m. In this wavelength range, \cite{deugenio_fast-rotator_2024} estimated a FWHM Point Spread Function (PSF) of $\sim$ 0.1\arcsec.

We reduced the data following the same procedure as described in \cite{perna_ultradense_2023}; here we summarize the most relevant details. We used the STScI pipeline v1.8.2 with CRDS context 1063. A patch was incorporated to address specific pipeline problems that affect this specific version of the pipeline; $1/f$ noise and outliers were corrected following the approaches described in \cite{perna_ultradense_2023} and \cite{deugenio_fast-rotator_2024}, respectively. The combination of exposures for the four dither positions utilized a drizzle weighting technique, as detailed in \citealt{law_3d_2023}, which enabled sub-sampling of detector pixels. This process yielded cube spaxels measuring $0.05\arcsec$ each, approximately corresponding to 25 pc/spaxel and covering a field of view (FoV) of $3.7\arcsec\, \times \, 3.7\arcsec$ ($1.9 \, \times \, 1.9$ kpc$^2$). 
The reduced datacubes showed the so-called "wiggles", sinusoidal modulations due to the PSF undersampling by the NIRSpec detectors, affecting mostly the brightest spaxels in the nuclear regions. We corrected them using the same procedure as described in \cite{perna_ultradense_2023}, applying it on the two nuclear regions separately.

For NGC 6240 we measured a redshift of z = 0.02474 $\pm$ 0.00011 (D = 106 Mpc) from a spectrum extracted from an aperture of 30-spaxel radius, centered on the NIRSpec FoV. It was computed as the centroid of the brightest molecular line, that is H$_2$ 1-0 S(1), modeled using 2 Gaussian profiles. This value is compatible with z = 0.0245 by \cite{muller-sanchez_two_2018}. We will use our measurement in the rest of this paper.

\begin{figure*}[!]
    \centering
    \includegraphics[width=1.\linewidth]{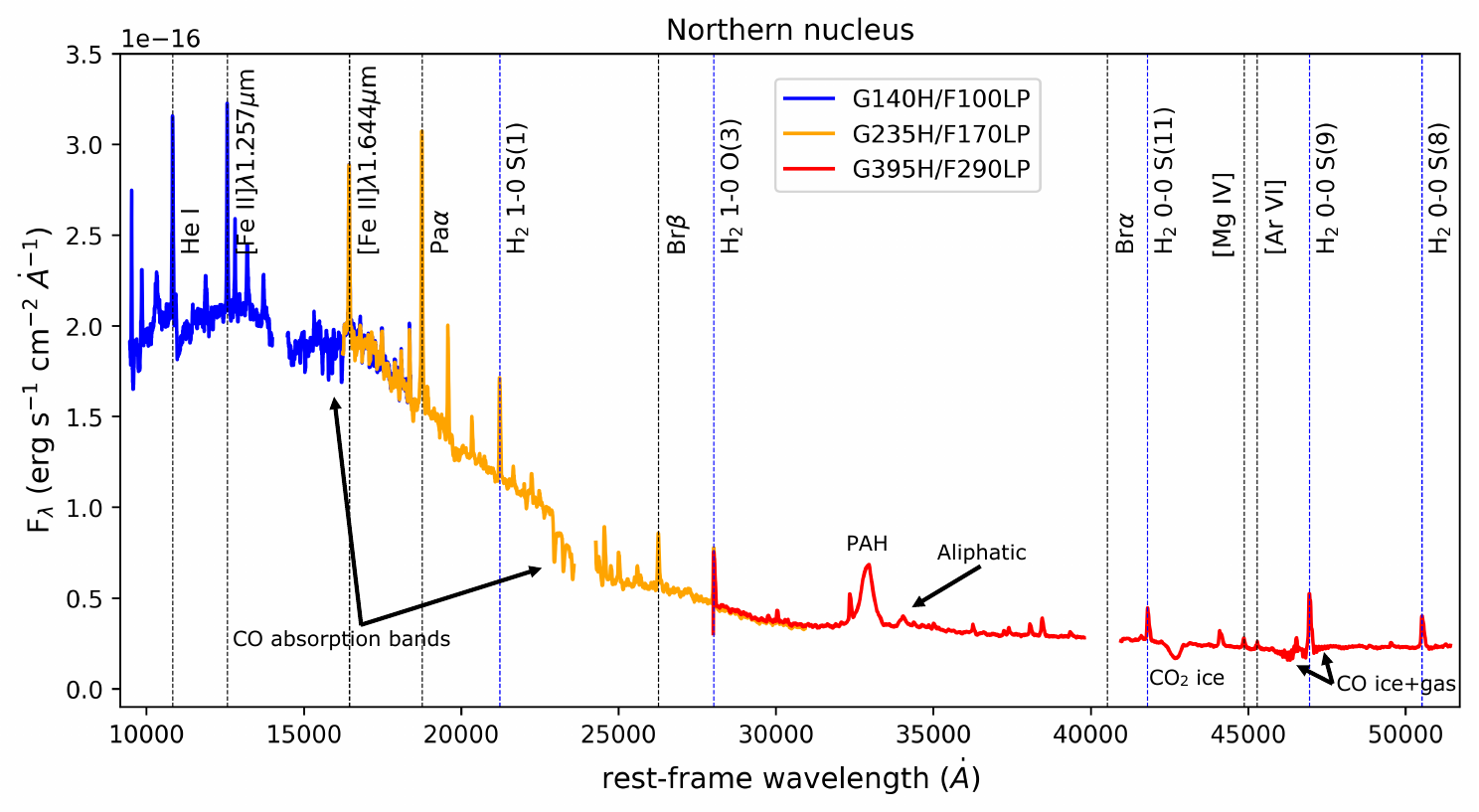}
    \caption{Spectrum of the northern nucleus, integrated over an aperture of 5-spaxel radius (r\,=\,0.25$\arcsec$ corresponding to $\sim$130 pc).  The blue curve is the spectrum from G140H/F100LP (cube1), the orange one from G235H/F170LP (cube2), and the red one from G395H/F290LP (cube3). Each cube spectrum shows a discontinuity in the middle of the band, which is due to the gap between the NIRSpec detectors. We indicated the brightest emission lines with vertical dashed lines (in blue those of molecular hydrogen, in black the others) and the more noticeable features, as the CO absorption bands, polycyclic aromatic hydrocarbons (PAH) and aliphatic emission, and  CO$_2$ ice absorption.}
    \label{fig: tot spectrum entire N}
    \vspace{0.2cm}
\end{figure*}
\begin{figure*}[!]
    \centering
    \includegraphics[width=1.\linewidth]{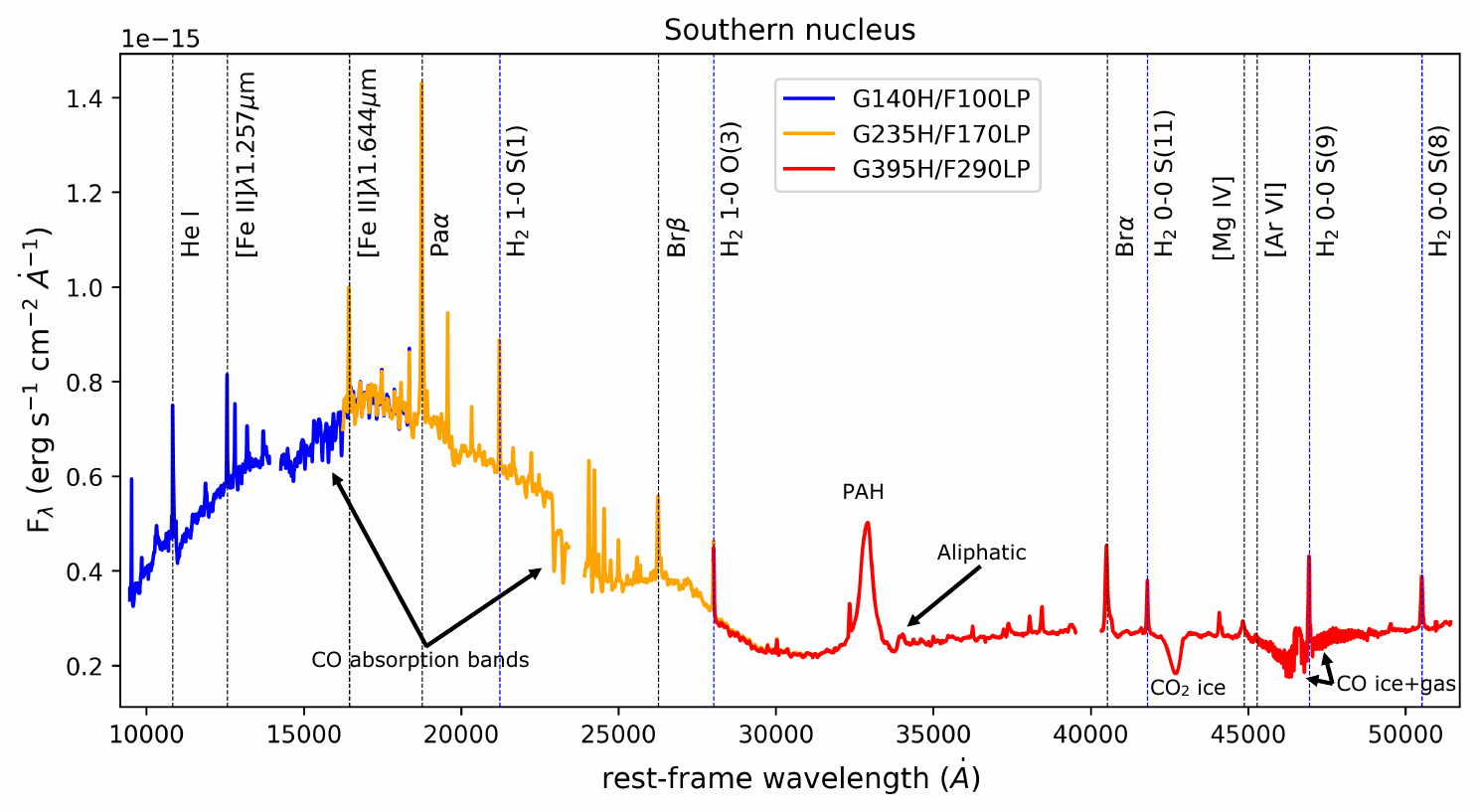}
    \caption{Spectrum of the southern nucleus integrated over an aperture of 5-spaxel radius (r\,=\,0.25$\arcsec$ corresponding to $\sim$130 pc). See Fig. \ref{fig: tot spectrum entire N} for further details.}
    \label{fig: tot spectrum entire S}
\end{figure*}

To correct the datacubes for astrometry, we assume that the peak of Pa$\alpha$ in the southern nucleus to be coincident with the southern radio source observed by \cite{gallimore_parsec-scale_2004}, because we expect the emission line to peak at the position of the AGN. We applied astrometric corrections to the three NIRSpec cubes, resulting in shifts of $\Delta$RA $= -0.2252\arcsec$ and $\Delta$DEC $= +0.1750\arcsec$. These offsets are consistent with the expected accuracy of JWST pointing in the absence of a dedicated target acquisition procedure (\citealt{boker_-orbit_2023}).

To illustrate the quality of the spectral information contained in the NIRSpec datacubes, we show respectively in Figs. \ref{fig: tot spectrum entire N} and \ref{fig: tot spectrum entire S} spectra of the northern and southern nucleus integrated over an aperture of radius 5 spaxels (r\,=\,0.25$\arcsec$ corresponding to $\sim$130 pc), with labels of the main emission lines and features. The discontinuities in the middle of the three bands are due to the gap between the NIRSpec detectors. The overlapping wavelength ranges at $\sim$1.85 $\mu$m and $\sim$3 $\mu$m show a good match between the different spectra, demonstrating a consistent flux calibration. 

\subsection{Narrow-Field Mode MUSE data}
In order to compute electron density from the optical [S II] doublet (see Section \ref{sect: density}), we also used data obtained and reduced by \cite{kollatschny_ngc6240_2020}\footnote{The reduced datacubes are available at the CDS via anonymous ftp to \hyperlink{blue}{cdsarc.u-strasbg.fr (130.79.128.5)} or via \hyperlink{blue}{http://cdsarc.u-strasbg.fr/viz-bin/cat/J/A+A/633/A79}} on April 22, 2018, as part of the commissioning run of the MUSE (Multi Unit Spectroscopic Explorer) instrument in its Narrow-Field Mode (NFM) using the four-laser adaptive optics system of ESO’s Very Large Telescope (VLT) unit telescope four (“Yepun”). They took four 500 s science exposures centered on NGC 6240 and another 500 s exposure on an offset sky field. The NFM covers a field of 7.5\arcsec $\times$ 7.5\arcsec on the sky. Between each on-target exposure, they rotated the FoV by 90 degrees. Spatial offsets of about 1.5\arcsec provided significantly larger coverage of $\sim$ 11.5\arcsec $\times$ 11.5\arcsec with some gaps at the field edges. The wavelength range covered is 4725 to 9350 $\AA$, with a spectral resolution of approximately 2.5  $\AA$. The spectra are sampled at 1.25  $\AA$ in the dispersion direction and 0.0253\arcsec in the spatial direction. However, there is a gap in the spectrum between 5780 and 6050 $\AA$ due to the notch filter to exclude the laser-guide sodium emission. The data reduction is described in detail in \cite{kollatschny_ngc6240_2020}. Similar to NIRSpec data, we used peaks of H$\alpha$ to correct MUSE observations for astrometry. Corrections result in shifts of $\Delta$RA $= +1.3944\arcsec$ and $\Delta$DEC $= -1.7892\arcsec$.

\section{Analysis}
\label{sec:Analysis}
In this section we describe the methods used to analyze the kinematics and properties of the emitting gas, following \cite{marasco_galaxy-scale_2020}, \cite{tozzi_connecting_2021}, and \cite{cresci_bubbles_2023}. We divided the fitting procedure into three steps: firstly, we subtracted the stellar contribution from the cubes; then we applied a multiple-component fitting procedure, dividing the gas species into two kinematically independent families; at the end we separated the emission of the narrow component (for example, disk and/or merger debris) from the broad component (outflow and/or turbulent gas). These three steps are described in detail in the following sections.

We applied two consecutive spectral fittings: firstly, we fit the full datacubes aiming to isolate and remove stellar and AGN continuum emission (see Section \ref{sec:stellar fit}); secondly, we modeled the gas emission lines in the continuum-subtracted cubes (see Section \ref{sec:lines fitting}) via a refined modeling with more freedom on the choice of line profiles and model constraints.

\subsection{Modeling of the stellar continuum}
\label{sec:stellar fit}
We applied the penalized PiXel-Fitting (pPXF; see \citealt{cappellari_parametric_2004}) package to model the continuum and gas emission, convolving a linear combination of stellar templates with a Gaussian velocity distribution. First, to increase the signal to noise ratio (S/N) of our datacubes, we used the Voronoi binning algorithm by \cite{cappellari_adaptive_2003}. The applied S/N threshold per wavelength channel for each bin was 5 for both cube1 and cube2.
Regarding the stellar contribution, we adopted the single stellar population models of \cite{maraston_stellar_2011}, which have a resolution of R = 20000 and a spectral range of [0.1 $\mu$m, 2.5 $\mu$m]. To take into account also the continuum from AGN, we added to the model a 3-degree polynomial. We also fit all the emission lines listed in Table \ref{tab: fitted lines}. We built a model for the emission lines, with a variable number of Gaussian components (from 1 to 3, because along the line of sight, we can have different kinematic features) and we added it to the previous stellar model, thus obtaining our total model. The kinematic parameters (v and $\sigma$) are the same for all  emission lines. We fit each spectrum three times (with 1, 2, or 3 Gaussian components), and then applied a Kolmogorov-Smirnov (KS) test on the residuals to choose the optimal number of Gaussian components in each spaxel, with  a threshold p-value of 0.5 (\citealt{marasco_galaxy-scale_2020}). In practice, we test whether adding an extra component (n + 1) results in residuals that are statistically different from those of a model with n components, starting with n = 1. If there is a significant difference, we increase n by 1 and repeat the test. If not, we consider n to be the optimal number of components for that specific spaxel. This method is independent of any particular likelihood estimator and focuses solely on the relative improvements that an additional component brings to the model. The tolerance for accepting a new component can be adjusted by changing the p-value threshold of the KS test.

\begin{figure*}[t!]
    \centering
    \includegraphics[width=1.\linewidth]{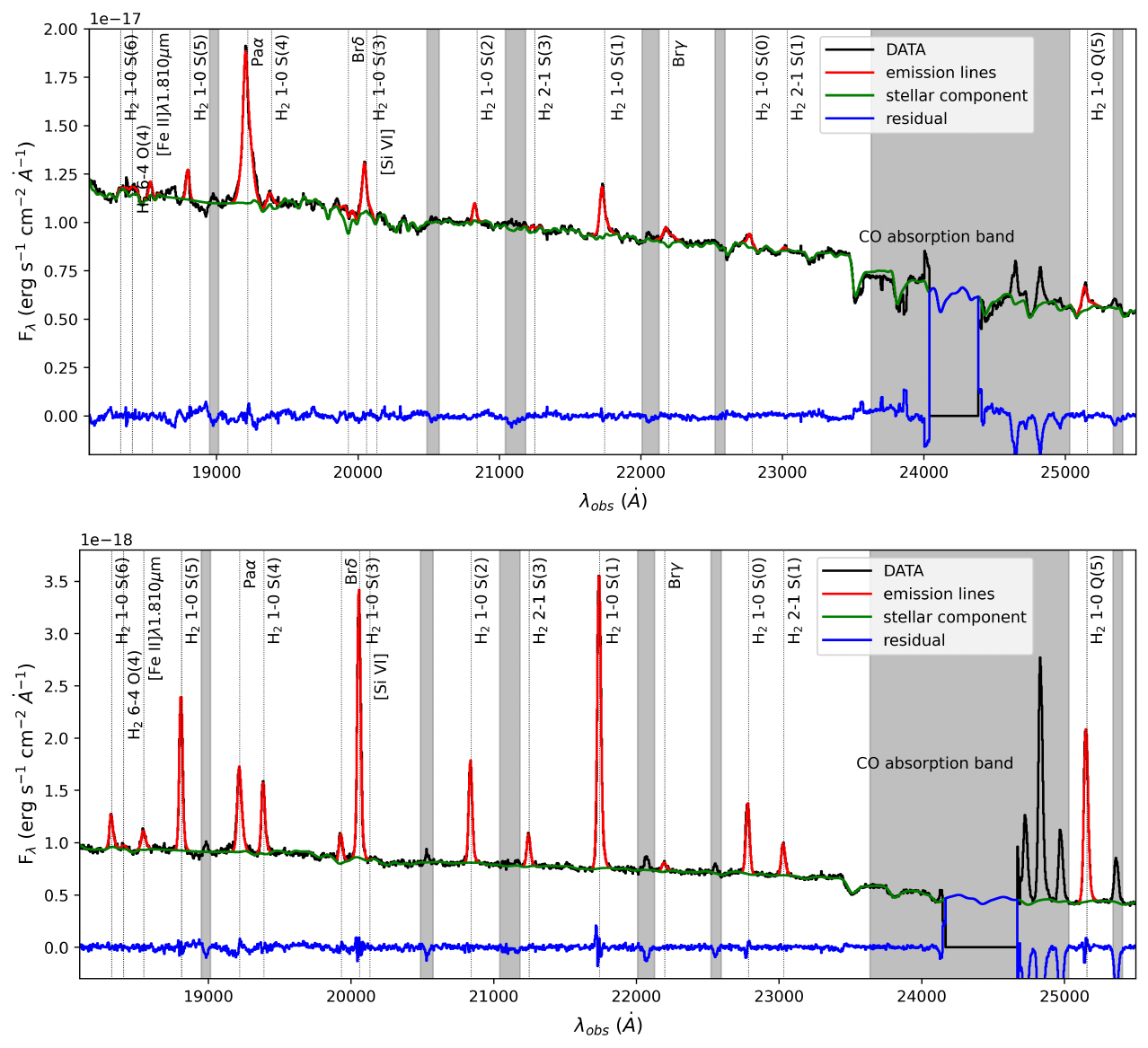}
    \caption{Examples of continuum and emission lines fitting in a spaxel from the southern nucleus (top panel) and one from the region between the nuclei (bottom panel) with three Gaussian components for each line. The data, the emission lines model, the stellar model, and the residuals are shown in black, red, green, and blue respectively. Gray bands indicate spectral regions excluded from the fit. Dashed vertical lines mark the emission lines modeled with pPXF fitting (listed in Table \ref{tab: fitted lines}).}
    \label{fig: ppxf example}
\end{figure*}

Examples of fit stellar component are shown in Fig. \ref{fig: ppxf example}, where we plot the fit cube2 spectra of a spaxel centered on the southern nucleus (upper panel) and another located in the region between the two nuclei (lower panel). In both, we show data (in black), the total model of emission lines and stellar component (in red), the model of only stars (in green), and the residuals between data and total model (in blue). The vertical lines mark all features modeled with pPXF procedure, while the gray bands show the spectral regions we masked during the fitting procedure, corresponding to the gap between the detectors and covering a few emission lines not relevant to this work. We focused on fitting only the most relevant emission lines to ensure a reasonable computational time. We note that the detector gap causes many wavelength channels to be lost and thus prevents a complete fit of the CO band-head absorption features longward of 2.4 $\mu$m. In addition, an imperfect wiggles correction has probably spoiled these features. To be more conservative, we decided to include in the fit only the first band-head ($\sim$2.35 $\mu$m in the observed frame), that is the one furthest from the spectral discontinuity due to the gap between the NIRSpec detectors, the one with fewer residuals by wiggles correction (see Fig. \ref{fig: ppxf example}). The main differences between the spectra of these two regions are the following. In the nucleus, the continuum flux is about one order of magnitude higher, and the predominant emission line is Pa$\alpha$, while in the region between the nuclei the molecular hydrogen emission lines are the strongest. The somewhat higher residuals in the nuclear spaxel are likely due to an imperfect wiggles correction. The cube1 spectra of the same spaxels, which have lower S/N, are shown in Appendix \ref{sec: ppxf cube1}.

\begin{table}[b!]
    \begin{center}   
    \caption{ List of all emission lines used during the fitting procedure in NIRSpec datacubes.}

    \tabulinesep=1.2mm
    \tabcolsep 4.2pt
    \begin{tabu}{|lcc|lcc|}
    \hline
    line & $\lambda_{vac}$ ($\AA$) & group & line & $\lambda_{vac}$ ($\AA$) & group \\
    \hline
    $[$C I$]$   &  9827          & 1 & H$_2$ 1-0 S(6) & 17880           & \\
    $[$C I$]$   &  9850          & 1 & H$_2$ 6-4 O(4) & 17960           & \\
    Pa$\delta$     &   10052      &   &  $[$Fe II$]$ &  18099            & \\
    He I   &     10314           &   & H$_2$ 1-0 S(5) & 18358           & \\
    $[$Fe III$]$     &  10806     &   & Pa$\alpha$ & 18756              & 1 \\
    He I     &   10832           &   & H$_2$ 1-0 S(4) & 18920           & 2 \\             
    H$_2$ 2-0 S(5)    &  10851    &   & Br$\delta$ & 19451              & 1 \\
    Pa$\gamma$     & 10941        & 1 & H$_2$ 1-0 S(3) & 19576           & 2 \\
    $[$P II$]$   &  11886        & 1 & $[$Si VI$]$ & 19625              & 1 \\ 
    $[$Fe II$]$   &   12567      & 1 & H$_2$ 1-0 S(2) & 20338           & 2 \\
    Pa$\beta$     & 12821        & 1 & H$_2$ 2-1 S(3) & 20735           & \\
    $[$Fe II$]$     &  15335      &   & H$_2$ 1-0 S(1) &  21218          & 2 \\
    $[$Fe II$]$     &  16440      & 1 & Br$\gamma$ & 21661              & 1 \\
    $[$Fe II$]$     &  16773      & 1 & H$_2$ 1-0 S(0) & 22235           & 2 \\
    H$_2$ 1-0 S(9)   &  16877     & 2 & H$_2$ 2-1 S(1) & 22477           & 2 \\
    H$_2$ 1-0 S(7)   &  17480     & 2 &  H$_2$ 1-0 Q(5) & 24548          & \\ 
                 
    \hline
    \end{tabu}
    \label{tab: fitted lines}
    \end{center}
\tablefoot{See Section \ref{sec:stellar fit} for details of the pPXF modeling. The "group" label defines the two line species used in the fitting procedure part (see Section \ref{sec:lines fitting}).}
\end{table}

\begin{figure*}
    \centering
    \includegraphics[width=.95\linewidth]{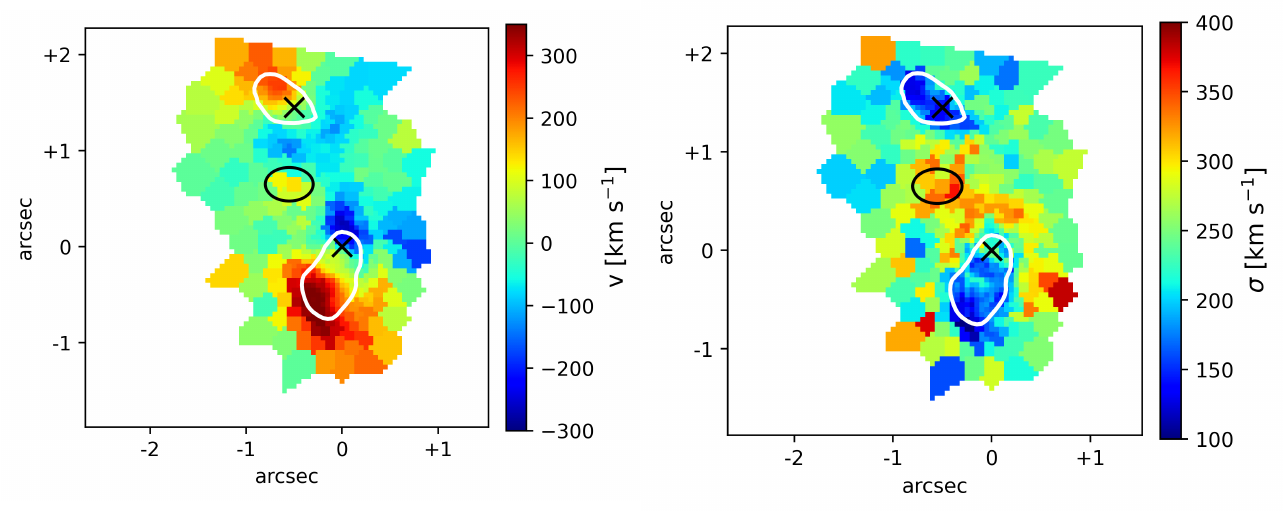}
    \caption{Velocity (left panel) and velocity dispersion (right panel) maps of the stellar component derived from cube2 with the pPXF fitting. White contours tracing continuum are overplotted as a spatial reference (for the southern nucleus we used 19$\%$ of the 1.2 $\mu$m continuum maximum and for the northern nucleus 6$\%$ of the 2.3 $\mu$m continuum maximum). Crosses and the ellipse indicate respectively the nuclei and the high-velocity molecular hydrogen blob positions (see Section \ref{sect: H2 pallocchio}).
    From the velocity map, we can distinguish the two stellar rotating disks in the proximity of the two nuclei.}
    \label{fig: star maps}
\end{figure*}
We plot in Fig. \ref{fig: star maps} the velocity and velocity dispersion maps of the stellar component, derived from cube2 with pPXF fitting. For these and the other maps in the rest of the paper, north is up and east is left. The velocities are calculated with respect to the systemic one, derived from our fiducial redshift (see Section \ref{sec:Obs}). From the stellar velocity map we can distinguish velocity gradients likely associated with the rotating stellar disks of the two nuclei. The major axis of the southern disk has a position angle\footnote{The position angle is measured counterclockwise from the north direction on the approaching part of the disk.} (PA) of 321.4$^\circ \pm 6.4^\circ$, estimated by fitting the star velocity map with the FIT KINEMATIC PA code (\citealp{cappellari_sauron_2007}, \citealp{krajnovic_atlas3d_2011}). The disk has positive velocities - which means stars moving away from the observer -  reaching a measured value of +350 km s$^{-1}$ in the south-east side, while the side approaching the observer is the north-west one, with a maximum of -250 km s$^{-1}$ with respect to the systemic velocity. The northern nucleus has more moderate velocities: with PA of 218.6$^\circ \pm 6.4^\circ$, it has the redder part in the NE direction and the bluer one in the SW direction. Respectively, the velocities peaks are +250 km s$^{-1}$ and -150 km s$^{-1}$.
In addition, the southern disk shows asymmetric velocity curve along the major axis, and in both disks the minor axes seem to be curved. These departures from regular rotation are likely related to gravitational interactions associated to the merging process.

Also the velocity dispersion map is not simply compatible with the presence of two disks, for which we would expect to observe the highest dispersion in the rotation center. Indeed, we find the lowest velocity dispersion of 100 - 150 km s$^{-1}$ south of the southern nucleus and east of the northern nucleus, while the highest values of about 350 - 400 km s$^{-1}$ are found between the nuclei and in a region to the west of the southern nucleus. These anomalously high dispersions is expected in a merger system. Indeed, these features could be attributed to emission from stars recently formed due to the interaction, which is superimposed on the light from the nucleus and leads to an overestimate of the velocity dispersion. These maps are derived from the cube2 data, but similar and compatible results are found using cube1. 

The stellar kinematics of the inner region of NGC 6240 have already been investigated with NIR IFS data from VLT/SINFONI and Gemini (e.g., \citealt{engel_ngc6240_2010,ilha_origin_2016}). Fitting the CO absorptions band-heads at $\sim$ 2.3 $\mu$m, \cite{engel_ngc6240_2010} and  \cite{ilha_origin_2016} found a velocity amplitude of $\sim$ 600 km s$^{-1}$ for the southern nucleus and $\sim$ 350 km s$^{-1}$ for the northern one, comparable to our findings  of 600 km/s and 400 km/s for the northern and southern nuclei, respectively. Their velocity dispersion maps are similar to ours with values in a range of $\sim$ 150 km s$^{-1}$ and $\sim$ 400 km s$^{-1}$ in both papers. Regarding the PA of the stellar disks, \cite{engel_ngc6240_2010} derived 329$^\circ \pm 2^\circ$ and 229$^\circ \pm 2^\circ$ for the southern and northern nucleus, respectively. Instead, \cite{ilha_origin_2016} estimated PA $\sim$ 330$^\circ$ and PA $\sim$ 225$^\circ$. Both estimations are comparable with ours within the 2-sigma uncertainties.

Having the stellar model for every spaxel in hand, we subtracted it from the original spaxel spectra, in order to obtain cubes containing only emission lines from the ISM of the galaxy. More precisely, we subtracted the stellar emission from the original unbinned datacube by rescaling the fit stellar model (constant within each Voronoi bin) to the original unbinned spectrum of each spaxel before subtracting it.

\subsection{Multi-Gaussian emission line fit}
\label{sec:lines fitting}
After the subtraction of the stellar component, we have datacubes containing just the emission of the ionized, neutral, and molecular gas.
In Section \ref{sec:stellar fit}, we used Voronoi binning to combine spaxels and trade spatial resolution to increase S/N. For our emission line analysis, we instead choose to smooth each spatial plane of both our cubes with a Gaussian kernel of size $\sigma_{smooth}$ = 1 pixel, corresponding to 0.05$\arcsec$. This kernel is smaller than the PSF (on the order of 0.1$\arcsec$, \citealt{deugenio_fast-rotator_2024}), and degrades the resulting spatial resolution only minimally.

At this point, we performed the emission line fitting spaxel-by-spaxel. The spectra in the continuum-subtracted datacube were refit for the following reasons: although we previously performed the emission line fitting using the pPXF package in Section \ref{sec:stellar fit}, we are now applying a more refined fitting process with additional constraints and line profile options.
We adopted again the strategy of a multiple components fitting procedure, to better reproduce the more asymmetric line profiles. In addition, we separated emission lines into two families, in order to probe potential kinematic differences between the molecular gas (traced by the roto-vibrational transitions of H$_2$), and the ionized and neutral gas. 
The advantage of this method is that it overcomes the low S/N of some individual lines by fitting all lines within the same group together.
The main disadvantages are that having a relatively high fraction of lines with low S/N may increase the scatter in the results and that we are fitting a large number of lines together, which will be computationally heavy with more components.
The obvious work-around would be to fit the two groups in separate runs, but, as we can see in Fig. \ref{fig: ppxf example}, some spectra contain blended lines with components from different line species that have to be fit simultaneously.

We performed the fit for cube1 including the molecular family of H$_2$ 1-0 S(7) and H$_2$ 1-0 S(9), and the other family of  [C I]$\lambda$0.983$\mu$m, [C I]$\lambda$0.985$\mu$m, [P II]$\lambda$1.189$\mu$m, [Fe II]$\lambda$1.257$\mu$m,  [Fe II]$\lambda$1.644$\mu$m, Pa$\beta$ and Pa$\gamma$. For cube2, we instead fit the group of molecular lines H$_2$ 1-0 S(2), H$_2$ 1-0 S(3), H$_2$ 1-0 S(1), H$_2$ 1-0 S(4), H$_2$ 1-0 S(0) and H$_2$ 2-1 S(1), and the one of [Si VI], Br$\delta$, Pa$\alpha$ and Br$\gamma$. These emission lines are listed in Table \ref{tab: fitted lines}, where the separation in the two kinematically independent groups of both datacubes is indicated. For all emission lines in either group, the fit parameters for the Gaussian profile are the flux, the velocity $v$ with respect to the systemic velocity, and the velocity dispersion $\sigma$. Within each line family and for every spaxel, the fit velocity and velocity dispersion were forced to be identical, while only the line flux was allowed to vary freely between lines.
We added also an additive polynomial for cube1 and cube2, respectively of degree 2 and 3, because, there was some residual continuum contribution in the nuclear spatial regions affected by wiggles.
We carried out the fit using from one to four Gaussian components, to reproduce all the kinematic components along the line of sight of every spaxel, using for each parameter physical and reasonable boundary conditions\footnote{|v| $\leq$ 1000 km s$^{-1}$ and $\sigma$ greater or equal than the NIRSpec spectral resolution.}.

After the fitting procedure, we have to select the optimal number of components in every spaxel (as done in Section \ref{sec:stellar fit}). 
We performed again the KS test, changing this time the p-value. We increased it to 0.9 to reproduce the line profiles with a higher accuracy. As expected from the chaotic kinematics, 3-4 Gaussian components were necessary in the majority of the spaxels to reproduce the highly asymmetric line profiles. Moreover, there is not any particular shape in the spatial distribution of individual Gaussian components that can be used to point out some kinematic structure, such as an outflow cone.

After modeling the emission line gas, we aim to decompose the line profiles. This is done by separating the narrow and broad components based on some reasonable criteria.

\subsection{Decomposition of narrow and broad components}
In this section we describe the method used to distinguish between different kinematic components. Achieving a detailed and precise identification of each component's contribution to the overall line emission spaxel-by-spaxel proves challenging, given the complexity of the velocity structure in the FoV. Hence, we executed a decomposition between broader components possibly linked to outflowing gas, which we refer to as "broad" components, and the others we simply call "narrow". The criterion we applied is as follows: we defined the wings of the total line profile as the tails exceeding a predefined velocity threshold with respect to the velocity peak of the total line profile. Then, we calculated for each Gaussian component the fraction of flux within these wings relative to the total flux of the Gaussian component. If this fraction exceeds a fixed threshold, the component is classified as broad. Otherwise, it is classified as narrow. The parameters of this method are the velocity threshold that defines the wings of the line, and the fraction of the total flux of the component. We used v$_{wings}$ = 200 km s$^{-1}$ and f$_{wings}$ = 0.5 for both datacubes. This means that a component is classified as 'broad' if more than 50\% of its flux is at velocities $>$ 200 km s$^{-1}$ w.r.t. the peak of the total line profile.
Our classification criterion relies on the fact that outflows typically generate broader and higher-velocity line emission (\citealt{marasco_galaxy-scale_2020}, \citealt{tozzi_connecting_2021}), contrasting with the other potential components, which usually produce narrow line emission near the galaxy's systemic velocity.

\begin{figure*}[t!]
    \centering
    \includegraphics[width=1.\linewidth]{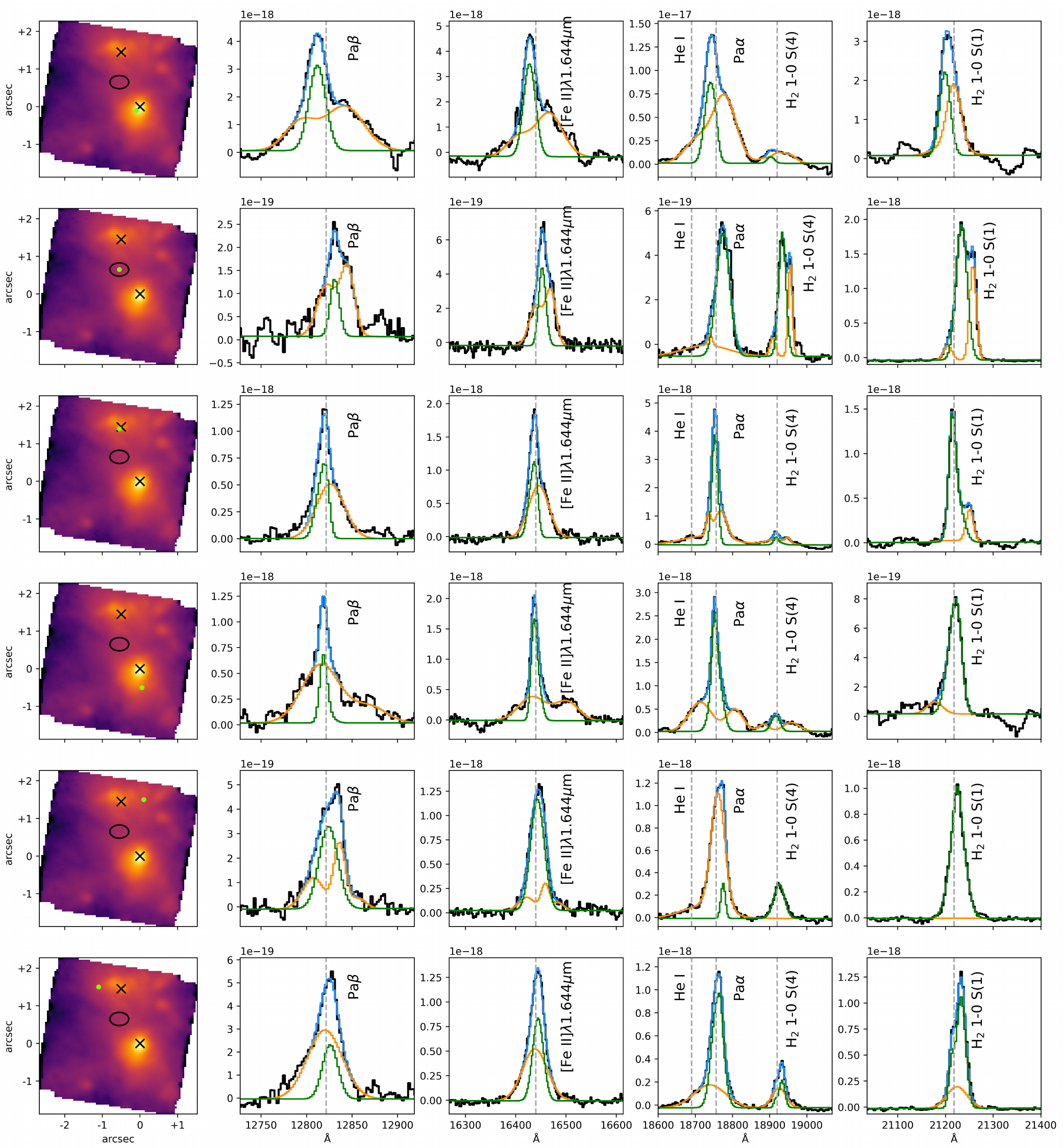}
    \caption{Examples of the decomposition of narrow and broad components at the position of some spaxels. The spaxels are shown with a green point in the left panels (which are Pa$\alpha$ maps in logarithmic scale). Black crosses and the black ellipse indicate respectively the nuclei and the high-velocity molecular hydrogen blob positions. The other panels show the results of the decomposition for the Pa$\beta$, [Fe II]$\lambda$1.644$\mu$m, Pa$\alpha$, H$_2$ 1-0 S(4), and H$_2$ 1-0 S(1). Black curve shows the observed emission lines after the subtraction of stellar continuum, the blue curve shows the total result of the fit, the green and orange curves mark the narrow and broad components, respectively. The systemic velocity is represented by a gray dashed line. The wavelengths on the x-axis are in the rest-frame. The y-axis units are is erg s$^{−1}$ cm$^{-2}$ \AA$^{-1}$ }.
    \label{fig: NvsB spectra example}
\end{figure*}
We present some examples of decomposition in Fig. \ref{fig: NvsB spectra example}. In every row we show the spectrum of the spaxel marked with a light green point in the first panel on the left, that is the image of the Pa$\alpha$ flux in logarithmic scale. 
In the other panels, we plot the continuum-subtracted spectrum, the total model, the model of the narrow components, and the model of the broad components (in black, blue, green, and orange, respectively) of the brightest emission lines. The narrow and broad sub-components are grouped to better analyze the global properties of the low- and high-velocity gas. 
The vertical lines represent line wavelength corresponding to the galaxy systemic velocity. As we can see, the shape and relative strength of the two components vary substantially across the FoV. We also find that the ionized gas always has a broad component, while the molecular gas sometimes has only narrow components. As can be seen in Fig. \ref{fig: NvsB spectra example} the Pa$\alpha$ is sometimes blended with He I. We decided not to fit this latter because it is negligible and when 3-4 Gaussian components are used, some He I components are classified as Pa$\alpha$ outflow (and vice versa).

\subsection{Nuclei positions}

As shown by \cite{max_core_2005}, we also observed that the apparent location of the nuclei shifts as longer wavelengths enable us to peer deeper into the dust to reveal the true continuum centers. Using adaptive optics facilities from the ground, they concluded that in the K' band the AGN in the southern nucleus is still enshrouded in dust and they tentatively identified the northern nucleus position. Our longer wavelength JWST/NIRSpec data enable us to identify the true position of both AGN.

\begin{figure*}[t!]
    \centering
    \includegraphics[width=.9\linewidth]{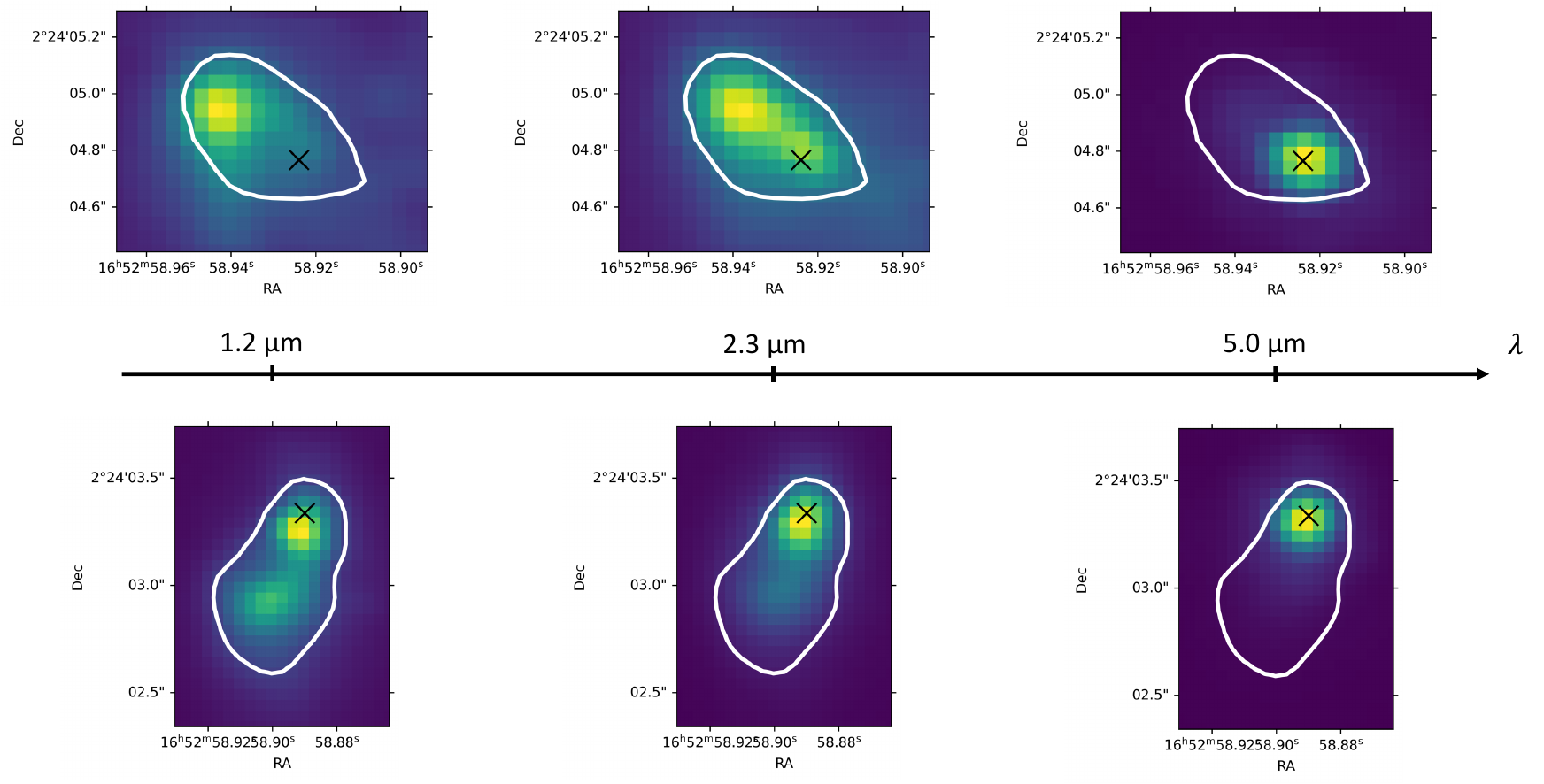}
    \caption{Zoomed-in continuum maps of the northern (top) and southern (bottom) nuclear region for different rest-frame wavelengths, as indicated by the axis. Crosses indicate the positions of the nuclei. White contours of the continuum flux are overplotted as a spatial reference (for the southern nucleus we used 19$\%$ of the 1.2 $\mu$m continuum maximum and for the northern nucleus 6$\%$ of the 2.3 $\mu$m continuum maximum). For every panel, the flux is normalized in the zoomed FoV.}
    \label{fig: 4 sources}
\end{figure*}

In each column of Fig. \ref{fig: 4 sources}, we show the continuum images of the datacubes, zooming in on the nuclei positions. White contours of the continuum are overlaid as spatial references, using the 1.2 $\mu$m continuum for the southern nucleus and the 2.3 $\mu$m continuum for the northern nucleus. The northern nucleus is shown in the top panels and the southern nucleus in the bottom. From left to right, the columns present the continuum at 1.2 $\mu$m, 2.3 $\mu$m, and 5.0 $\mu$m (rest frame) obtained with a window of $\pm 0.01 \mu$m. 
We can easily observe that the continuum emission peaks shift across the NIR spectral range. Looking at the northern nucleus, we show that the peak position moves from the northeastern position - observable in the optical band (\citealt{gerssen_hubble_2004}) - to the southwestern one, which matches the radio position (\citealt{gallimore_parsec-scale_2004}). Regarding the southern nucleus, in the visible band two sub-components are present in the nuclear region. As we move to redder wavelengths, the southern component fades, while the northern component shifts slightly to the north until it reaches the position of the radio nucleus at 5.0 $\mu$m.
We thus confirm the variation of the position of the continuum peaks through infrared wavelengths, as also reported in previous works (e.g., \citealt{gerssen_hubble_2004,max_core_2005,max_locating_2007}; see also \citealt{scoville_arp220_1998} and \citealt{perna_no_2024} for a similar behavior in Arp 220). 
We verified that the continuum peaks at the longest NIR wavelengths correspond to the positions of X-ray and radio sources, confirming the results of \cite{max_locating_2007} and the validity of our initial choice to use the Pa$\alpha$ peak to mark the AGN positions.

\subsection{Extinction correction}
\label{sec: extinction}

\begin{figure}[t!]
    \centering
    \includegraphics[width=.88\linewidth]{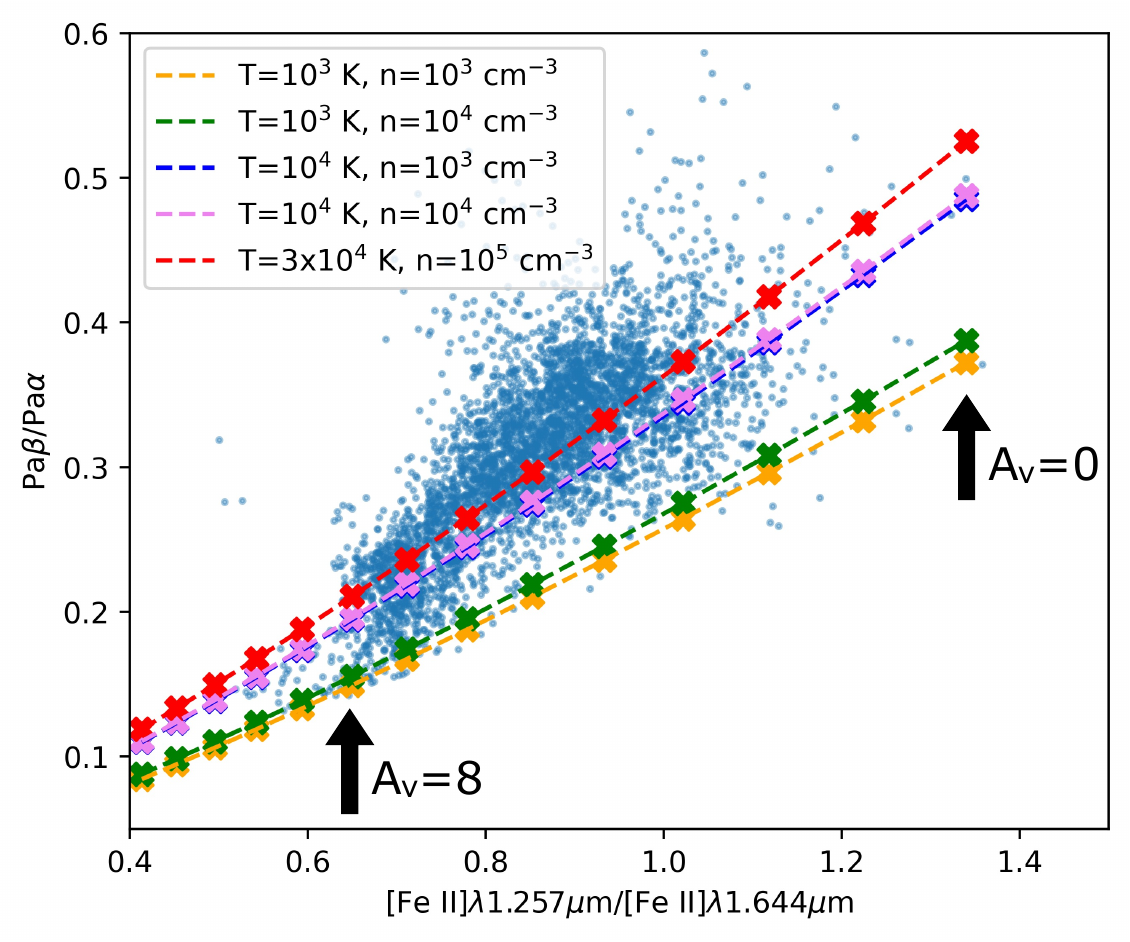}
    \caption{Line ratios diagram used to compute the extinction. Blue points are the observed line ratios of Pa${\beta}$/Pa${\alpha}$ and [Fe II]$\lambda$1.257$\mu$m/[Fe II]$\lambda$1.644$\mu$m. Dashed lines fit extinction values computed assuming theoretical line ratios based on the \cite{cardelli_relationship_1989} extinction law and different electronic temperatures and densities (as indicated). The symbols along the model lines mark optical extinction values in steps of A$_\text{V}$=1, increasing from right to left.} 
    \label{fig: extinc calcul}
\end{figure}

\begin{figure}[t!]
    \centering
    \includegraphics[width=1\linewidth]{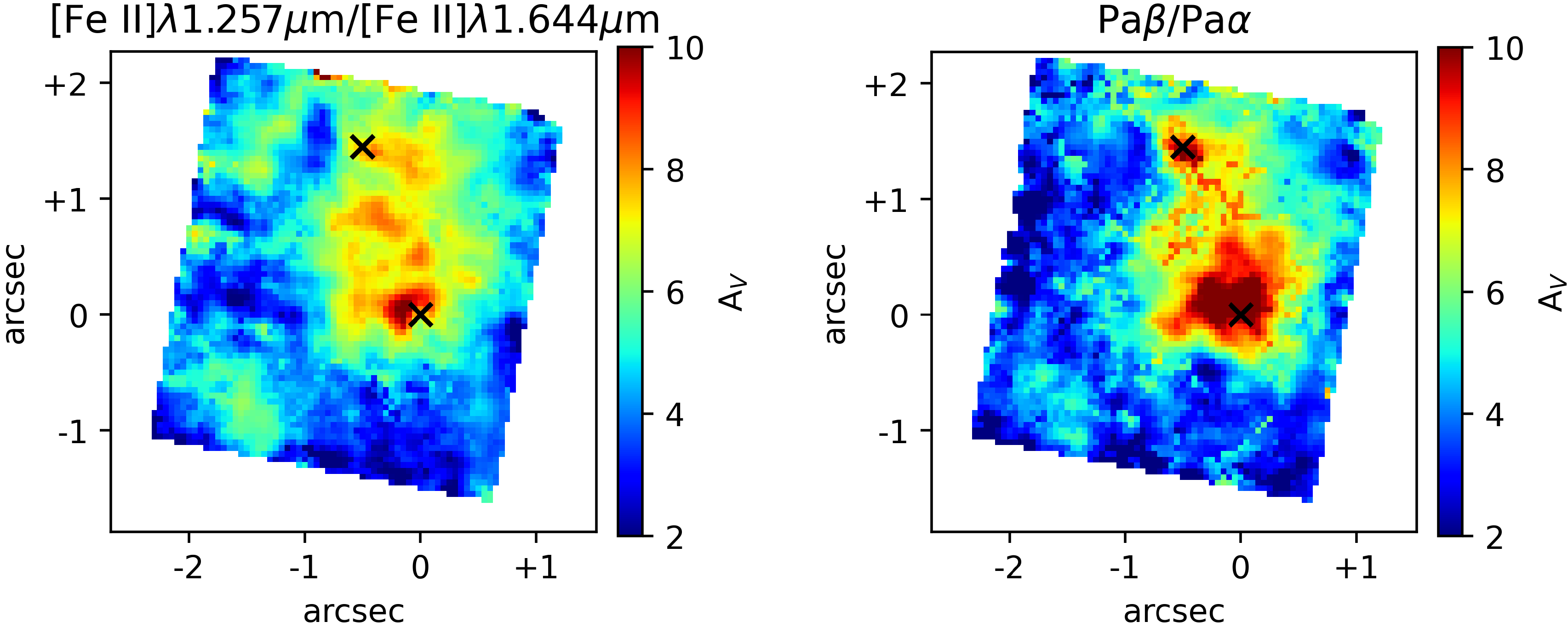}
    \caption{Extinction maps obtained from the iron line ratio (\textit{left panel}) and the Paschen decrement (\textit{right panel}) with the extinction law of \cite{cardelli_relationship_1989}. Crosses indicate the positions of the nuclei.} 
    \label{fig: extinc maps Fe Pa}
\end{figure}

\begin{figure}[t!]
    \centering
    \includegraphics[width=1\linewidth]{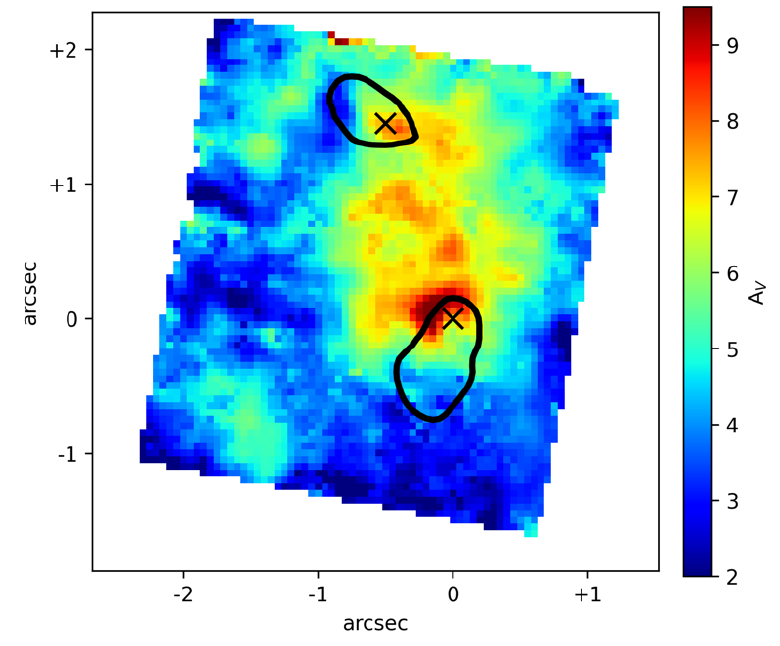}
    \caption{Extinction map used for dust correction of data. It was obtained as a combination of the extinction maps of the iron line ratio and the Paschen decrement of Fig. \ref{fig: extinc maps Fe Pa} (see text in Section \ref{sec: extinction} for details).
    Crosses indicate the positions of the nuclei. Black contours are the same as those in Fig. \ref{fig: 4 sources}, and indicate the regions where most of the continuum emission originates.}
    \label{fig: extinction map}
\end{figure}

Following \cite{almeida_near-infrared_2009}, we estimated the extinction using the Pa${\beta}$/Pa${\alpha}$ and the [Fe II]$\lambda$1.257$\mu$m/[Fe II]$\lambda$1.644$\mu$m line ratios. In Fig. \ref{fig: extinc calcul}, we plot the observed Paschen and [Fe II] line ratios and then we overplot inferred values for fixed A$_V$, assuming a range of physically reasonable values of gas density and temperature. We used the \cite{cardelli_relationship_1989} extinction law, valid in the range 0.9\,$\mu$m\,$\leq$\,$\lambda$\,$\leq$ 3.3 $\mu$m,
\begin{equation*}
    A(\lambda) = A_V \, (a(\lambda) + b(\lambda)/R_V),
\end{equation*}
where
\begin{align*}
    a(\lambda) & = 0.574 \, \lambda^{-1.61},\\
    b(\lambda) & = 0.527 \, \lambda^{-1.61},
\end{align*}
and R$_V$ = 3.1, which is the standard value for the diffuse ISM. As shown in the plot, while iron ratios are not very sensitive to line-emitting gas density and temperature, this is not true for the recombination lines (\citealt{almeida_near-infrared_2009}). The theoretical values of the two line ratios  were taken from \cite{Hummer&Storey_recombination_1987} for Paschen lines and \cite{bautista_ionization_1998} for [Fe II] lines. As shown in the figure, the model with T$_e$ = $3 \times 10^4$ K and n$_e$ = 10$^5$ cm$^{-3}$ (i.e., red curve) reproduces best the data.
In Fig. \ref{fig: extinc maps Fe Pa} we plot the extinction maps obtained from the Paschen decrement and the [Fe II] lines. As can be noticed, there are discrepancies between the reddening estimations derived with the two different line ratios, as discussed in \cite{riffel_0824_2006}: given that the intrinsic ratio of the recombination lines is likely influenced by high density and radiative transfer effects, it is difficult to determine which species more accurately reflects the actual extinction in the galaxy. Therefore, we used both line ratios to estimate the extinction. After observing that the dispersion from the red curve is due to the noise of the spaxels with the lowest S/N, we projected all the observed points perpendicularly to the model curve and calculated the extinction for every spaxel. We show the extinction map used in the following in Fig. \ref{fig: extinction map}.

We observed the maximum extinction of A$_V$ $\approx$ 10 near the southern nucleus. This indicates the highest concentration of dust, which extends from there to north and east directions. Regarding the northern nucleus, we found A$_V$ $\approx$ 8. We observed patchy dust located between the nuclei, which leads to a mean value of  A$_V$ $\approx$ 7 in that region. Instead, in the more external regions, we find an A$_V$ range of 2-5. The fact that the higher values of extinction are measured at the locations of the AGN confirms that the variation of continuum peaks with wavelength is (at least in part) due to obscuration effects observed in Fig. \ref{fig: 4 sources}. We point out the connection between the extinction gradient inside the continuum contours and the variation of continuum peaks with wavelength (observable in Fig. \ref{fig: 4 sources}). Indeed, focusing on these regions, the less extincted parts correspond to the location of the peak in the blue wavelengths, while the more reddened zones coincide with the AGN. 

Consistent extinction values have been obtained from the stellar continuum from pPXF fitting. We used spectra from 1 to 3 $\mu$m, masking out all the gas emission lines and fitting only the continuum without any additive/multiplicative polynomial, finding A$_V$ $\approx$ 8 on the nuclei position and A$_V$ $\approx$ 6 in the region between them. These extinction values are consistent with those provided by the gas.

\cite{lutz_iso_2003} estimated the dust absorption toward NGC~6240 in the MIR-FIR spectral range from ISO SWS data with a FoV of 20\arcsec$\,\times\,$20\arcsec, and found values of A$_V$ $\approx$ 15-20. The discrepancy with our results is probably due to the different spectral range used, rather than the different spatial regions. Our measurements only probe to the depth at which the nuclei become optically thick in the K band. If we are not probing all the way into the nuclei, measurements at longer wavelengths will result in higher inferred values of A$_V$. Using B-K colors from HST, \cite{gerssen_hubble_2004} estimated A$_V$ $\geq$ 3-5 at the position of the northern peak and the two southern peaks of the continuum at 1.2 $\mu$m in Fig. \ref{fig: 4 sources} (which they call N1, N2, and N3, respectively; see their Fig. 3). Another estimation is that by \cite{pollack_circumnuclear_2007}, who, with Keck II observation in K and H bands, measured extinction from stellar clusters scattered across the galaxy. They found A$_V$ = 8 near the southern nucleus, and A$_V$ = 5 at the location of the northern peak of the bluest continuum map in Fig. \ref{fig: 4 sources}. \cite{kollatschny_ngc6240_2020} estimated extinction with the Balmer decrement and obtained values of A$_V$ $\simeq$ 10 in the nuclear regions. All these estimates are consistent with ours, despite the different estimation methods used. 

After the estimation of the extinction for every spaxel, we can now correct all observed fluxes for reddening. Going forward, we will use extinction-corrected flux values.

\section{Results}
\label{sec:Results}

\begin{figure*}[t]
    \centering
    \includegraphics[width=1\linewidth]{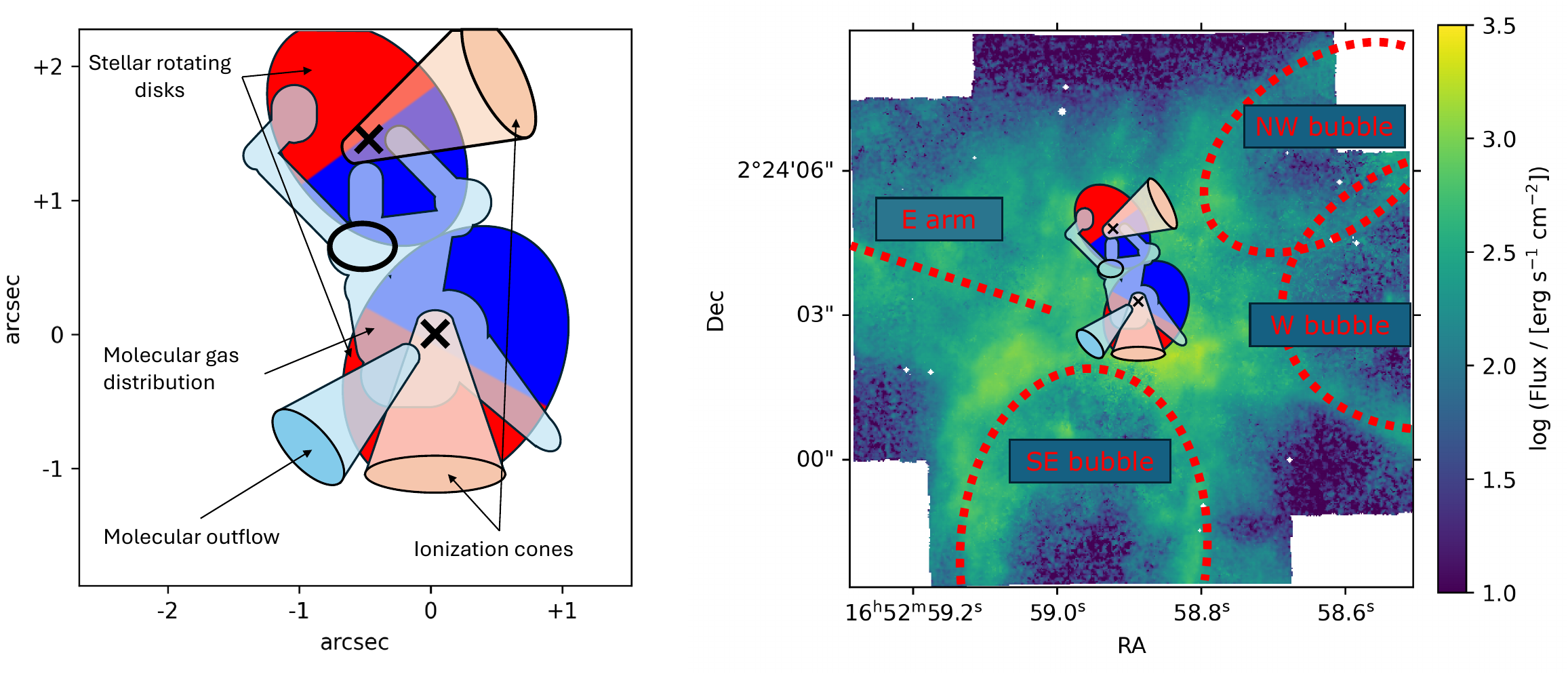}
    \caption{
    Main kinematical features and gas distribution observed in the NIRSpec FoV (\textit{left panel}),  over-imposed on the flux map of H$\alpha$ from NFM MUSE data (\textit{right panel}) with schematic identification of arms and bubbles already known from the literature. The two ellipses indicate the rotating disk of stars, where the approaching region is in blue and the receding one is in red (Section \ref{sec:stellar fit}). The molecular gas distribution is shown in light blue (Section \ref{sect: moment maps} and \ref{sect: H2 pallocchio}). The ionization cones of the two AGN (Section \ref{sect: NIR diagnostic diagram} and \ref{sect: CL}) are indicated in orange. The light blue cone represents the molecular outflow near the southern nucleus (Section \ref{sect: H2 pallocchio}). Black crosses and the black ellipse indicate respectively the nuclei and the high-velocity molecular hydrogen blob positions.}
    
    \label{fig: kin_structures}
\end{figure*}

In this section we present the results obtained from these NIRSpec data, divided into moment maps, ISM properties and kinematics of gas. 
Through these sections, we identify some structures with distinct kinematical features. In particular, the present NIRSpec data show the presence of outflows associated with the nuclei, two rotating disks, and a complex molecular gas distribution. For sake of clarity in Fig. \ref{fig: kin_structures} (left panel) we sketch these structures that will discussed in more detail in the following.

\subsection{Moment maps}
\label{sect: moment maps}
\begin{figure*}
    \centering
    \includegraphics[width=1\linewidth]{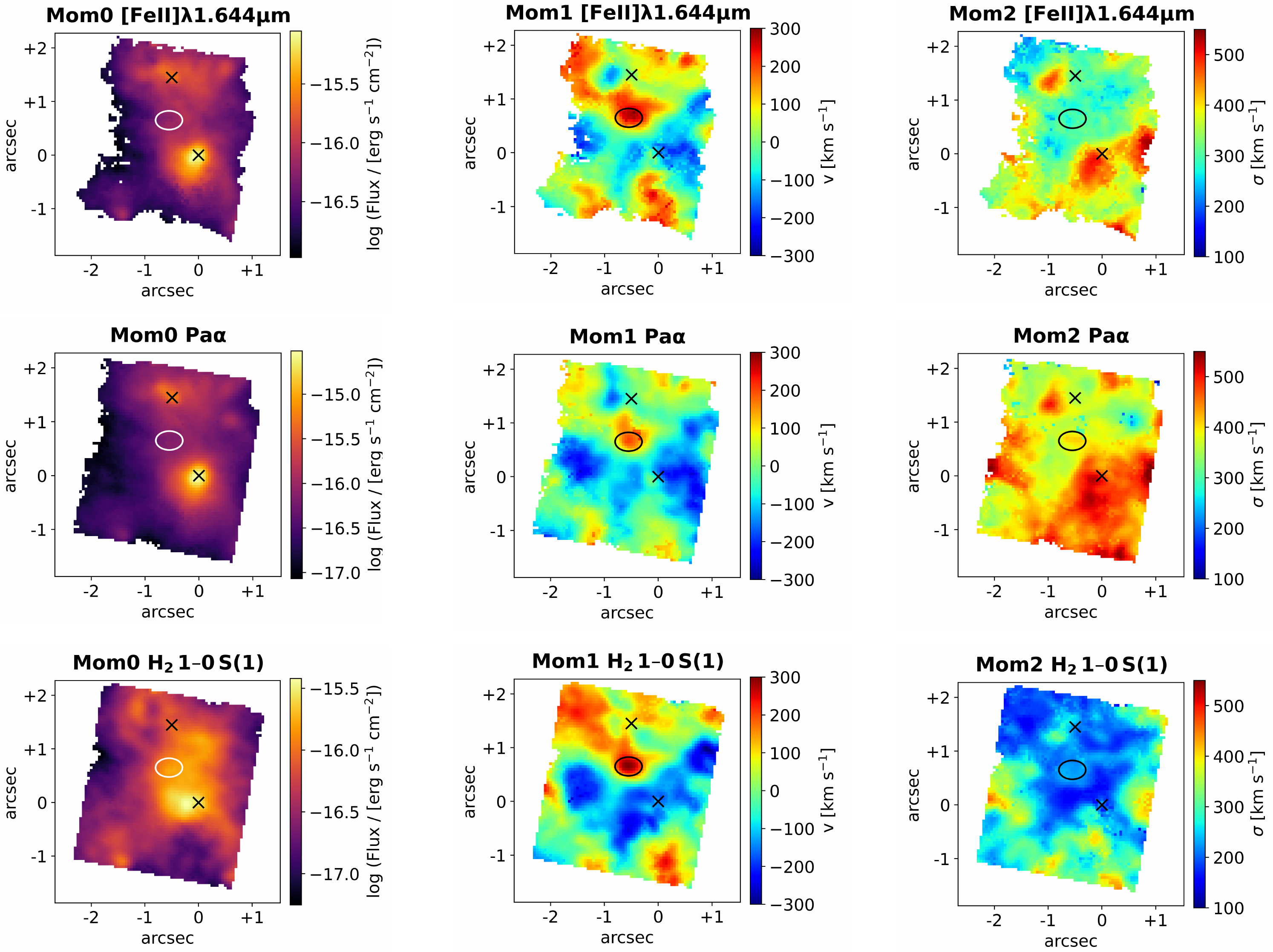}
    \caption{Moment maps of [Fe II]$\lambda$1.644$\mu$m, Pa$\alpha$, and H$_2$ 1-0 S(1) total emission lines with a threshold of S/N > 15. From left to right, we plot the flux map, the velocity map, and the velocity dispersion map. Crosses and the ellipse indicate, respectively, the nuclei and the high-velocity molecular hydrogen blob positions.}
    \label{fig: mom maps}
\end{figure*}

We show the moment maps of [Fe II]$\lambda$1.644$\mu$m, Pa$\alpha$, and H$_2$ 1-0 S(1), obtained with our fit model, in Fig. \ref{fig: mom maps}. The zero moment maps  - the left column of panels - show clear differences between the ionized and molecular gas components. This supports the notion that the molecular gas follows different kinematics than the ionized gas, as already hypothesized in our fitting procedure. Regarding the ionized lines, they have a similar distribution, even if [Fe II]$\lambda$1.644$\mu$m seems to be less concentrated around the nuclei. The brightest line is Pa$\alpha$ and its emission is mostly located at the nuclei, with the other lines being fainter by about one order of magnitude.
However, for each species, the extended distribution of the gas is observed in the region between the nuclei, to the NW of the northern nucleus, and the SE and SW of the southern nucleus (corresponding to the bridge, bubble, Ext 1, and Ext 2 regions in \citealp{MIRI_6240}, respectively).
The velocity maps have values ranging between -300 km s$^{-1}$ and +300 km s$^{-1}$. We point out the presence of a region in all the maps with high redshifted velocities in the order of 200-300 km s$^{-1}$, in the center of the FoV (see Section \ref{sect: H2 pallocchio} for a detailed discussion). The differences between ionized and molecular species are more evident looking at the most blueshifted velocities in the  molecular gas map, that is dominant in the bottom half of the image. 
Regarding the velocity dispersion maps, the molecular hydrogen features the lowest $\sigma$, with a median value of 250 km s$^{-1}$ in the entire FoV, against the more turbulent other species with a median $\sigma$ of 400 km s$^{-1}$.
We also recognized part of the high-$\sigma$ ``V''-like shape structure observed by \cite{MIRI_6240}, which is starting from the SE of the southern nucleus and it extends toward the NE direction the NW one, passing through the southern nucleus (see their Fig. 6).

\begin{figure*}
    \centering
    \includegraphics[width=1\linewidth]{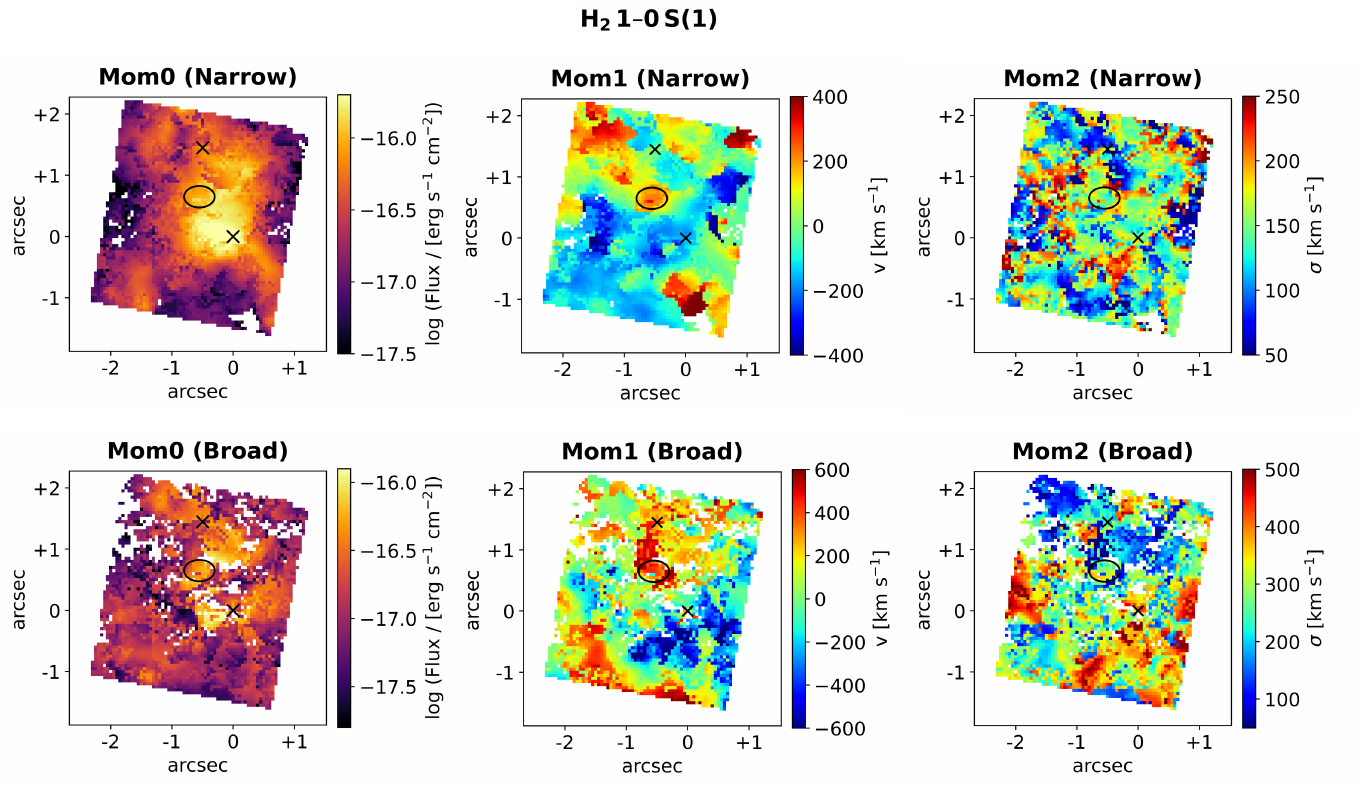}
    \caption{Moment maps of H$_2$ 1-0 S(1) narrow and broad components. 
    The panels in the upper (lower) row correspond to the moment maps for the flux, velocity, and velocity dispersion of the narrow (broad) component. Crosses and the ellipse indicate respectively the nuclei and the high-velocity molecular hydrogen blob positions. We masked spaxels with S/N < 5.}
    \label{fig: H2 NvsB}
\end{figure*}

In Fig. \ref{fig: H2 NvsB} we plot the moment maps for the H$_2$ 1-0 S(1) emission line, but now separated into its narrow (top row) and broad (bottom row) components. In Appendix \ref{app moment maps}, we plot the same moment maps for [Fe II]$\lambda$1.644$\mu$m and Pa$\alpha$ in Figs. \ref{fig: FeII NvsB}, \ref{fig: PaA NvsB}, respectively.
The flux maps of the broad component of [Fe II]$\lambda$1.644$\mu$m and Pa$\alpha$ show notable features near both nuclei. While the broad emission component of both lines is brightest at the southern nucleus, both maps show an extended emission toward the S direction. Additionally, near the northern nucleus, there is an extended structure in the E direction.
The respective velocity dispersion maps also show
elevated values at these locations, especially near the southern nucleus where the dispersion reaches a maximum of $\approx$600 km s$^{-1}$, and forms a structure that extends along the N-S direction, coinciding with an ionization cone defined in Section \ref{sect: NIR diagnostic diagram} and depicted in Fig. \ref{fig: exc diag diagram}.
These observed properties, characterized by extended structures, high relative velocities, and significant velocity dispersions, can be naturally explained by the presence of outflows. As we will discuss in Section \ref{sect: NIR diagnostic diagram}, the coincidence of the high $\sigma$ structure with an ionization cone in the southern nucleus supports this interpretation, as observed in other local Seyfert galaxies (e.g., \citealt{mingozzi_magnum_2019}).

The molecular hydrogen maps show a more complex kinematic behavior in the broad component. We point out two high velocity features: from the southern nucleus an approaching (blueshifted) outflow with v $\approx$ - 600 km s$^{-1}$ in the SE direction, and from the northern nucleus a vertical component at $\sim$ 700 km s$^{-1}$, extending toward the S direction.

\subsection{Interstellar medium properties}
In this section we explore the properties of the ISM in the nuclear region of NGC 6240. We exploit the most used NIR diagnostics in the literature.

\subsubsection{Near-infrared excitation diagnostic diagram}
\label{sect: NIR diagnostic diagram}
To understand the nature of the nuclear activity responsible for the excitation of the observed emission lines, we used the line ratios H$_2$ 1-0 S(1)/Br${\gamma}$ and [Fe II]$\lambda$1.257$\mu$m/Pa$\beta$ (respectively H$_2$/Br${\gamma}$ and [Fe II]/Pa$\beta$, from now on), following \cite{larkin_nearinfrared_1998}, who observed a strong linear correlation in the log–log plot of the two line ratios.
\begin{figure*}[t!]
    \centering
    \includegraphics[width=1\linewidth]{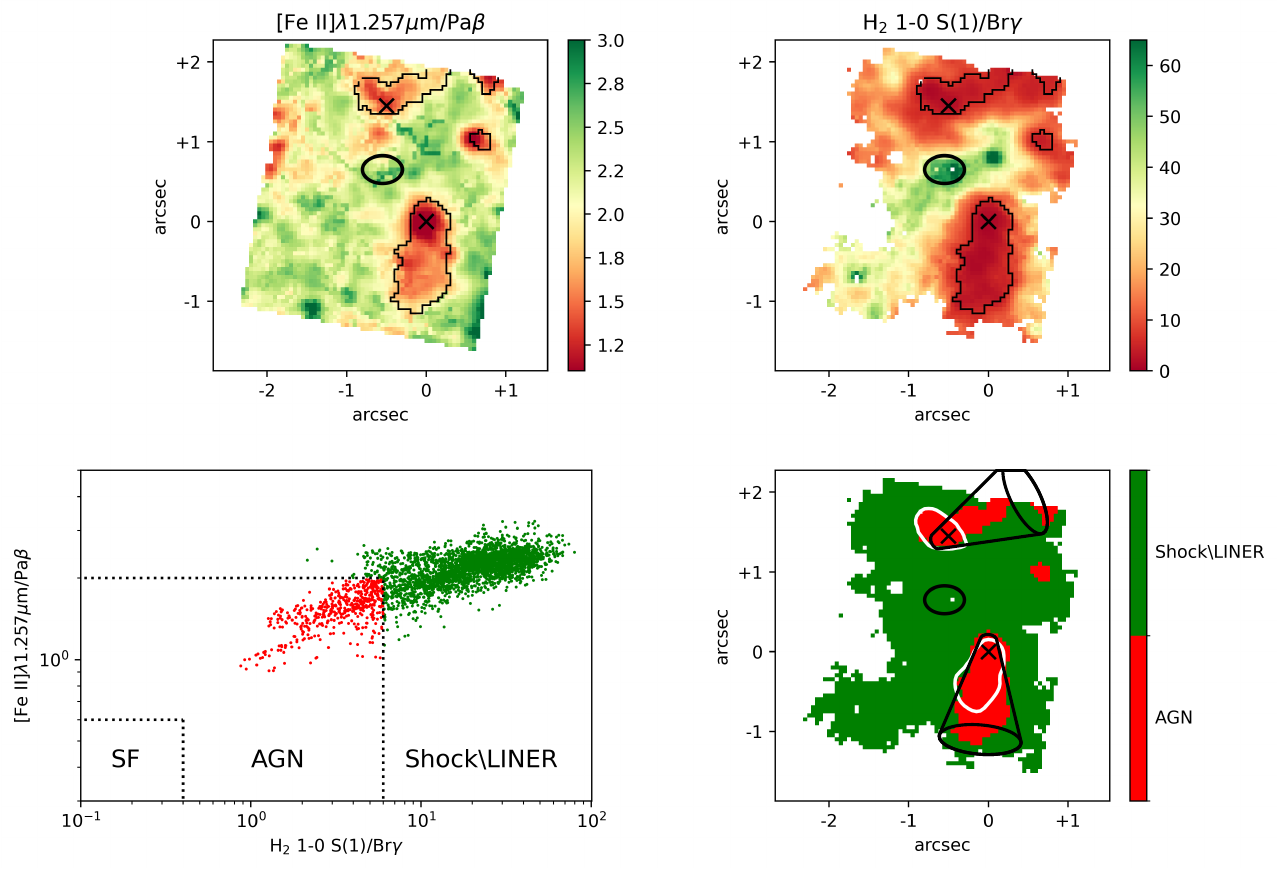}
    \caption{
    Line ratio maps and diagnostic diagram for NIR emission lines. \textit{Top panels:} [Fe II]$\lambda$1.257$\mu$m/Pa$\beta$ and H$_2$ 1-0 S(1)/Br$\gamma$ maps. Black contours delimit AGN-excited spaxels.
    \textit{Bottom-left panel:} NIR diagnostic diagram. The dotted lines are the limit values dividing SF-, AGN-, and shock/LINER-excitation (see text for details). No SF-excitation points are detected.
    \textit{Bottom-right panel:} Excitation map color coded according to the diagnostic diagram. Shock/LINER-excited spaxels are in green, and AGN-excited spaxels in red. Crosses and the ellipse indicate respectively the nuclei and the high-velocity molecular hydrogen blob positions. White contours are the same as those in Fig. \ref{fig: 4 sources}, and indicate the regions where most of the continuum emission originates. Black cones indicate the approximate position of the two ionization cones. We used a S/N threshold of three, due to faint Br$\gamma$ emission.}
    
    \label{fig: exc diag diagram}
\end{figure*}
 To distinguish among the possible mechanisms that regulate the intensities of these lines, we use the threshold values derived by \cite{riffel_molecular_2013}, estimated using a large sample of galaxies previously classified: 
\begin{itemize}
    \item  Star-forming galaxies (SFGs): [Fe II]/Pa$\beta$ $\lesssim$ 0.6 and H$_2$/Br${\gamma}$ $\lesssim$ 0.4;
    \item AGN: 0.6 $\lesssim$ [Fe II]/Pa$\beta$ $\lesssim$ 2 and 0.4 $\lesssim$ H$_2$/Br${\gamma}$ $\lesssim$ 6;
    \item Low ionization nuclear emission-line regions (LINERs) or shock: [Fe II]/Pa$\beta$ $\gtrsim$ 2 and H$_2$/Br${\gamma}$ $\gtrsim$ 6.
\end{itemize}
We plot in the top panels of Fig. \ref{fig: exc diag diagram} the maps of these line ratios.
As we can see in the bottom left panel, spaxels are divided into shock-LINER and AGN excitation dominated regions, while SFG excitation is not detected in any spaxel.
To understand better the distribution and the morphology of these regions, we applied the color-coding of the excitation diagram to plot an excitation map, Fig. \ref{fig: exc diag diagram} (bottom right panel). Most of the FoV is dominated by shock-LINER excitation, as can be expected by a dusty merging system such as NGC 6240, except for the areas around the two nuclei, which are clearly dominated by AGN excitation. In particular, we distinguish the two elongated structures that, starting around each nucleus, are extended in the S direction for the southern nucleus and in the NW for the northern one. There is also some AGN ionization in the east direction of the northern nucleus. We identify those structures with the ionization cones of the two AGN, which are sketched in the panel (see also Fig. \ref{fig: kin_structures}). This result is in agreement with the excitation map of \cite{riffel_agnifs_2021}, who have a slightly wider FoV so they better appreciated and more clearly identified the northern structure (see their Fig. A1). Our H$_2$/Br${\gamma}$ map is also consistent with the H$_2$/Pa${\beta}$ map by \cite{medling_tracing_2021} (see their Fig. 5).

\cite{muller-sanchez_outflows_2011} and \cite{medling_tracing_2021} described the bubble and filamentary structures of NGC 6240 observed at a larger scale, in a FoV of almost 25\arcsec$\,\times\,$25\arcsec. The most important ones are indicated in Fig. \ref{fig: kin_structures} (right panel). \cite{muller-sanchez_two_2018} concluded that the NW bubble, detected in H$\alpha$ with no [O III] detection, was possibly a starburst-driven outflow. The SE bubble is observed in both H$\alpha$ and [O III], and they hypothesized a contribution of both AGN- and SF-ionization in this structure. \cite{medling_tracing_2021} found from BPT diagrams that all the arms and bubbles are in the LINER or shock-like ionization mechanism regime, except the E arm, which is more similar to the AGN-photo-ionization mechanism. Despite the differences in resolution and spatial scale between our data and \cite{muller-sanchez_outflows_2011} and \cite{medling_tracing_2021} ones, the ionization cone of the northern nucleus we detected with the NIR diagnostic diagram (bottom right panel of Fig. \ref{fig: exc diag diagram}) spatially aligns with the direction of the NW bubble. We are thus probably observing the launching region of this outflow, which would be AGN-driven in this scenario. This conclusion is also supported by the detection of coronal lines (CLs) in the same spatial direction (details in Section \ref{sect: CL}; see also \citealp{MIRI_6240}).
Similarly, the ionization cone observed near the southern nucleus and the molecular outflow in the SE direction (see Section \ref{sect: H2 pallocchio}) could be kinematically connected with the SE bubble. The connection of the kinematical structures discussed in this work and the ones observed on larger scale will be discussed by Tozzi et al. (in prep.) using a combination of IFU data from MUSE and KMOS on larger spatial scale. Regarding the other large-scale structures (i.e., the E arm and the W bubble), we did not find any correlation during the data analysis.

\subsubsection{Detection of coronal line emission} \label{sect: CL}
Coronal lines result from collisionally excited forbidden transitions within low lying states of highly ionized species with ionization potentials IP > 100 eV and can be formed either in gas photoionized by a hard UV continuum or in a very hot collisionally ionized plasma. Indeed, they can be considered as indicative of AGN activity (e.g., \citealt{oliva_size_1994}, \citealt{moorwood_origin_1997}). 
Most of these highly ionized lines have been observed in NIR spectra of nearby Seyfert galaxies (e.g., \citealt{rodriguez-ardila_near-infrared_2011}, \citealt{lamperti_bat_2017}, \citealt{den_brok_bass_2022}). While no comprehensive quantitative studies have been conducted on the prevalence of CLs in the AGN population, some evidence suggests they are more likely observed in unobscured type-1 sources than in obscured ones. This aligns with recent detections of CLs in the nearby Seyfert 1.5 nucleus of NGC7469 using NIRSpec and MIRI (\citealt{U_resolving_2022,armus_goals-jwst_2023, bianchin_goals-jwst_2023}), contrasted with their absence in the highly obscured active nuclei of VV114 and Arp 220 (\citealt{perna_no_2024}).

\cite{armus_detection_2006} detected the [Ne V]$\lambda$14.3$\mu$m emission with Spitzer/IRS integrated spectra of NGC 6240. The other CL detection in NGC 6240 reported in literature is that of [Ca VIII]$\lambda$2.32$\mu$m by \cite{ilha_origin_2016}. This CL is found at both nuclei as well as in the region between them, with the strongest emission observed at the southern nucleus.

In our NIRSpec data, we detected resolved emission of the following CLs only near the northern nucleus: [Si VI]$\lambda$1.96$\mu$m (IP $\sim$ 167 eV), [Si VII]$\lambda$2.48$\mu$m (IP $\sim$ 205 eV) and [Mg VIII]$\lambda$3.03$\mu$m (IP $\sim$ 225 eV). We fit these CLs with the same method presented in Section \ref{sec:Analysis}. Being Si lines blended with some H$_2$ lines, one of the two kinematically independent groups of the fitting procedure is the molecular family of H$_2$ 1-0 S(3), H$_2$ 1-0 Q(6) and H$_2$ 1-0 S(1)\footnote{We decided to fit also the H$_2$ 1-0 S(1) because it could help the fit to constrain the molecular line profile, being a strong and isolated emission line.} and the other one is the CLs group. Due to the low S/N of the CLs, we only use up to 2 Gaussian components.
\begin{figure*}[!]
    \centering
    \includegraphics[width=.8\linewidth]{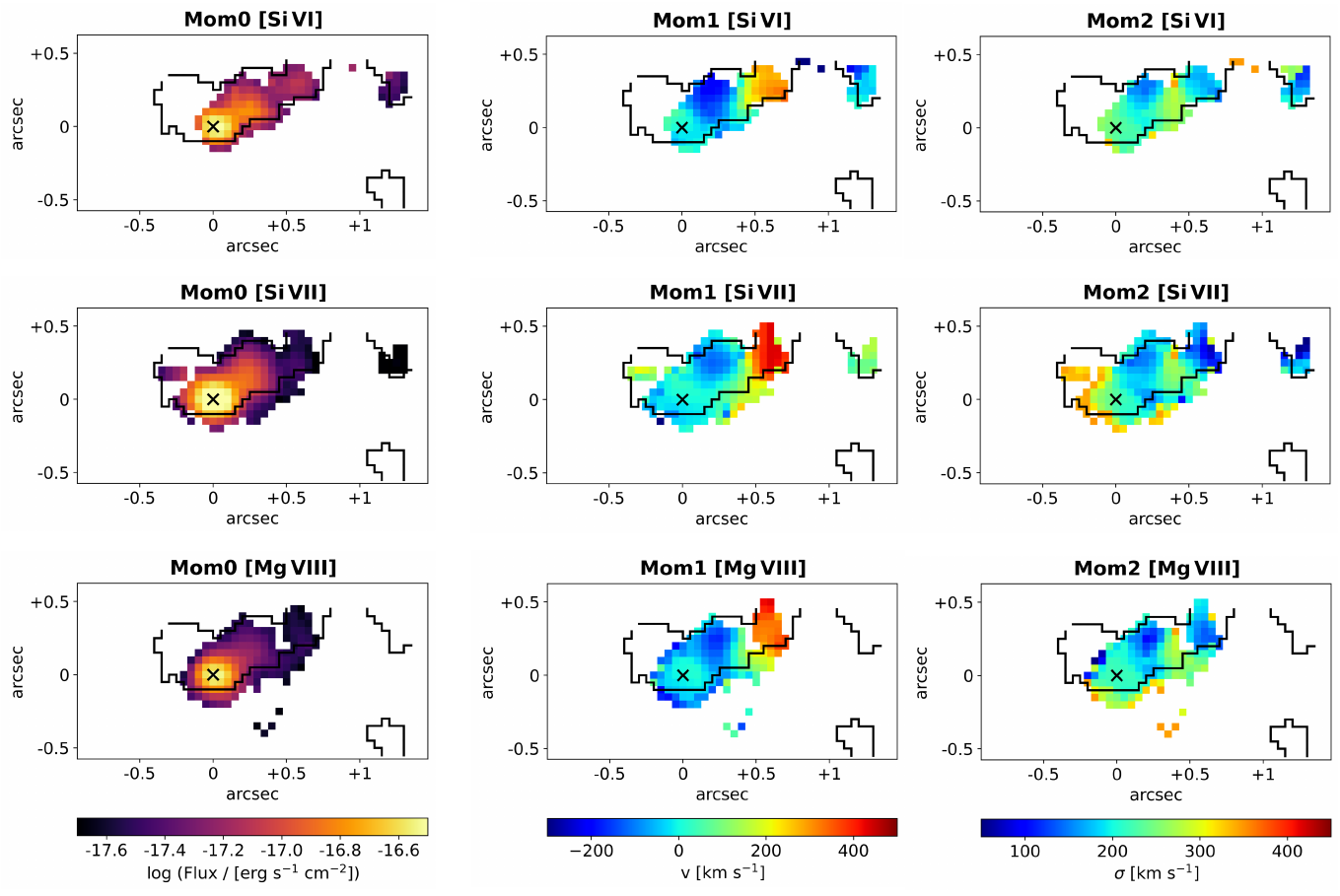}
    \caption{Moment maps around the northern nucleus of [Si VI]$\lambda$1.96$\mu$m , [Si VII]$\lambda$2.48$\mu$m, and [Mg VIII]$\lambda$3.03$\mu$m emission lines. From left to right, we plot the flux map, the velocity map, and the velocity dispersion map. The cross indicates the northern nucleus position. The black contour represents the ionization cone of the northern nucleus from Fig. \ref{fig: exc diag diagram}.  We masked spaxels with S/N < 5.}
    \label{fig: CL moment maps}
\end{figure*}
In Fig. \ref{fig: CL moment maps} we plot the moment maps of the CLs in the northern nuclear region. The flux maps reveal that these species are predominantly concentrated in the nuclear region, with an extension toward the NW direction. The flux peak correspond to the northern nucleus position, confirming our assumption that Pa$\alpha$ peaks at the AGN positions (see Section \ref{sec:Obs}).
All three CLs show elongated emission extending over $\sim$ 0.7$\arcsec$ ($\sim$ 350 pc) in the NW direction, which aligns closely with the ionizing cone contour identified in Section \ref{sect: NIR diagnostic diagram}. Notably, [Si VI]$\lambda$1.96$\mu$m, which has the lowest IP among the observed species, displays higher fluxes at greater distances from the nucleus. In contrast, [Mg VIII]$\lambda$3.03$\mu$m, characterized by a higher IP, shows comparatively lower brightness. Furthermore, silicon lines are detected also in the detached structure of the ionization cone contour at almost 1.2$\arcsec$ ($\sim$ 900 pc), in the top right angle of the FoV. This is consistent with the scenario that most energetic photons do not reach large distances from the AGN and are absorbed earlier.

The CLs velocity dispersion maps are similar to each other, with values ranging from 100 to 300 km s$^{-1}$.
Qualitatively, the velocity maps are similar: the species present a blueshifted structure at 0.3$\arcsec$ with $\sim$ - 100-200 km s$^{-1}$ and a redshifted one at 0.6$\arcsec$ with $\sim$ + 300-400 km s$^{-1}$. 

\begin{figure*}[!]
    \centering
    \includegraphics[width=0.8\linewidth]{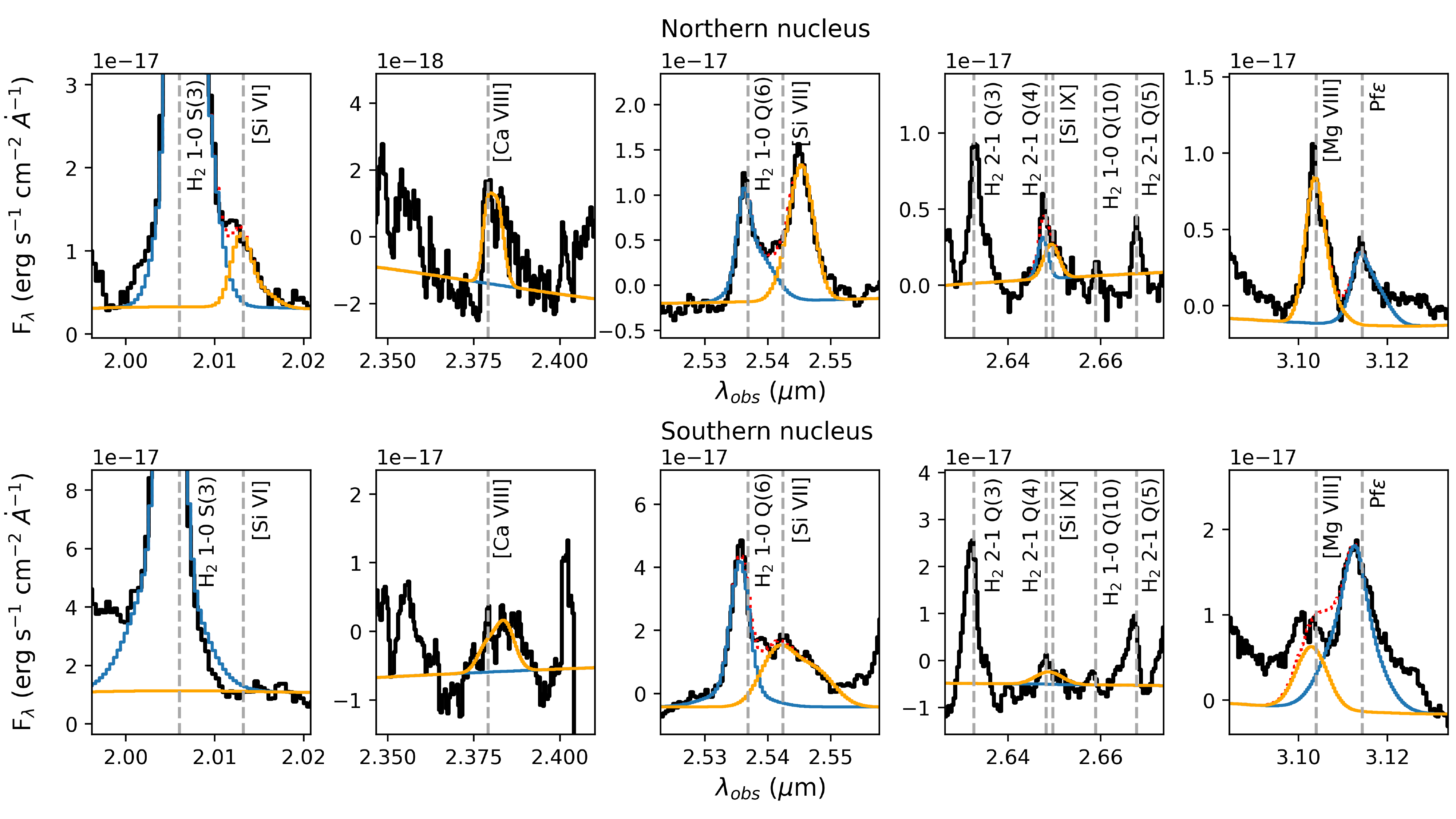}
    \caption{Continuum-subtracted integrated spectra of the northern and southern nucleus in the upper and bottom panel, respectively. The black curve represents data, the red dotted curve represents the total model, the orange curve is the model of CLs, and the blue curve is the model of other lines.}
    \label{fig: CLs spectra}
\end{figure*}

To detect these CLs in the southern nucleus - previously undetectable in individual spaxels around the nucleus - and to identify other faint CLs around both nuclei, we fit the spectra integrated over a radius of 5 spaxels on the nuclei position. This approach was used to enhance the signal-to-noise ratio (S/N).  In addition to the previous CLs, we detected [Ca VIII]$\lambda$2.32$\mu$m (IP $\sim$ 127 eV) and [Si IX]$\lambda$2.58$\mu$m (IP $\sim$ 303 eV) in both nuclei regions. Only [Si VI] was not detected in southern nucleus. In this case, we kinematically separate the CLs group from the other lines.\footnote{We also fit Pf$\epsilon$ because it's blended with [Mg VIII] in the integrated spectra, differently from the spatially resolved spectra.} The fit spectra are shown in Fig. \ref{fig: CLs spectra}, where we plot the integrated spectra of the northern and southern nuclei. The CLs' fit luminosities are listed in Table \ref{tab: CLs fit}.

\begin{table}[b!]
    \begin{center}    
    \tabcolsep 5pt
    \caption{High ionization lines detected in NGC 6240.}
    \begin{threeparttable}
    \begin{tabular}{l c c c}
    \hline
    line & $\lambda_{\text{vac}}$ & IP     & log L$_{\text{line}}$ \\
         & ($\AA$)                & (eV)   & (erg s$^{-1}$)        \\
    \hline
     $[$Si VI]   & 19625  & 167 & 39.07 $\pm$  0.05 (N)\\
     & & & not detected (S)\\
    $[$Ca VIII] & 23210  & 127 & 38.66 $\pm$ 0.06 (N)  \\
    & & & 38.8 $\pm$ 0.1 (S)\\
    $[$Si VII]  & 24807  & 205 & 39.21 $\pm$ 0.04 (N) \\
    & & & 39.42 $\pm$ 0.10 (S)\\
    $[$Si IX]   & 25839  & 303 & 38.5 $\pm$ 0.8 (N) \\
     & & & 38 $\pm$ 1 (S)\\
    $[$Mg VIII] & 30276  & 225 & 39.01 $\pm$ 0.13 (N) \\
     & & & 39.32 $\pm$ 0.12 (S)\\
    \hline

    \end{tabular}
    \end{threeparttable}
    \label{tab: CLs fit}
    \end{center}
\tablefoot{From left to right: Name of the CLs, rest-frame wavelengths, ionization potentials, and log luminosities in the northern (N) and southern (S) nucleus. See Section \ref{sect: CL} for details.}

\end{table}

We do not succeed in detecting a [Ca VIII]$\lambda$2.32$\mu$m spatially resolved emission as done by \cite{ilha_origin_2016} with Gemini NIFS data. 
In cube1 we do not detect any CLs. This is probably due to the higher absorption of dust in that datacube and the lower S/N of the spectra.
The analysis of cube3 is beyond the scope of this work, but from the reduced datacube we can state the presence of other CLs in the northern nucleus, including [Ca IV]$\lambda$3.21$\mu$m, [Al VI]$\lambda$3.66$\mu$m, [Si IV]$\lambda$3.93$\mu$m, [Mg IV]$\lambda$4.49$\mu$m and [Ar VI]$\lambda$4.53$\mu$m
(see Figs. \ref{fig: tot spectrum entire N} and \ref{fig: tot spectrum entire S}).

\cite{MIRI_6240} studied NGC 6240 with JWST/MIRI data. Thanks to its spectral range of 5-28 $\mu$m and larger FoV, they detected some high excitation lines (such as [Mg V], [Ne V], and [Fe VIII]) in the N nucleus and in a bubble-like shape toward the NW direction of almost 5.2$\arcsec$ ($\sim$ 2.7 kpc), outside of the NIRSpec FoV (see Fig. 6 in \citealp{MIRI_6240}). This extended emission results completely redshifted and with a relatively low $\sigma$ ($\sim$ 200 km s$^{-1}$), in accordance with our moment maps (see Fig. \ref{fig: CL moment maps}). This means that with NIRSpec we are looking at the launch region of this highly ionized bubble located out of our FoV. Regarding the southern nucleus, they detected [Mg V] and [Ne VI] in the integrated spectrum, after subtracting the strong PAH features.

\subsubsection{Excitation mechanisms and temperatures of H$_2$ lines}
\label{sect: H2 diagram}
\begin{figure}[t!]
    \centering
    \includegraphics[width=1\linewidth]{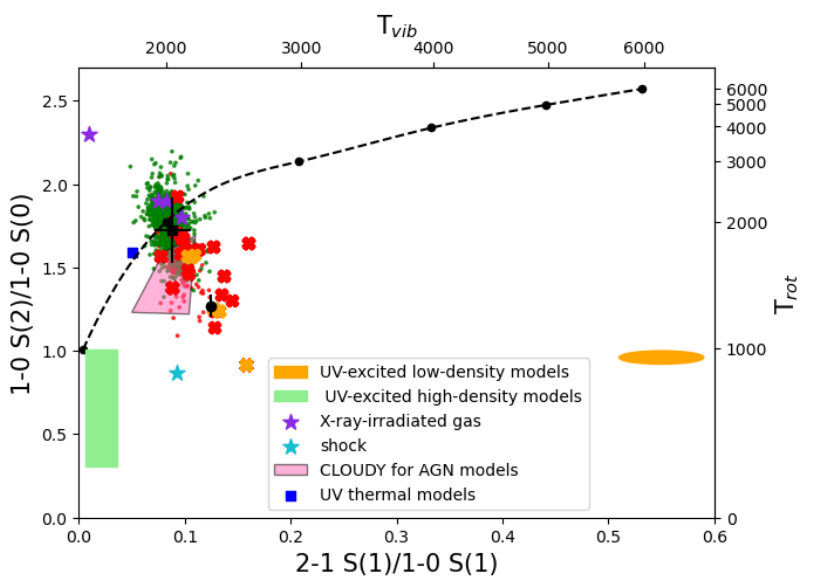}
    \caption{H$_2$ excitation diagram. Green and red points are respectively the shock/LINER-dominated and the AGN-dominated spaxels (from Section \ref{sect: NIR diagnostic diagram}). Red and orange crosses represent the rebinning regions near the southern and northern nuclei. Black square with error bars are the median and the standard deviation of all these points. Black circle with error bars are derived from the estimation of T$_{\rm rot}$ and T$_{\rm vib}$ by \cite{riffel_agnifs_2021} integrating NGC 6240 data in a slightly wider FoV. Black dashed curve represents the thermal emission and on the secondary axis, we have T$_{\rm rot}$ and T$_{\rm vib}$. Each point on the curve represents temperatures from 1000 to 6000 K, in steps of 1000 K, from left to right. The orange ellipse represents the region occupied by non-thermal UV excitation models of \cite{black_fluorescent_1987} and the pink polygon covers the region of the photoionization code \textit{CLOUDY} for AGN models of \cite{dors_jr_x-rays_2012}. The green filled rectangle covers the locus of the thermal UV excitation models of \cite{sternberg_infrared_1989}. The filled blue squares are the UV thermal models from \cite{davies_molecular_2003}. The purple stars are models of X-ray-irradiated gas from \cite{lepp_x-ray_1983} and \cite{draine_h2_1990} and the cyan star is from the shock model of \cite{kwan_molecular_1977}.}
    \label{fig:H2_diagram}
\end{figure}
Following \cite{mouri_molecular_1994}, we investigated the dominant excitation mechanism of H$_2$ lines. Using H$_2$ line ratios it is possible to spatially separate the physical process undergoing in the galaxy. Molecular hydrogen IR emission lines can be due to two mechanisms: 
\begin{itemize}
    \item thermal process: collisions, when molecular clouds are heated by processes such as shocks, X-rays, or UV photons;
    \item non-thermal process: radiative decay resulting from the absorption of a UV photon in the Lyman-Werner band (912-1108 $\AA$) or the collision with a fast electron due to X-ray ionization.
\end{itemize}
In each case, we expect a different emission-line spectrum, and therefore, the relative intensities among the H$_2$ emission lines may be used to discriminate between the powering mechanisms, keeping in mind that multiple emitting regions and multiple mechanisms may simultaneously be present.

\cite{mouri_molecular_1994} suggested the use of the ratios 2-1 S(1) 2.247 $\mu$m/1-0 S(1) 2.121 $\mu$m versus 1-0 S(2) 2.033 $\mu$m/1-0 S(0) 2.223 $\mu$m. Another possible plot includes the line ratios 2-1 S(1) 2.247 $\mu$m/1-0 S(1) 2.121 $\mu$m versus 1-0 S(3) 1.958 $\mu$m/1-0 S(1) 2.121 $\mu$m: we did not use this latter because in our spectra the 1-0 S(3) line is blended with the [Si VI]$\lambda$1.963$\mu$m, so its intensity cannot be reliable. In Fig. \ref{fig:H2_diagram}, we plot previous line ratios using the color-coding obtained in the previous section, which means that we separate the shock/LINER-dominated regions from the AGN-dominated ones. As one can expect, if we simply plot every spaxel we notice a high dispersion, assignable to spaxels with low S/N or those in the nuclear regions previously subjected to wiggles correction. Therefore, we performed a Voronoi binning to increase the S/N, using the faintest line, that is H$_2$ 1-0 S(0), to set a threshold at S/N = 4.
We fit these new spectra with the same procedure described in Section \ref{sec:Analysis} and we computed the H$_2$ line ratios to plot them in Fig. \ref{fig:H2_diagram} as red and orange crosses, respectively for the southern and northern nuclei. We computed the median and standard deviation of all the points and crosses (black square with error bars).
In the same panel, we plot a black curve representing the predictions for thermal emission, where the gas is in local thermodynamic equilibrium (LTE) and so the rotational and vibrational temperature have the same values (see next paragraphs for details). Each point on the curve represents temperature from 1000 to 6000 K, in steps of 1000 K, from left to right.

Fig. \ref{fig:H2_diagram} also shows values of model calculations found in the literature, to better constrain the excitation mechanisms involved. \cite{black_fluorescent_1987} calculated UV-excited low-density gas models that occupied the orange ellipse on the right, while UV-excited high-density models by \cite{sternberg_infrared_1989} are represented by the green filled box on the left. The purple stars are models of X-ray-irradiated gas from \cite{lepp_x-ray_1983} and \cite{draine_h2_1990}. \cite{dors_jr_x-rays_2012} calculated these ratios with the photoionization code \textit{CLOUDY} (\citealp{ferland_hazy_1993}) for AGN models, pink polygon in the plot. The filled blue square is a UV thermal model by \cite{davies_molecular_2003} and the cyan star is from the shock model of \cite{kwan_molecular_1977}. 

As we can see in the figure, points in the diagram are away from the region predicted by the non-thermal UV excitation models of \cite{black_fluorescent_1987} (identified by the orange ellipse), indicating that the H$_2$ emission originates mainly in thermal processes. Indeed, our data are concentrated around the thermal emission at 2000 K but with a mild dispersion that could be an effect of low S/N. The models that better reproduce them are the ones of X-ray-irradiated gas and the AGN photoionization \textit{CLOUDY} models. In particular, these latter represent the AGN-dominated regions - crosses and red points in Fig. \ref{fig:H2_diagram} -, as expected. We also observe that the ionization cones where we rebinned, indicated by crosses in the plot, exhibit the most extreme values. This suggests that these regions are slightly farther from the thermal emission and may contain some contamination from non-thermal emission.

According to \cite{rodriguez-ardila_molecular_2005}, an alternative way to
determine the mechanism driving the thermal or non-thermal origin of the H$_2$ excitation processes is to calculate the rotational temperature T$_{\rm rot}$ and the vibrational temperature T$_{\rm vib}$ of the molecular gas. Indeed, in a gas dominated by thermal excitation the two temperatures are quite similar, as expected for a gas in LTE. Non-thermal excitation, on the other hand, is characterized by a high vibrational temperature and a low rotational temperature; non-local UV photons, not characteristic of the local kinetic temperature, overpopulate the highest energy levels compared to that expected for a Maxwell–Boltzmann population. 

To estimate these temperatures we used the H$_2$ 2-1 S(1) 2.247 $\mu$m/H$_2$ 1-0 S(1) 2.121 $\mu$m and H$_2$ 1-0 S(2) 2.033 $\mu$m/H$_2$ 1-0 S(0) 2.223 $\mu$m line ratios, the same present in Fig. \ref{fig:H2_diagram}. We used $N_i/g_i$ = $F_i \lambda_i / (A_i g_i)$ = exp$[-E_i / (k_B T_{exc})]$, where $N_i$ are the column densities in the upper level, $g_i$ the statistical weights, $F_i$ and $\lambda_i$ are the line fluxes and wavelengths, $A_i$ are the transition probabilities, $E_i$ are the energies of the upper level, $T_{exc}$ is the excitation temperature and $k_B$ is the Boltzmann constant. Using the transition probabilities from \cite{turner_quadrupole_1977}, the rotational temperature is given by
\begin{equation}
    T_{rot} \simeq - \frac{1113}{\text{ln} \left( 0.323 \frac{F_{H_2 \, 1-0 \, S(2)}}{F_{H_2 \, 1-0 \, S(0)}}  \right)},
\end{equation}
and the vibrational temperature is given by
\begin{equation}
    T_{vib} \simeq \frac{5594}{\text{ln} \left( 1.355 \frac{F_{H_2 \, 1-0 \, S(1)}}{F_{H_2 \, 2-1 \, S(1)}}  \right)}.
\end{equation}

\begin{figure}
    \centering
    \includegraphics[width=1\linewidth]{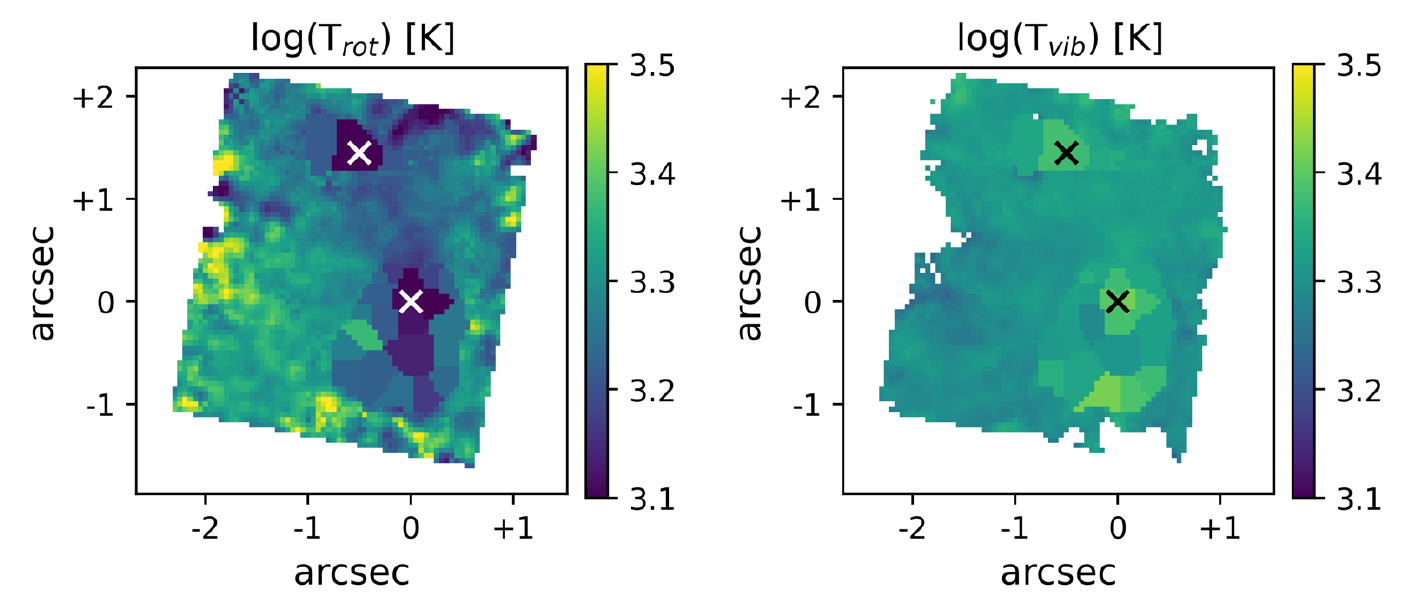}
    \caption{ 
    Logarithm of the rotational (\textit{left panel}) and vibrational (\textit{right panel}) temperatures computed from H$_2$ line ratios. Crosses indicate the positions of the nuclei. 
    }
    \label{fig: Ts}
\end{figure}

In Fig. \ref{fig: Ts} we show the maps of T$_{\rm rot}$ and T$_{\rm vib}$, including the Voronoi regions. 
As noted in the previous paragraphs, the lowest T$_{\rm rot}$ values and highest T$_{\rm vib}$ values in the nuclear regions indicate non-thermal emission contamination near the nuclei. Comparing Fig. \ref{fig:H2_diagram} with Fig. \ref{fig: Ts}, the dispersion of the green points in the diagram is likely caused by extreme T$_{\rm rot}$ values in spaxels near the FoV edges, which are probably affected by noise.
The median values and standard deviations of these maps are T$_{\rm rot}$ = (1894 $\pm $ 397) K  and T$_{\rm vib}$ = (2050 $\pm$ 141) K. Recent studies by \cite{riffel_agnifs_2021}, using data obtained with the NearInfrared Integral Field Spectrograph (NIFS; on the Gemini north Telescope) in the K band (slightly wider FoV and higher spectral resolution), computed T$_{\rm rot}$ = (1247 $\pm $ 74) K  and T$_{\rm vib}$ = (2339 $\pm$ 3) K. Apart from the slight discrepancy with these latter results - possibly due to our higher sensitivity - we found a rotational and vibrational temperatures that are consistent with each other, as they should be in a gas in LTE. We conclude again, as done by using the diagnostic diagram of Fig. \ref{fig:H2_diagram}, that what we are seeing originates from thermal emission.

\subsubsection{Mass of molecular gas}
\label{sect: molecular mass}
Following \cite{riffel_mapping_2008}, we used the fluxes of the H$_2$ 1-0 S(1) and Pa$\alpha$ emission lines to compute the mass of hot molecular and ionized hydrogen. The term "hot" depends on the fact that, compared to cold molecular gas (T $<$ 100 K) detected via sub-millimeter CO line emission, IR H$_2$ roto-vibrational transitions trace a hotter phase of molecular gas with typical temperatures of T $\sim$ 100 − 1000 K  (e.g., \citealt{garcia-burillo_galaxy_2021}). 

The mass of hot H$_2$ can be estimated, under the assumptions of local thermodynamic equilibrium and a T$_{\rm vib}$ = 2000 K, as proposed in \cite{scoville_velocity_1982}, as
\begin{equation} \label{eq: M_H2}
    \left( \frac{M_{H_2}}{M_{\odot}} \right) = 5.0776 \times 10^{13} \left( \frac{F_{\text{H$_2$ 1-0 S(1)}}}{\text{erg s$^{-1}$ cm$^{-2}$} } \right) {\left( \frac{D}{\text{Mpc}} \right)}^2,
\end{equation}
where F$_{\rm H_2 \, 1-0 \, S(1)}$ is the dereddered H$_2$ 1-0 S(1) emission-line flux and D = 106 Mpc the galaxy distance (see Section \ref{sec:Obs}). Integrating the flux over all the FoV, we obtained a value of M$_{\rm H_2}$ $\approx 1.3 \times 10^5$ M$_{\odot}$. This value is in agreement with \cite{riffel_agnifs_2021}, where M$_{\rm H_2}$ $\approx 1.1 \times10^5$ M$_{\odot}$.

\begin{figure}
    \centering
    \includegraphics[width=.92\linewidth]{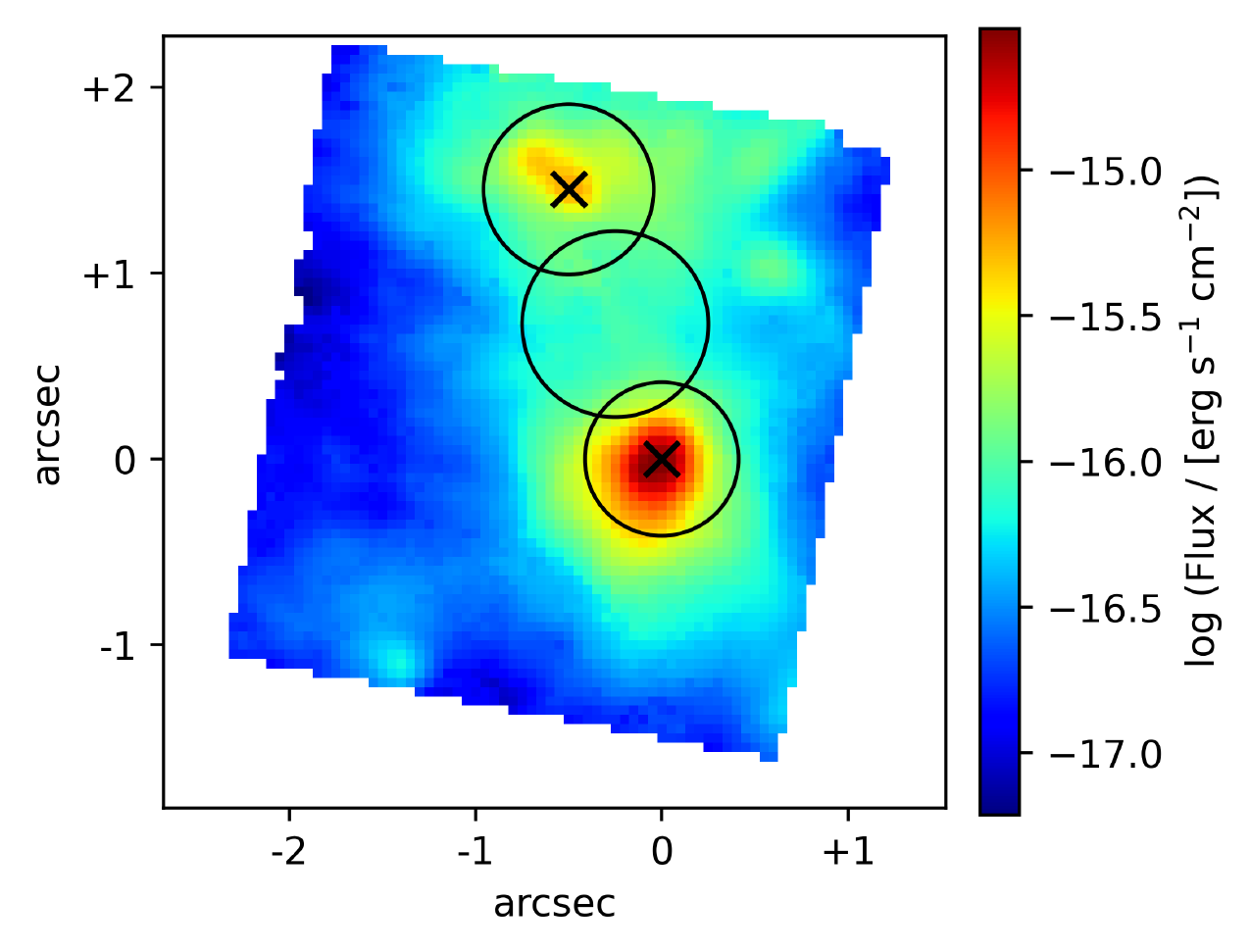}
    \caption{Flux of the Pa$\alpha$ emission line in logarithmic scale. Crosses indicate the positions of the nuclei. Circles indicate the spheres of influence of each SMBH and a 1” diameter aperture centered equidistantly along the line connecting the two nuclei used by \cite{treister_molecular_2020}.}
    \label{fig: Treister}
\end{figure}

We compared the hot-to-cold ratios of molecular gas with the previous estimation by \cite{medling_tracing_2021}. They found a ratio of $\sim$ 10$^{-6}$ in the nuclear region between the nuclei, and roughly 100 times higher in the external filaments of the galaxy (see their Fig. 11). We also considered the results of \cite{treister_molecular_2020}, who computed the cold molecular gas mass from CO (2-1) ALMA observations in three distinct regions indicated in Fig. \ref{fig: Treister}: the sphere of influence of the northern and southern SMBHs and 1\arcsec diameter aperture centered equidistantly along the line connecting the two nuclei. They found $7.4\times10^8 M_{\odot}$ for the northern nucleus, $3.3\times10^9 M_{\odot}$ for the southern nucleus and $8.6\times10^9 M_{\odot}$ for the central region. For a right comparison, we also calculated the hot molecular gas in those regions: we obtained $1.0\times10^4 M_{\odot}$ for the northern nucleus, $2.3\times10^4 M_{\odot}$ for the southern nucleus and $2.5\times10^4 M_{\odot}$ for the central region. The respectively hot-to-cold ratios of molecular gas are therefore $1.4\times10^{-5}$, $7.1\times10^{-6}$ and $2.9\times10^{-6}$. 
These integrated values follow the same trend of the spatially resolved ratio of \cite{medling_tracing_2021}, with lower values in the region between nuclei and higher values on the nuclei. Our estimations are well within the 10$^{-7}$-10$^{-5}$ range found for molecular gas in starburst galaxies and AGN by \cite{dale_warm_2005}. In particular, the highest ratio of the northern nucleus is slightly lower but comparable to the extreme ratio of $6-7\times10^{-5}$ observed in the outflows of local LIRGs, for example by \cite{emonts_outflow_2014} and \cite{pereira-santaella_high-velocity_2016}.

\subsubsection{Electron density} \label{sect: density}
In order to compute the mass of ionized gas, we needed to estimate the electron density (see Section \ref{sec: ionized mass} and Eq. \ref{eq: M_HII PaA}). 
From a spectrum integrated into the same FoV of  NIRSpec data, we fit the [S II]$\lambda$6716$\AA$ and [S II]$\lambda$6731$\AA$ emission lines observed in the MUSE NFM datacube,\footnote{To reduce the degeneracy in the doublet, we fit this latter spectrum with H${\rm \alpha}$, H${\rm \beta}$, [S III]$\lambda$9069$\AA$ lines and doublets of [O III]$\lambda\lambda$5007,4959$\AA$, [N II]$\lambda\lambda$6548,83$\AA$, and [O I]$\lambda\lambda$6300,64$\AA$.} with the same multi-Gaussian fitting procedure described in Section \ref{sec:Analysis}. 
We estimated a value of n$_{\rm e}$ = (359 $\pm$ 52) cm$^{-3}$ using the module \textit{PyNeb} (with T$_{\rm e}$ = 10$^4$K).

Another estimation of density can be achieved using some NIR [Fe II] lines, as described in \cite{koo_infrared_2016}, such as [Fe II]$\lambda$1.534$\mu$m, [Fe II]$\lambda$1.600$\mu$m, [Fe II]$\lambda$1.644$\mu$m, [Fe II]$\lambda$1.664$\mu$m, and [Fe II]$\lambda$1.677$\mu$m. Indeed, these are multiple lines of comparable excitation energies, with ratios that are mainly a function of electron density n$_{\rm e}$ and depend weakly on temperature T$_{\rm e}$.  
We needed to significantly increase the S/N of the [Fe II] emission lines. In fact, except for the strong [Fe II]$\lambda$1.644$\mu$m, other lines are almost undetected at spaxel level. We decide to extract an integrated spectrum from an aperture located in the FoV center with a radius of 30 spaxels - or 1.5\arcsec -, deriving [Fe II]$\lambda$1.677$\mu$m and [Fe II]$\lambda$1.664$\mu$m emission lines fluxes. After applying the usual fitting procedure, we computed the electron density with the equations
\begin{equation} \label{eq: Fe 1.664}
\log n_e = 3.14 + 13.7 \frac{ F_{1.664}}{F_{1.644}}, \quad \text{with} \quad 0.03 \leq \frac{F_{1.664}}{F_{1.644}} \leq 0.13;
\end{equation}
\begin{equation} \label{eq: Fe 1.677}
\log n_e = 2.93 + 8.14 \frac{ F_{1.677}}{F_{1.644}}, \quad \text{with} \quad 0.07 \leq \frac{F_{1.677}}{F_{1.644}} \leq 0.23.
\end{equation}
The two [Fe II] diagnostics gave comparable values, with a mean of log(n$_{\rm e}/$cm$^{-3}$) = 4.12. This density exceeds that derived from optical [S II] lines, around 359 cm$^{−3}$.
A plausible interpretation of this is that the two methods trace different gas regions. This would mean that [Fe II] lines, with critical densities higher than [S II] (n$_{\rm c} \sim 1600$ cm$^{-3}$ for [S II]$\lambda$6716$\AA$; n$_{\rm c} \sim 10^4$ cm$^{-3}$ for iron lines), delineate post-shock regions experiencing greater compression. At the same time, due to different ionization energy (E$_{\rm ion}$ = 7.9 eV for iron and E$_{\rm ion}$ = 10.4 eV for sulfur), we expect that [S II] identifies regions with a higher degree of ionization. 
For comparison, we note that \cite{MIRI_6240} found a density of n$_{\rm e} \sim 1800$ cm$^{-3}$, using the ratio of [Ne V]$\lambda$14$\mu$m and [Ne V]$\lambda$24$\mu$m obtained from an integrated MIRI spectrum on the northern nucleus.

\subsubsection{Mass of ionized gas} \label{sec: ionized mass}
Similar to \cite{riffel_agnifs_2021}, we calculated the mass of ionized gas M$_{\rm HII}$, assuming an electron temperature T$_{\rm e}$ = 10$^4$ K, with the formula 
\begin{equation}\label{eq: M_HII PaA}
     \left( \frac{M_{HII}}{\text{M$_{\odot}$}} \right) = 2.30 \times 10^{18} \left( \frac{F_{\text{Pa${\alpha}$}}}{\text{erg s$^{-1}$ cm$^{-2}$} } \right) {\left( \frac{D}{\text{Mpc}} \right)}^2 {\left( \frac{N_e}{\text{cm$^{-3}$}} \right)}^{-1}.
\end{equation}
where F$_{\text{Pa${\alpha}$}}$ is the Pa${\alpha}$ integrated flux and n$_{\rm e}$ is the electron density. 

Using the value of n$_{\rm e}$ $\sim$ 359 cm$^{-3}$, obtained with the [S II] line ratio from MUSE NFM data (see Section \ref{sect: density}), we obtained M$_{\rm HII}$ $\approx 2.8 \times 10^7$ M$_{\odot}$, similar to the one obtained by \cite{riffel_agnifs_2021} of M$_{\rm HII}$ $\approx 1.0 \times 10^7$ M$_{\odot}$ (they assumed n$_{\rm e}$ = 500 cm$^{-3}$). Using a higher density in equation \ref{eq: M_HII PaA}, we expect lower mass values. Indeed, using density from iron lines (see Section \ref{sect: density}) we computed M$_{\rm HII}$ $\approx 7.7 \times 10^5$ M$_{\odot}$. Given that we do not know which tracer is better suited to measure the density of the bulk of the gas, we will consider both extremes in the following.

\begin{figure}[t!]
    \centering
    \includegraphics[width=1\linewidth]{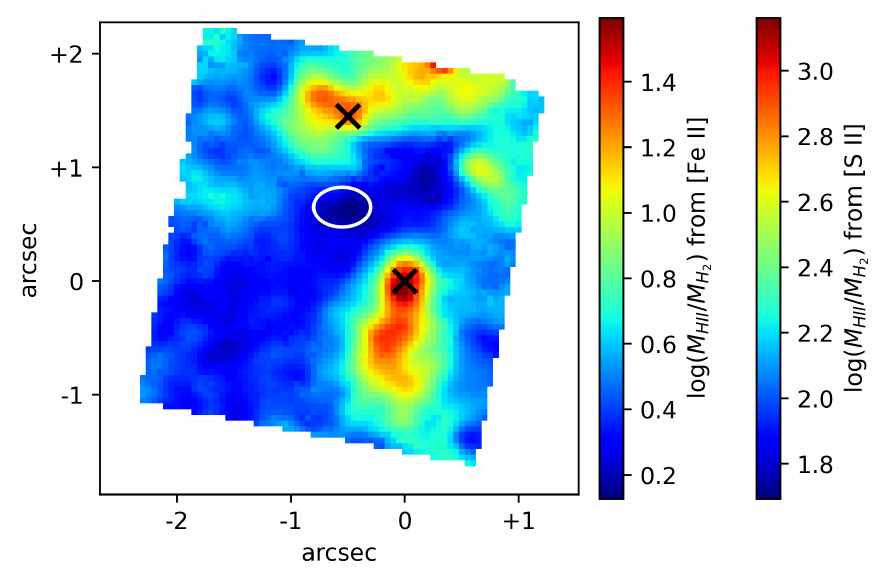}
    \caption{Logarithm ratio between the ionized gas mass and the hot molecular gas mass. The color bars depict the two values obtained with different proxy for the density estimation, namely [Fe II] and [S II] line ratios (see text for details). Crosses and the ellipse indicate respectively the nuclei and the high-velocity molecular hydrogen blob positions.}
    
    \label{fig: MH2/MHII}
\end{figure}

We show in Fig. \ref{fig: MH2/MHII} the logarithm ratio between the ionized gas mass and the hot molecular gas mass. For the ionized mass, we used both the density estimations obtained above. The figure displays the ratio calculated with both iron lines and sulfur lines, as it is indicated by the color bars. We computed it considering all the components (narrow+broad) of the emission lines. Here it is clear the enhancement of ionized gas with respect to the molecular one in the two ionization cones, defined with the NIR diagnostic diagram in Section \ref{sect: NIR diagnostic diagram}. Furthermore, the highest concentration of molecular gas with respect to the ionized one is located in the position of the highly redshifted blob of molecular hydrogen (see Section \ref{sect: H2 pallocchio}) indicated with a white ellipse.
In the hypothesis that all the outflow is in the broad component, we calculated the same ratio M$_{\rm H_2}$/M$_{\rm HII}$ in the outflow present in our FoV. We found a logarithm value of 1.0 with iron lines density and 2.6 with sulfur lines density. 

\subsection{Gas kinematics}

\subsubsection{Molecular gas connecting the two nuclei}
\label{sect: H2 pallocchio}
\begin{figure*}
    \centering
    \includegraphics[width=0.8\linewidth]{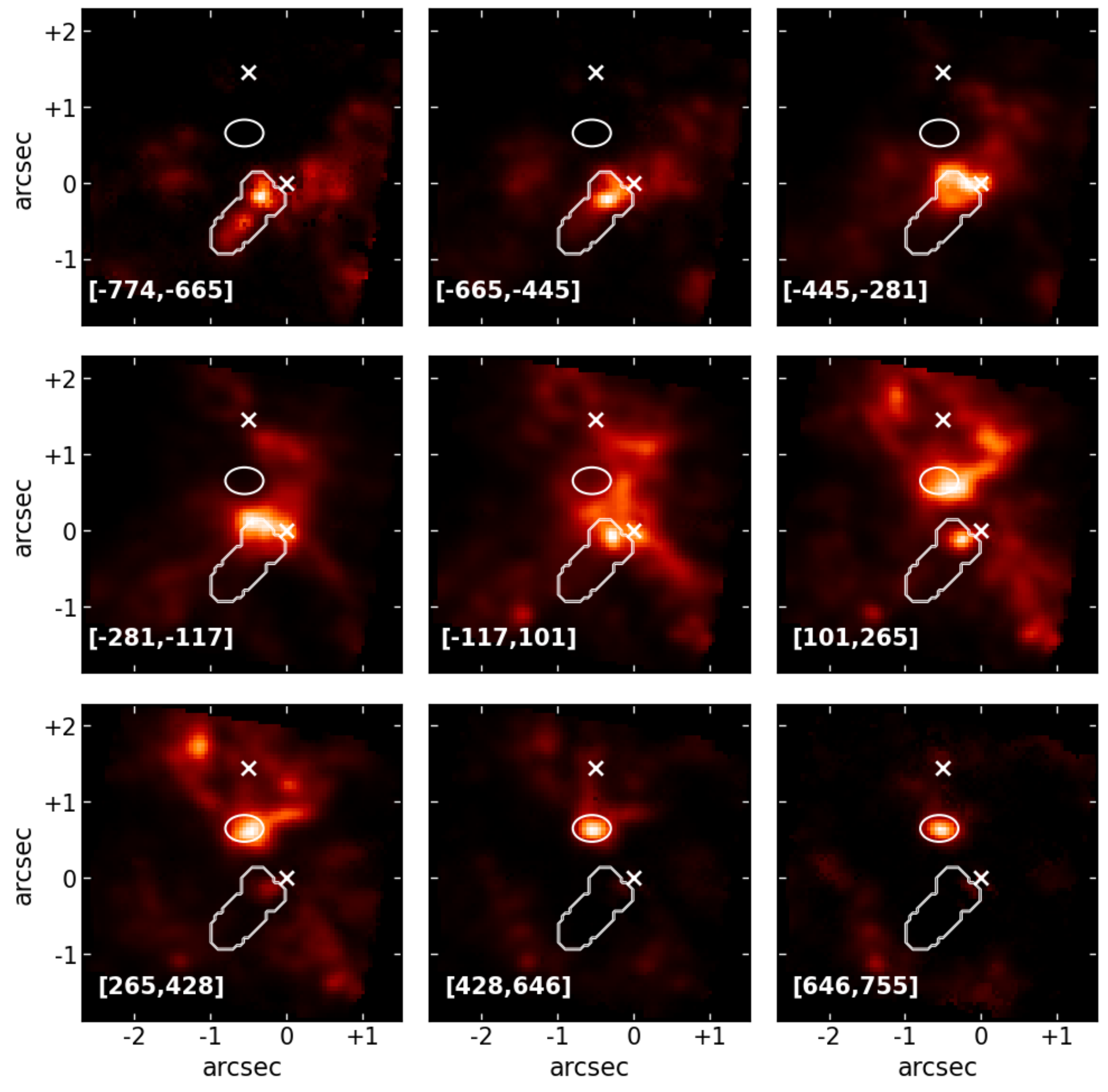}
    \caption{Channel maps of H$_2$ 1-0 S(1) emission lines. Velocity bins are indicated in the bottom left of every panel (in km s$^{-1}$). Crosses and the ellipse indicate respectively the nuclei and the high-velocity molecular hydrogen blob positions. White contour is the outflow launched from the southern nucleus.
    }
    \label{fig: H2 channel maps}
\end{figure*}

In Fig. \ref{fig: H2 channel maps} we plot channel maps of the emission line H$_2$ 1-0 S(1), starting from about -800 km s$^{-1}$ until 800 km s$^{-1}$\footnote{A movie showing the flux for every velocity channel is available at \hyperlink{blue}{https://www.youtube.com/shorts/Uh2g3QybRE0}}. This figure contains a lot of interesting features: in the first row, in the bluer maps, an outflow can be recognized - highlighted with a white contour in the figure - starting from the south nucleus, pointing in the SE direction. In the central row, a filamentary structure with three lobes can be observed, near the systemic velocity. From the right panel in the central row, the most interesting structure of the gas molecular kinematics appears increasingly in the channel maps [101 km s$^{-1}$, 265 km s$^{-1}$], where it seems there are two arms connecting the north nucleus and a region between the two nuclei. This region then continues to be connected to the north nucleus through an NS-directed arm and stays visible until about 800 km s$^{-1}$.

\begin{figure*}
    \centering
    \includegraphics[width=.9\linewidth]{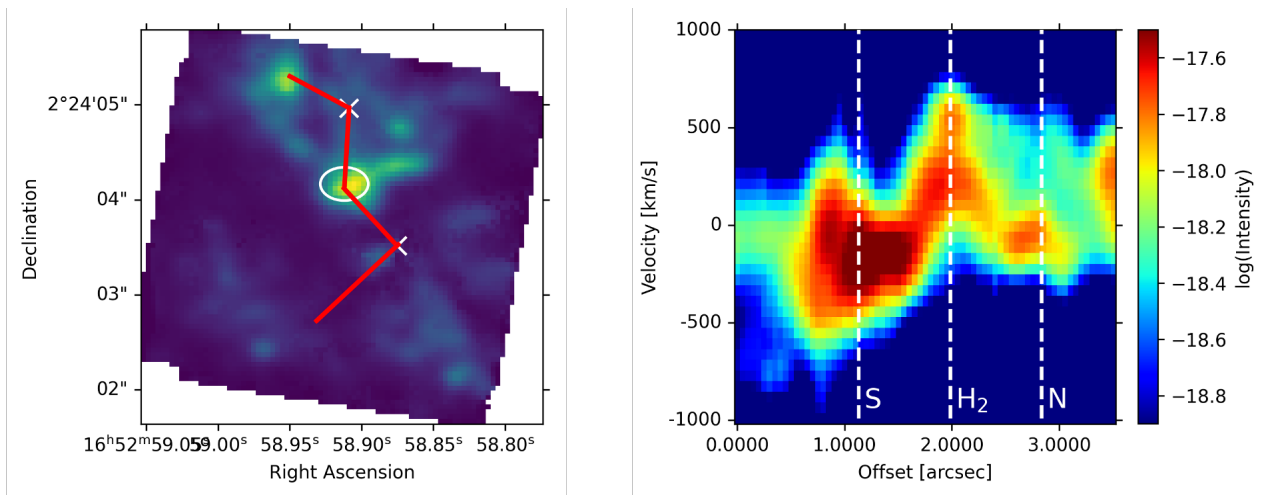}
    \caption{Position-velocity diagram on the H$_2$ 1-0 S(1) emission. \textit{Left panel:} Velocity channel map of H$_2$ 1-0 S(1) of $\sim$ 400 km s$^{-1}$. In red we show the path used in the p-v diagram in the right panel. Crosses and the ellipse indicate respectively the nuclei and the high-velocity molecular hydrogen blob positions. \textit{Right panel:} Position-velocity diagram of the H$_2$ 1-0 S(1) emission line. Vertical dashed lines indicate the position of the landmarks along the path (S is the southern nucleus, H$_2$ is the molecular blob, and N is the northern nucleus). The flux is in log scale.}
    \label{fig:pv_H2}
\end{figure*}

A significant reservoir of high-velocity molecular gas between the nuclei has been already seen in the cold molecular component (see Section \ref{sect: intro}): it was studied and discussed by \cite{cicone_alma_2018} with [C I] (1-0) line and by \cite{treister_molecular_2020} with CO (2-1) transition, both with ALMA data. Also in \cite{fyhrie_molecular_2021} they observed this redshifted component, thanks to the channel maps of CO (3-2) and CO (6-5) obtained with ALMA (see their Figs. 4-5).
To date, given its high velocity, it has been speculated by \cite{treister_molecular_2020} that this clumpy concentration of material is an outflow of molecular gas expelled from the nuclear region of the merger system.
From the channel map of Fig. \ref{fig: H2 channel maps}, this reservoir of molecular gas seems to be connected to both nuclei with some gas bridges. To better understand the nature of this clump and the kinematic structures of the molecular stream connecting the nuclei, we have used a position-velocity (p-v) diagram on the H$_2$ 1-0 S(1) emission. As shown in the left panel of Fig. \ref{fig:pv_H2}, the path of the diagram follows the direction of the previously mentioned southern outflow, and then in order it passes through the south nucleus, the reservoir of H$_2$ and the north nucleus. In the right panel of the Fig. \ref{fig:pv_H2}, we display the p-v diagram, with vertical lines indicating the position of the landmarks along the path - namely, in increasing order of the offset, the southern nucleus, the high-velocity H$_2$ reservoir, and the northern nucleus. The molecular hydrogen is mainly concentrated on the southern nucleus, as already stated in other works (e.g., \citealt{ilha_origin_2016}). Near the southern nucleus, we observe the molecular outflow component at a maximum velocity of about $\sim$ -- 800 km s$^{-1}$. This feature corresponds to the structure with high blueshifted velocity and high velocity dispersion observed in the moment maps of the molecular broad component (see Fig. \ref{fig: H2 NvsB}). In Fig. \ref{fig: kin_structures}, we compare it with other observed kinematical structures, showing that the molecular outflow is not aligned with the ionization cone of the southern nucleus. This misalignment could have the same physical origin as the observed spatial anticorrelation between H$\alpha$ and CO(2-1) observed by \cite{medling_tracing_2021}. They suggest that the hot ionized gas is flowing outwards along the path of least resistance (i.e., where denser molecular gas is not present). It is also interesting that our hot molecular outflow has the same direction of the CO flux map extension in the SE direction (see their Fig. 10). 
Such behavior is anyhow not surprising in a ULIRG, as it can also be observed, for example, in the merger galaxy Mrk 231, where the molecular and the neutral outflows have different axes (\citealp{rupke_mrk231_2011, feruglio_mrk231_2015}).
These misalignment between different outflow gas phases point toward a complex interplay of the two phenomena during the merger, or the presence of different outflow episodes, or a combination of the two.
Another possibility is that the molecular outflow and the ionization cone are part of the same structure, with the molecular outflow forming a shell surrounding the ionized gas dominated cone but with only one side clearly visible, possibly due to a projection effect. Since we have no evidence of the other side of the molecular outflow, we consider only the detected outflow in the following discussion.

Another interesting feature is a gradient in velocity, which links the H$_2$ reservoir (which has a bulk velocity of $\sim$ 500 km s$^{-1}$) and the two nuclei, with well-defined, steep behavior. Furthermore, between the H$_2$ reservoir and the northern nucleus, there is a fainter component, representing the arm visible in the central bottom panel of Fig. \ref{fig: H2 channel maps} linking the two structures. Although the highly redshifted molecular reservoir has been observed since \cite{tecza_stellar_2000}, to date, nobody has ever clarified its nature. This unprecedented view with higher sensitivity and resolution using NIRSpec still does not allow us to reach a definite conclusion. However, we suggest two possible explanations. 

First, it could be an outflow launched by the northern nucleus and oriented in the S direction. 
Indeed, the filaments in the [-117,101] km s$^{-1}$ channel and the arms in the redshifted channels, point to what is expected in an expanding bubble. A similar structure was observed by \cite{cresci_bubbles_2023} in the z = 1.59 prototypical obscured AGN XID2028, where, using the modeling outflows kinematics in AGN 3D (MOKA3D, see \citealt{marconcini_moka3d_2023}) on the [O III] emission line, they assessed that the observed two emission-line arms and the cavity between them are attributable to an expanding bubble and a collimated outflow. This scenario has a critical point: from the p-v diagram in Fig. \ref{fig:pv_H2} it is clear that this cloud is also connected to the southern nucleus, so we should conclude that the two outflows interact in the blob position. But in this case, we would expect an enhanced excitation from shock in that region traceable for example by [Fe II] emission, which is not observed (see Fig. \ref{fig: exc diag diagram}). 
    \cite{ohyama_superwind-driven_2003} proposed a scenario where the kinematics of H$_2$ and the CO and H$_2$ intensity peaks might be explained by a collision between the off-nuclear molecular gas concentration and an external superwind outflow from the southern nucleus (see their Fig. 6). This is not consistent with our results, indeed they asserted a link between the off-nuclear molecular gas and the southern nucleus only, excluding any role of the northern nucleus.
    
Another plausible phenomenon is an inflow. Indeed, molecular gas inflows toward the central region in interacting and/or merging galaxies are expected, as observed in some ULIRGs by \cite{herrera-camus_molecular_2020}, to periodically replenish the molecular gas that is consumed by star formation or the SMBH, and/or ejected by outflows. With this idea, the high-redshift blob can be thought of as the result of the interaction between the two progenitor galaxies: during the merging, the molecular gas has been concentrated in the gravitational center of the system and now is directly feeding the two AGN. The main problem of this scenario is understanding how this blob can have such a high velocity relative to the nuclei.

Without other information, the outflow and inflow are in principle indistinguishable. Thus, as already stated, we cannot draw any definite conclusions.

\subsubsection{Outflow energetics of molecular and ionized gas}
Mass-outflow rates are often calculated based on physical parameters such as outflow mass ($M_{\text{out}}$), radius ($R_{\text{out}}$), and velocity ($v_{\text{out}}$). This calculation typically assumes a spherical or (multi-)cone geometry with a constant velocity throughout the volume, under the simplifying assumption that there is no temporal or spatial acceleration or deceleration of the flow. While this assumption may oversimplify the dynamics, neglecting potential effects such as ballistic deceleration, radiation pressure acceleration, or interaction with other winds, it represents a simplified but useful framework for interpreting current observations. Therefore, even though gas kinematics is clearly much more complex in the nuclear region of NGC 6240, we made these simplifying assumptions using a radius of almost half of the FoV of our observations. The outflow rate is then derived via 
\begin{equation} \label{eq: dMout/dt}
    \dot{M}_{out} = C \frac{M_{out} v_{out}}{R_{out}}.
\end{equation}
The value of C depends on the adopted outflow history (\citealt{lutz_molecular_2020}): assuming that the outflow has started at a point in the past at −t$_{\rm flow}$ = −R$_{\rm out}$/v$_{\rm out}$ and has continued with a constant mass outflow rate until now directly leads to C = 1. 
\begin{table*}[!]

    \begin{center}
    \caption{Properties of the ionized and molecular outflow in NGC 6240.}
     \tabulinesep=1.2mm
    \begin{tabu}{|c|c|c|c|c|}
        \hline
         Species and spatial region & v$_{out}$ (km s$^{-1}$) & $\dot{M}_{out}$ ($M_{\odot}$ yr$^{-1}$) & $E_{out}$ (erg) & $\dot{E}_{out}$ (erg s$^{-1}$) \\
         \hline
         Pa$\alpha$ (with n$_{\rm e}^{[S\,II]}$), circular region & 914.5 & 9.4 & 6.3 $\times$ 10$^{55}$  & 2.5 $\times$ 10$^{42}$ \\
         \hline
         Pa$\alpha$ (with n$_{\rm e}^{[Fe\,II]}$), circular region & 914.5 & 0.26 & 1.7 $\times$ 10$^{54}$  & 6.8 $\times$ 10$^{40}$ \\
         \hline
        H$_2$ 1-0 S(1), circular region & 796.5 & 0.020 & 1.1 $\times$ 10$^{53}$  & 3.9 $\times$ 10$^{39}$ \\
        \hline
        H$_2$ 1-0 S(1), SE outflow & 787.3 & 0.003 & 1.4 $\times$ 10$^{52}$ & 6.4 $\times$ 10$^{38}$ \\
        \hline

    \end{tabu}
    \label{tab: outflow properties}
    \end{center}
\tablefoot{Velocity, mass rate, kinetic energy, and power of the outflow for ionized and molecular gas. These values were calculated for a circular region centered in our FoV. In addition, we calculated the outflow energetics for the outflow near the southern nucleus in the SE direction, discussed in Section \ref{sect: H2 pallocchio}.}
\end{table*}
We computed the average outflow mass rate, the outflow energy, and the outflow power as
\begin{equation} \label{eq: all}
\begin{aligned}
    \dot{M}_{out} &= \frac{M_{out} v_{out}}{R_{out}}, \\
    E_{out} &= \frac{1}{2} M_{out} v_{out}^2, \\
    \dot{E}_{out} &= \frac{1}{2} \dot{M}_{out} v_{out}^2.
\end{aligned}
\end{equation}
We applied equations \ref{eq: all} in a circular region with radius R$_{\rm out}$ of 30 spaxels, that is 1.50\arcsec, which represents almost all NIRSpec FoV - excluding bad spaxels at the edge of the image. We calculated the M$_{\rm out}$ for the broad component of Pa$\alpha$ (with both sulfur lines density and iron lines density) and H$_2$ 1-0 S(1) emission line in the following way: we used the broad component flux of the two species, described in Section \ref{sect: moment maps}, and we computed the masses with equation \ref{eq: M_HII PaA} and \ref{eq: M_H2} (see Section \ref{sec: ionized mass} and \ref{sect: molecular mass}). In the literature, there is not a single definition of the bulk velocity of the outflow (see the review by \citealp{harrison_review_2018}). 
We computed the outflow velocity as
\begin{equation} \label{eq: v_out}
    v_{out} = \text{max} \left( \left|v_{10}^{max} - v_{sys} \right|, \, \left|v_{90}^{max} - v_{sys}\right| \right),
\end{equation}
where v$_{\rm sys}$ is the systemic velocity of the galaxy (estimated from the redshift in Section \ref{sec:Obs} and set to 0 km s$^{-1}$) and v$_{10}^{\rm max}$ and v$_{90}^{\rm max}$ are the maximum value, respectively, of  10th and 90th velocity percentiles of the overall emission line profile (i.e., narrow + broad components). The necessity of this definition, widely adopted in the literature (e.g., \citealt{canodiaz_observational_2012, cresci_magnum_2015,carniani_ionised_2015, marasco_galaxy-scale_2020, tozzi_connecting_2021, tozzi_super_2024}), arises from the unknown geometry and orientation of the outflow relative to the line of sight. Due to our lack of knowledge regarding the true angle of the outflow with respect to the observer's line of sight, and considering that the majority of the outflow is unlikely to be directed toward the observer, we assumed that the most accurate representation of the outflow speed is provided by the velocity "tail" of the line profile, that is v$_{10}$ and v$_{90}$ as described in Eq. \ref{eq: v_out}. 
As discussed in other studies (e.g., \citealt{cresci_magnum_2015}, \citealt{tozzi_super_2024}), these values are considered more appropriate for representing the outflow velocity than the mean or median velocity of the line, which can be significantly influenced by projection effects and dust absorption, this latter being very strong in NGC\,6240.
Thus we used v$_{\rm out}$  $\approx$ 915 km s$^{-1}$ for Pa$\alpha$ emission and v$_{\rm out}$  $\approx$ 800 - 900 km s$^{-1}$ for H$_2$ 1-0 S(1) emission. The results are presented in Table \ref{tab: outflow properties}. For the ionized phase, we found a  mass outflow rate in the range  0.2-9 $M_{\odot}$ yr$^{-1}$, depending on the assumption on the electron density. For the hot molecular gas, we found outflow rate of $\sim$ 0.02 $M_{\odot}$ yr$^{-1}$.

These values are mostly indicative because we did not recognize a classical outflow conical shape across all FoV, but we calculated the properties of all the broad components in our FoV, associating it with a pure outflow.
As stated in the previous section in Fig. \ref{fig: H2 channel maps}, in the blue channel maps of H$_2$ 1-0 S(1) emission line we could outline a collimated outflow in the SE direction starting from the southern nucleus. So we calculated the outflow energetics also for this structure, using R$_{\rm out}$ = 550 pc and v$_{\rm out}$  $\approx$ 790 km s$^{-1}$, with the resulting values in Table \ref{tab: outflow properties}. \cite{cicone_alma_2018} estimated the outflow mass rate of the cold molecular component on a scale of 6 kpc, using the flux of CO (1-0) and obtaining a value of almost 2500 M$_\odot$ yr$^{−1}$. Despite the extension of the outflow, they stated that the largest contribution is given by the central region. Indeed, \cite{treister_molecular_2020} - with a FoV of about 6\arcsec$\,\times\,$6\arcsec, more similar to ours - also found a compatible value with CO (2-1) ALMA observations. It is interesting to note that if we use our most conservative estimation of hot-to-cold ratio (assuming this ratio is similar for the broad component; see Section \ref{sect: molecular mass}) for molecular gas of $\sim$ 10$^{-5}$, the estimation of the outflow mass rate for the hot molecular gas of 0.020 M$_\odot$ yr$^{−1}$ is in agreement with the cited literature's values for the cold component. Furthermore, using the smaller value of $\sim$ 10$^{-6}$, we found that the outflow mass rate for the cold molecular component should reach $\sim$ 10$^4$ M$_\odot$ yr$^{−1}$.   
Regarding the ionized component, its outflow rate is greater than the hot molecular one by 1-3 orders of magnitude, depending on the used value of electron density (see Section \ref{sect: density}). However, we confirmed that the total molecular phase accounts for the majority of the outflowing gas once accounting for the cold phase, featuring masses approximately two orders of magnitude larger than those observed in ionized outflows, as observed by \cite{fiore_agn_2017}.

\begin{figure}[!]
    \centering
    \includegraphics[width=.95\linewidth]{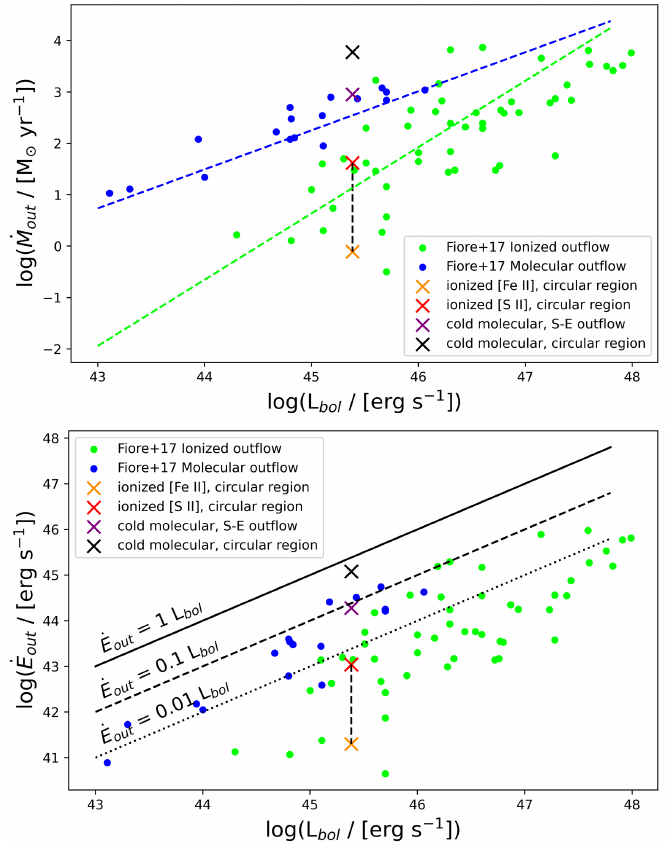}
    \caption{
    Outflow energetics relations compared with the literature.
    \textit{Top panel:} Mass outflow rate as a function of the AGN bolometric luminosity. Green points and green dashed line are the outflow values and correlation for the ionized gas from \cite{fiore_agn_2017}, while blue points and blue dashed line are those for the molecular gas. Black cross is the cold molecular outflow in the circular region and purple cross that in the SE outflow. Orange and red crosses are the ionized outflow in the circular region computed with the [Fe II] method and the [S II] method, respectively. \textit{Bottom panel:} Outflow power as a function of the AGN bolometric luminosity. Solid, dashed and dotted line represent the correlations $\dot{E}_{out}$ = 1, 0.1, 0.01 $L_{bol}$.}    
    
    \label{fig: Fiore+17}
\end{figure}

\cite{fiore_agn_2017} identified a strong correlation between mass outflow rate, outflow power, and bolometric luminosity in AGN. As shown in Fig. \ref{fig: Fiore+17}, they reported $\dot{M} \propto L_{bol}^{0.76\pm0.06}$ and $\dot{E}_{out} \propto L_{bol}^{1.29\pm0.08}$ for molecular winds (blue dashed lines in the top and bottom panel, respectively), while for ionized winds the scaling is $\dot{M} \propto L_{bol}^{1.29\pm0.38}$ and $\dot{E}_{out} \propto L_{bol}^{1.48\pm0.37}$ (green dashed lines in top and bottom panel, respectively).
\cite{yamada_comprehensive_2021} found an X-ray luminosity of 1.37 $\times$ 10$^{10}$ L$_\odot$ and 5.21 $\times$ 10$^{9}$ L$_\odot$ for the southern and northern nucleus, respectively. Applying the bolometric correction from \cite{duras_universal_2020}, we obtained L$_{bol}$ = 2.38 $\times$ 10$^{11}$ L$_\odot$ for the southern nucleus and  L$_{bol}$ = 8.53 $\times$ 10$^{10}$ L$_\odot$ for the northern one.
To calculate the total bolometric luminosity, we summed the values of the two nuclei. Using these values, along with those listed in Table \ref{tab: outflow properties}\footnote{\cite{fiore_agn_2017} calculated outflow properties with C = 3 (see Eq. \ref{eq: dMout/dt}), so we applied this factor to the values in the table.}, we plot in Fig. \ref{fig: Fiore+17} the outflow properties of the cold molecular gas within the circular region (black crosses) and the SE outflow (purple crosses), as well as the ionized gas within the circular region, employing both density estimations (orange crosses for the [Fe II] method and red crosses for the [S II] method). For the calculation of the cold molecular gas mass, we assumed a conservative hot-to-cold ratio of $10^{-5}$. 
The ionized outflow values are consistent with both relations. Regarding the cold molecular ones, it seems that the values of the circular region are too high. This discrepancy is likely due to the inclusion of features in the broad component of H$_2$ that are associated with the merger rather than the outflow. This could suggest that the H$_2$ blob (discussed in Section \ref{sect: H2 pallocchio}) may not be an outflow but rather a tidal feature.  Another possibility is that the hot-to-cold molecular gas ratio derived for the bulk of the gas is too low for the outflow phase (e.g., see \citealt{ulivi_arp220_2024}, who found in Arp 220 hot-to-cold molecular gas ratios of 2$\times$10$^{-4}$ and ~9$\times$10$^{-5}$  in the outflows of the east and west nuclei, respectively), where a larger part of the cold gas could be destroyed or is undetectable in the NIR transitions probed by NIRSpec.

\section{Conclusion}
\label{sec:Conclusions}
In this work, we have analyzed high spectral resolution data obtained from JWST/NIRSpec of the ULIRG NGC 6240, a local prototypical merger known to host at least two AGN. Despite the high sensitivity and spatial resolution of our NIRSpec observations, we did not find any evidence of the presence of the third nucleus proposed by \cite{kollatschny_ngc6240_2020}. 
In Fig. \ref{fig: kin_structures}, we show a graphical summary of the more interesting features observed with this data. In the following, we summarize our main findings.

We first subtracted the stellar continuum and then conducted an emission line multi-component fitting, dividing the emission species into two families, namely molecular and ionized lines, under the hypothesis that the two gas phases follow different kinematics. At the end of the fitting procedure, we also separated the broad and the narrow components in order to distinguish between different kinematic structures in the galaxy. Then we performed the dust extinction correction on the datacubes.

As for the physical properties of the ISM, our main results are the following:
\begin{itemize}
    \item We studied the properties of the ISM using NIR emission-line diagnostics. From the H$_2$ 1-0 S(1)/Br${\gamma}$ and [Fe II]$\lambda$1.257$\mu$m/Pa$\beta$ line ratios, we concluded that the circumnuclear regions of this galaxy are mostly excited from shock-LINER excitation, apart from the AGN excitation that delineates two ionization cones starting from the two nuclei (orange cones in Fig. \ref{fig: kin_structures}).

    \item We detected spatially resolved emission of CLs around the northern nucleus. The emission peaks on the nucleus position and is extended in the NW direction, aligning closely with the ionizing cone found with NIR diagnostics. Using integrated spectra, we detected CLs also in the southern nucleus.

    \item We investigated the excitation mechanisms of the molecular hydrogen lines using specific H$_2$ line ratios. We found that the nuclear region is mainly characterized by thermal emission at $\sim$ 2000 K, with a non-thermal contribution near the two nuclei. We confirmed this result by computing the vibrational and rotational temperatures in our FoV, which are equal to T$_{\rm rot}$ = (1894 $\pm $ 397) K  and T$_{\rm vib}$ = (2050 $\pm$ 141) K. 

    \item We computed the mass of hot H$_2$ in the FoV, assuming local thermodynamic equilibrium and a T$_{\rm vib}$ = 2000 K and obtaining a mass of $\sim$ 1.3 $\times  \, 10^5$ M$_{\odot}$. Using previous literature results, we estimated the hot-to-cold ratio in some spatial regions of our FoV, finding values of 10$^{-6}$-10$^{-5}$, which are within the range found in the literature. Regarding the ionized gas mass, the estimation depends on the electron density. For this, we adopted the value obtained with the [S II] line ratio and that with the [Fe II]. We obtained M$_{\rm HII}$ $\approx 2.8 \times 10^7$ M$_{\odot}$ and M$_{\rm HII}$ $\approx 7.7 \times 10^5$ M$_{\odot}$, respectively.
\end{itemize}

Regarding the kinematics of the stellar and ISM components our findings are as follows:
\begin{itemize}
    \item We studied in detail the kinematics of the ISM and in particular that of the ionized and hot molecular gas components. We presented the moment maps of the brightest emission lines for both the total component and the separated narrow and broad components. We highlighted the different kinematics of the molecular hydrogen with respect to the other lines, in particular its higher velocity values and different spatial distribution (light blue region in Fig. \ref{fig: kin_structures}). We also found a match around the southern nucleus between the velocity map of the stars and that of the narrow component of Pa$\alpha$, meaning that the narrow component of this line follows the rotating disk motion in that region. 
    \item From the channel maps, we discovered features never observed before with this resolution. A blue outflow at high velocity, $\sim$ -800 km s$^{-1}$, is observable in the SE direction near the southern nucleus (light blue outflow in Fig. \ref{fig: kin_structures}). Around the systemic velocity, there are some filaments that, with increasing velocity until $\sim$ 400 km s$^{-1}$, assume a more complex structure, with some arms connecting the northern nucleus and a blob located between the nuclei. The most peculiar characteristic is that this blob remains observable up to $\sim$ 800 km s$^{-1}$. Although its true nature remains ambiguous, we considered both the possibility of an outflow and an inflow.
    \item In the end, we estimated the energetics of the outflow in our FoV. We calculated the outflow mass rate, its kinetic energy, and its power for both the ionized gas and the molecular gas, finding that the molecular outflow largely dominates the outflow rate. In addition, we estimated these properties for the blue outflow observed near the southern nucleus.

\end{itemize}

In the future, we will analyze cube3, the less extinction-affected datacube, to better constrain the physical conditions in NGC 6240 by exploiting, for instance, PAH and CL emission. Observations in other bands with NIRSpec-like spatial and spectral resolution might allow us to fully understand the nature of the H$_2$ cloud with a parallel theoretical approach where simulations could explain its current kinematics.
To conclude, these NIRSpec IFU observations have allowed us to characterize with unprecedented detail a complex system such as NGC 6240 and have demonstrated the fundamental role of JWST in the study of local nearby galaxies.

\begin{acknowledgements}

We thank the anonymous referees for their feedback and suggestions, which have helped improve this manuscript.
GC, AM, MP and LU acknowledge the support of the INAF Large Grant 2022 "The metal circle: a new sharp view of the baryon cycle up to Cosmic Dawn with the latest generation IFU facilities". GC and AM acknowledge support from PRIN-MUR project "PROMETEUS" (202223XPZM) and INAF Large Grant 2022 "Dual and binary SMBH in the multi-messenger era".  
SA and MP acknowledge support from the research project PID2021-127718NB-I00 of the Spanish Ministry of Science and Innovation/State Agency of Research (MCIN/AEI/10.13039/501100011033). MP acknowledges support through the grant RYC2023-044853-I, funded by  MICIU/AEI/10.13039/501100011033 and FSE+.
KF acknowledges support through the ESA research fellowship programme.
IL acknowledges support from PID2022-140483NB-C22 funded by AEI 10.13039/501100011033 and BDC 20221289 funded by MCIN by the Recovery, Transformation and Resilience Plan from the Spanish State, and by NextGenerationEU from the European Union through the Recovery and Resilience Facility, and from  PRIN-MUR project “PROMETEUS”  financed by the European Union -  Next Generation EU, Mission 4 Component 1 CUP B53D23004750006.
AJB acknowledges funding from the “FirstGalaxies” Advanced Grant from the European Research Council (ERC) under the European Union’s Horizon 2020 research and innovation program (Grant agreement No. 789056).
GT acknowledges financial support from the European Research Council (ERC) Advanced Grant under the European Union's Horizon Europe research and innovation programme (grant agreement AdG GALPHYS, No. 101055023). 

\end{acknowledgements}

\bibliographystyle{aa}
\bibliography{NGC6240} 

\begin{thebibliography}{110}
\expandafter\ifx\csname natexlab\endcsname\relax\def\natexlab#1{#1}\fi

\bibitem[{{Alexander} \& {Hickox}(2012)}]{alexander_what_2012}
{Alexander}, D.~M. \& {Hickox}, R.~C. 2012, \nar, 56, 93

\bibitem[{{Armus} {et~al.}(2006){Armus}, {Bernard-Salas}, {Spoon}, {Marshall}, {Charmandaris}, {Higdon}, {Desai}, {Hao}, {Teplitz}, {Devost}, {Brandl}, {Soifer}, \& {Houck}}]{armus_detection_2006}
{Armus}, L., {Bernard-Salas}, J., {Spoon}, H.~W.~W., {et~al.} 2006, \apj, 640, 204

\bibitem[{{Armus} {et~al.}(2023){Armus}, {Lai}, {U}, {Larson}, {Diaz-Santos}, {Evans}, {Malkan}, {Rich}, {Medling}, {Law}, {Inami}, {Muller-Sanchez}, {Charmandaris}, {van der Werf}, {Stierwalt}, {Linden}, {Privon}, {Barcos-Mu{\~n}oz}, {Hayward}, {Song}, {Appleton}, {Aalto}, {Bohn}, {B{\"o}ker}, {Brown}, {Finnerty}, {Howell}, {Iwasawa}, {Kemper}, {Marshall}, {Mazzarella}, {McKinney}, {Murphy}, {Sanders}, \& {Surace}}]{armus_goals-jwst_2023}
{Armus}, L., {Lai}, T., {U}, V., {et~al.} 2023, \apjl, 942, L37

\bibitem[{{Baan} {et~al.}(2007){Baan}, {Hagiwara}, \& {Hofner}}]{baan_h_2007}
{Baan}, W.~A., {Hagiwara}, Y., \& {Hofner}, P. 2007, \apj, 661, 173

\bibitem[{{Bautista} \& {Pradhan}(1998)}]{bautista_ionization_1998}
{Bautista}, M.~A. \& {Pradhan}, A.~K. 1998, \apj, 492, 650

\bibitem[{{Bennett} {et~al.}(2014){Bennett}, {Larson}, {Weiland}, \& {Hinshaw}}]{bennett_1_2014}
{Bennett}, C.~L., {Larson}, D., {Weiland}, J.~L., \& {Hinshaw}, G. 2014, \apj, 794, 135

\bibitem[{{Bianchin} {et~al.}(2024){Bianchin}, {U}, {Song}, {Lai}, {Remigio}, {Barcos-Mu{\~n}oz}, {D{\'\i}az-Santos}, {Armus}, {Inami}, {Larson}, {Evans}, {B{\"o}ker}, {Kader}, {Linden}, {Charmandaris}, {Malkan}, {Rich}, {Bohn}, {Medling}, {Stierwalt}, {Mazzarella}, {Law}, {Privon}, {Aalto}, {Appleton}, {Brown}, {Buiten}, {Finnerty}, {Hayward}, {Howell}, {Iwasawa}, {Kemper}, {Marshall}, {McKinney}, {M{\"u}ller-S{\'a}nchez}, {Murphy}, {van der Werf}, {Sanders}, \& {Surace}}]{bianchin_goals-jwst_2023}
{Bianchin}, M., {U}, V., {Song}, Y., {et~al.} 2024, \apj, 965, 103

\bibitem[{{Black} \& {van Dishoeck}(1987)}]{black_fluorescent_1987}
{Black}, J.~H. \& {van Dishoeck}, E.~F. 1987, \apj, 322, 412

\bibitem[{{B{\"o}ker} {et~al.}(2022){B{\"o}ker}, {Arribas}, {L{\"u}tzgendorf}, {Alves de Oliveira}, {Beck}, {Birkmann}, {Bunker}, {Charlot}, {de Marchi}, {Ferruit}, {Giardino}, {Jakobsen}, {Kumari}, {L{\'o}pez-Caniego}, {Maiolino}, {Manjavacas}, {Marston}, {Moseley}, {Muzerolle}, {Ogle}, {Pirzkal}, {Rauscher}, {Rawle}, {Rix}, {Sabbi}, {Sargent}, {Sirianni}, {te Plate}, {Valenti}, {Willott}, \& {Zeidler}}]{boker_near-infrared_2022}
{B{\"o}ker}, T., {Arribas}, S., {L{\"u}tzgendorf}, N., {et~al.} 2022, \aap, 661, A82

\bibitem[{{B{\"o}ker} {et~al.}(2023){B{\"o}ker}, {Beck}, {Birkmann}, {Giardino}, {Keyes}, {Kumari}, {Muzerolle}, {Rawle}, {Zeidler}, {Abul-Huda}, {Alves de Oliveira}, {Arribas}, {Bechtold}, {Bhatawdekar}, {Bonaventura}, {Bunker}, {Cameron}, {Carniani}, {Charlot}, {Curti}, {Espinoza}, {Ferruit}, {Franx}, {Jakobsen}, {Karakla}, {L{\'o}pez-Caniego}, {L{\"u}tzgendorf}, {Maiolino}, {Manjavacas}, {Marston}, {Moseley}, {Ogle}, {Perna}, {Pe{\~n}a-Guerrero}, {Pirzkal}, {Plesha}, {Proffitt}, {Rauscher}, {Rix}, {Rodr{\'\i}guez del Pino}, {Rustamkulov}, {Sabbi}, {Sing}, {Sirianni}, {te Plate}, {{\'U}beda}, {Wahlgren}, {Wislowski}, {Wu}, \& {Willott}}]{boker_-orbit_2023}
{B{\"o}ker}, T., {Beck}, T.~L., {Birkmann}, S.~M., {et~al.} 2023, \pasp, 135, 038001

\bibitem[{{Cano-D{\'\i}az} {et~al.}(2012){Cano-D{\'\i}az}, {Maiolino}, {Marconi}, {Netzer}, {Shemmer}, \& {Cresci}}]{canodiaz_observational_2012}
{Cano-D{\'\i}az}, M., {Maiolino}, R., {Marconi}, A., {et~al.} 2012, \aap, 537, L8

\bibitem[{{Cappellari} \& {Copin}(2003)}]{cappellari_adaptive_2003}
{Cappellari}, M. \& {Copin}, Y. 2003, \mnras, 342, 345

\bibitem[{{Cappellari} \& {Emsellem}(2004)}]{cappellari_parametric_2004}
{Cappellari}, M. \& {Emsellem}, E. 2004, \pasp, 116, 138

\bibitem[{{Cappellari} {et~al.}(2007){Cappellari}, {Emsellem}, {Bacon}, {Bureau}, {Davies}, {de Zeeuw}, {Falc{\'o}n-Barroso}, {Krajnovi{\'c}}, {Kuntschner}, {McDermid}, {Peletier}, {Sarzi}, {van den Bosch}, \& {van de Ven}}]{cappellari_sauron_2007}
{Cappellari}, M., {Emsellem}, E., {Bacon}, R., {et~al.} 2007, \mnras, 379, 418

\bibitem[{{Cardelli} {et~al.}(1989){Cardelli}, {Clayton}, \& {Mathis}}]{cardelli_relationship_1989}
{Cardelli}, J.~A., {Clayton}, G.~C., \& {Mathis}, J.~S. 1989, \apj, 345, 245

\bibitem[{{Carniani} {et~al.}(2015){Carniani}, {Marconi}, {Maiolino}, {Balmaverde}, {Brusa}, {Cano-D{\'\i}az}, {Cicone}, {Comastri}, {Cresci}, {Fiore}, {Feruglio}, {La Franca}, {Mainieri}, {Mannucci}, {Nagao}, {Netzer}, {Piconcelli}, {Risaliti}, {Schneider}, \& {Shemmer}}]{carniani_ionised_2015}
{Carniani}, S., {Marconi}, A., {Maiolino}, R., {et~al.} 2015, \aap, 580, A102

\bibitem[{{Cicone} {et~al.}(2018){Cicone}, {Severgnini}, {Papadopoulos}, {Maiolino}, {Feruglio}, {Treister}, {Privon}, {Zhang}, {Della Ceca}, {Fiore}, {Schawinski}, \& {Wagg}}]{cicone_alma_2018}
{Cicone}, C., {Severgnini}, P., {Papadopoulos}, P.~P., {et~al.} 2018, \apj, 863, 143

\bibitem[{{Cresci} {et~al.}(2015){Cresci}, {Marconi}, {Zibetti}, {Risaliti}, {Carniani}, {Mannucci}, {Gallazzi}, {Maiolino}, {Balmaverde}, {Brusa}, {Capetti}, {Cicone}, {Feruglio}, {Bland-Hawthorn}, {Nagao}, {Oliva}, {Salvato}, {Sani}, {Tozzi}, {Urrutia}, \& {Venturi}}]{cresci_magnum_2015}
{Cresci}, G., {Marconi}, A., {Zibetti}, S., {et~al.} 2015, \aap, 582, A63

\bibitem[{{Cresci} {et~al.}(2023){Cresci}, {Tozzi}, {Perna}, {Brusa}, {Marconcini}, {Marconi}, {Carniani}, {Brienza}, {Giroletti}, {Belfiore}, {Ginolfi}, {Mannucci}, {Ulivi}, {Scholtz}, {Venturi}, {Arribas}, {{\"U}bler}, {D'Eugenio}, {Mingozzi}, {Balmaverde}, {Capetti}, {Parlanti}, \& {Zana}}]{cresci_bubbles_2023}
{Cresci}, G., {Tozzi}, G., {Perna}, M., {et~al.} 2023, \aap, 672, A128

\bibitem[{{Dale} {et~al.}(2005){Dale}, {Sheth}, {Helou}, {Regan}, \& {H{\"u}ttemeister}}]{dale_warm_2005}
{Dale}, D.~A., {Sheth}, K., {Helou}, G., {Regan}, M.~W., \& {H{\"u}ttemeister}, S. 2005, \aj, 129, 2197

\bibitem[{{Davies} {et~al.}(2003){Davies}, {Sternberg}, {Lehnert}, \& {Tacconi-Garman}}]{davies_molecular_2003}
{Davies}, R.~I., {Sternberg}, A., {Lehnert}, M., \& {Tacconi-Garman}, L.~E. 2003, \apj, 597, 907

\bibitem[{{den Brok} {et~al.}(2022){den Brok}, {Koss}, {Trakhtenbrot}, {Stern}, {Cantalupo}, {Lamperti}, {Ricci}, {Ricci}, {Oh}, {Bauer}, {Riffel}, {Rodr{\'\i}guez-Ardila}, {B{\"a}r}, {Harrison}, {Ichikawa}, {Mej{\'\i}a-Restrepo}, {Mushotzky}, {Powell}, {Boissay-Malaquin}, {Stalevski}, {Treister}, {Urry}, \& {Veilleux}}]{den_brok_bass_2022}
{den Brok}, J.~S., {Koss}, M.~J., {Trakhtenbrot}, B., {et~al.} 2022, \apjs, 261, 7

\bibitem[{{D'Eugenio} {et~al.}(2024){D'Eugenio}, {P{\'e}rez-Gonz{\'a}lez}, {Maiolino}, {Scholtz}, {Perna}, {Circosta}, {{\"U}bler}, {Arribas}, {B{\"o}ker}, {Bunker}, {Carniani}, {Charlot}, {Chevallard}, {Cresci}, {Curtis-Lake}, {Jones}, {Kumari}, {Lamperti}, {Looser}, {Parlanti}, {Rix}, {Robertson}, {Rodr{\'\i}guez Del Pino}, {Tacchella}, {Venturi}, \& {Willott}}]{deugenio_fast-rotator_2024}
{D'Eugenio}, F., {P{\'e}rez-Gonz{\'a}lez}, P.~G., {Maiolino}, R., {et~al.} 2024, Nature Astronomy [\eprint[arXiv]{2308.06317}]

\bibitem[{{Dors} {et~al.}(2012){Dors}, {Riffel}, {Cardaci}, {H{\"a}gele}, {Krabbe}, {P{\'e}rez-Montero}, \& {Rodrigues}}]{dors_jr_x-rays_2012}
{Dors}, Oli~L., J., {Riffel}, R.~A., {Cardaci}, M.~V., {et~al.} 2012, \mnras, 422, 252

\bibitem[{{Draine} \& {Woods}(1990)}]{draine_h2_1990}
{Draine}, B.~T. \& {Woods}, D.~T. 1990, \apj, 363, 464

\bibitem[{{Duras} {et~al.}(2020){Duras}, {Bongiorno}, {Ricci}, {Piconcelli}, {Shankar}, {Lusso}, {Bianchi}, {Fiore}, {Maiolino}, {Marconi}, {Onori}, {Sani}, {Schneider}, {Vignali}, \& {La Franca}}]{duras_universal_2020}
{Duras}, F., {Bongiorno}, A., {Ricci}, F., {et~al.} 2020, \aap, 636, A73

\bibitem[{{Emonts} {et~al.}(2014){Emonts}, {Piqueras-L{\'o}pez}, {Colina}, {Arribas}, {Villar-Mart{\'\i}n}, {Pereira-Santaella}, {Garcia-Burillo}, \& {Alonso-Herrero}}]{emonts_outflow_2014}
{Emonts}, B.~H.~C., {Piqueras-L{\'o}pez}, J., {Colina}, L., {et~al.} 2014, \aap, 572, A40

\bibitem[{{Engel} {et~al.}(2010){Engel}, {Davies}, {Genzel}, {Tacconi}, {Hicks}, {Sturm}, {Naab}, {Johansson}, {Karl}, {Max}, {Medling}, \& {van der Werf}}]{engel_ngc6240_2010}
{Engel}, H., {Davies}, R.~I., {Genzel}, R., {et~al.} 2010, \aap, 524, A56

\bibitem[{{Fabbiano} {et~al.}(2020){Fabbiano}, {Paggi}, {Karovska}, {Elvis}, {Nardini}, \& {Wang}}]{fabbiano_revisiting_2020}
{Fabbiano}, G., {Paggi}, A., {Karovska}, M., {et~al.} 2020, \apj, 902, 49

\bibitem[{{Fahrion} {et~al.}(2024){Fahrion}, {B{\"o}ker}, {Perna}, {Beck}, {Maiolino}, {Arribas}, {Bunker}, {Charlot}, {Ceci}, {Cresci}, {De Marchi}, {L{\"u}tzgendorf}, \& {Ulivi}}]{fahrion_growing_2024}
{Fahrion}, K., {B{\"o}ker}, T., {Perna}, M., {et~al.} 2024, \aap, 687, A83

\bibitem[{{Ferland}(1993)}]{ferland_hazy_1993}
{Ferland}, G.~J. 1993, {Hazy, A Brief Introduction to Cloudy 84}

\bibitem[{{Feruglio} {et~al.}(2015){Feruglio}, {Fiore}, {Carniani}, {Piconcelli}, {Zappacosta}, {Bongiorno}, {Cicone}, {Maiolino}, {Marconi}, {Menci}, {Puccetti}, \& {Veilleux}}]{feruglio_mrk231_2015}
{Feruglio}, C., {Fiore}, F., {Carniani}, S., {et~al.} 2015, \aap, 583, A99

\bibitem[{{Fiore} {et~al.}(2017){Fiore}, {Feruglio}, {Shankar}, {Bischetti}, {Bongiorno}, {Brusa}, {Carniani}, {Cicone}, {Duras}, {Lamastra}, {Mainieri}, {Marconi}, {Menci}, {Maiolino}, {Piconcelli}, {Vietri}, \& {Zappacosta}}]{fiore_agn_2017}
{Fiore}, F., {Feruglio}, C., {Shankar}, F., {et~al.} 2017, \aap, 601, A143

\bibitem[{{Fyhrie} {et~al.}(2021){Fyhrie}, {Glenn}, {Rangwala}, {Wheeler}, {Beck}, \& {Bally}}]{fyhrie_molecular_2021}
{Fyhrie}, A., {Glenn}, J., {Rangwala}, N., {et~al.} 2021, \apj, 922, 208

\bibitem[{{Gallimore} \& {Beswick}(2004)}]{gallimore_parsec-scale_2004}
{Gallimore}, J.~F. \& {Beswick}, R. 2004, \aj, 127, 239

\bibitem[{{Garc{\'\i}a-Burillo} {et~al.}(2021){Garc{\'\i}a-Burillo}, {Alonso-Herrero}, {Ramos Almeida}, {Gonz{\'a}lez-Mart{\'\i}n}, {Combes}, {Usero}, {H{\"o}nig}, {Querejeta}, {Hicks}, {Hunt}, {Rosario}, {Davies}, {Boorman}, {Bunker}, {Burtscher}, {Colina}, {D{\'\i}az-Santos}, {Gandhi}, {Garc{\'\i}a-Bernete}, {Garc{\'\i}a-Lorenzo}, {Ichikawa}, {Imanishi}, {Izumi}, {Labiano}, {Levenson}, {L{\'o}pez-Rodr{\'\i}guez}, {Packham}, {Pereira-Santaella}, {Ricci}, {Rigopoulou}, {Rouan}, {Shimizu}, {Stalevski}, {Wada}, \& {Williamson}}]{garcia-burillo_galaxy_2021}
{Garc{\'\i}a-Burillo}, S., {Alonso-Herrero}, A., {Ramos Almeida}, C., {et~al.} 2021, \aap, 652, A98

\bibitem[{{Genzel} {et~al.}(1998){Genzel}, {Lutz}, {Sturm}, {Egami}, {Kunze}, {Moorwood}, {Rigopoulou}, {Spoon}, {Sternberg}, {Tacconi-Garman}, {Tacconi}, \& {Thatte}}]{genzel_what_1998}
{Genzel}, R., {Lutz}, D., {Sturm}, E., {et~al.} 1998, \apj, 498, 579

\bibitem[{{Gerssen} {et~al.}(2004){Gerssen}, {van der Marel}, {Axon}, {Mihos}, {Hernquist}, \& {Barnes}}]{gerssen_hubble_2004}
{Gerssen}, J., {van der Marel}, R.~P., {Axon}, D., {et~al.} 2004, \aj, 127, 75

\bibitem[{{Harrison} {et~al.}(2018){Harrison}, {Costa}, {Tadhunter}, {Fl{\"u}tsch}, {Kakkad}, {Perna}, \& {Vietri}}]{harrison_review_2018}
{Harrison}, C.~M., {Costa}, T., {Tadhunter}, C.~N., {et~al.} 2018, Nature Astronomy, 2, 198

\bibitem[{{Heckman} {et~al.}(1990){Heckman}, {Armus}, \& {Miley}}]{heckman_nature_1990}
{Heckman}, T.~M., {Armus}, L., \& {Miley}, G.~K. 1990, \apjs, 74, 833

\bibitem[{{Hermosa Mu{\~n}oz} {et~al.}(2024){Hermosa Mu{\~n}oz}, {Alonso-Herrero}, {Labiano}, {Guillard}, {Pantoni}, {Buiten}, {Dicken}, {Baes}, {B{\"o}ker}, {Colina}, {Donnan}, {Garc{\'\i}a-Bernete}, {{\"O}stlin}, {van der Werf}, {Ward}, {Brandl}, {Walter}, {Wright}, {G{\"u}del}, {Henning}, {Lagage}, \& {Ray}}]{MIRI_6240}
{Hermosa Mu{\~n}oz}, L., {Alonso-Herrero}, A., {Labiano}, A., {et~al.} 2024, arXiv e-prints, arXiv:2412.14707

\bibitem[{{Herrera-Camus} {et~al.}(2020){Herrera-Camus}, {Sturm}, {Graci{\'a}-Carpio}, {Veilleux}, {Shimizu}, {Lutz}, {Stone}, {Gonz{\'a}lez-Alfonso}, {Davies}, {Fischer}, {Genzel}, {Maiolino}, {Sternberg}, {Tacconi}, \& {Verma}}]{herrera-camus_molecular_2020}
{Herrera-Camus}, R., {Sturm}, E., {Graci{\'a}-Carpio}, J., {et~al.} 2020, \aap, 633, L4

\bibitem[{{Hopkins} {et~al.}(2006){Hopkins}, {Hernquist}, {Cox}, {Di Matteo}, {Robertson}, \& {Springel}}]{hopkins_unified_2006}
{Hopkins}, P.~F., {Hernquist}, L., {Cox}, T.~J., {et~al.} 2006, \apjs, 163, 1

\bibitem[{{Hummer} \& {Storey}(1987)}]{Hummer&Storey_recombination_1987}
{Hummer}, D.~G. \& {Storey}, P.~J. 1987, \mnras, 224, 801

\bibitem[{{Ilha} {et~al.}(2016){Ilha}, {Bianchin}, \& {Riffel}}]{ilha_origin_2016}
{Ilha}, G. d.~S., {Bianchin}, M., \& {Riffel}, R.~A. 2016, \apss, 361, 178

\bibitem[{{Jakobsen} {et~al.}(2022){Jakobsen}, {Ferruit}, {Alves de Oliveira}, {Arribas}, {Bagnasco}, {Barho}, {Beck}, {Birkmann}, {B{\"o}ker}, {Bunker}, {Charlot}, {de Jong}, {de Marchi}, {Ehrenwinkler}, {Falcolini}, {Fels}, {Franx}, {Franz}, {Funke}, {Giardino}, {Gnata}, {Holota}, {Honnen}, {Jensen}, {Jentsch}, {Johnson}, {Jollet}, {Karl}, {Kling}, {K{\"o}hler}, {Kolm}, {Kumari}, {Lander}, {Lemke}, {L{\'o}pez-Caniego}, {L{\"u}tzgendorf}, {Maiolino}, {Manjavacas}, {Marston}, {Maschmann}, {Maurer}, {Messerschmidt}, {Moseley}, {Mosner}, {Mott}, {Muzerolle}, {Pirzkal}, {Pittet}, {Plitzke}, {Posselt}, {Rapp}, {Rauscher}, {Rawle}, {Rix}, {R{\"o}del}, {Rumler}, {Sabbi}, {Salvignol}, {Schmid}, {Sirianni}, {Smith}, {Strada}, {te Plate}, {Valenti}, {Wettemann}, {Wiehe}, {Wiesmayer}, {Willott}, {Wright}, {Zeidler}, \& {Zincke}}]{jakobsen_near-infrared_2022}
{Jakobsen}, P., {Ferruit}, P., {Alves de Oliveira}, C., {et~al.} 2022, \aap, 661, A80

\bibitem[{{Joseph} {et~al.}(1984){Joseph}, {Wade}, \& {Wright}}]{joseph_detection_1984}
{Joseph}, R.~D., {Wade}, R., \& {Wright}, G.~S. 1984, \nat, 311, 132

\bibitem[{{Kollatschny} {et~al.}(2020){Kollatschny}, {Weilbacher}, {Ochmann}, {Chelouche}, {Monreal-Ibero}, {Bacon}, \& {Contini}}]{kollatschny_ngc6240_2020}
{Kollatschny}, W., {Weilbacher}, P.~M., {Ochmann}, M.~W., {et~al.} 2020, \aap, 633, A79

\bibitem[{{Komossa} {et~al.}(2003){Komossa}, {Burwitz}, {Hasinger}, {Predehl}, {Kaastra}, \& {Ikebe}}]{komossa_discovery_2003}
{Komossa}, S., {Burwitz}, V., {Hasinger}, G., {et~al.} 2003, \apjl, 582, L15

\bibitem[{{Koo} {et~al.}(2016){Koo}, {Raymond}, \& {Kim}}]{koo_infrared_2016}
{Koo}, B.-C., {Raymond}, J.~C., \& {Kim}, H.-J. 2016, Journal of Korean Astronomical Society, 49, 109

\bibitem[{{Krajnovi{\'c}} {et~al.}(2011){Krajnovi{\'c}}, {Emsellem}, {Cappellari}, {Alatalo}, {Blitz}, {Bois}, {Bournaud}, {Bureau}, {Davies}, {Davis}, {de Zeeuw}, {Khochfar}, {Kuntschner}, {Lablanche}, {McDermid}, {Morganti}, {Naab}, {Oosterloo}, {Sarzi}, {Scott}, {Serra}, {Weijmans}, \& {Young}}]{krajnovic_atlas3d_2011}
{Krajnovi{\'c}}, D., {Emsellem}, E., {Cappellari}, M., {et~al.} 2011, \mnras, 414, 2923

\bibitem[{{Kwan} {et~al.}(1977){Kwan}, {Gatley}, {Merrill}, {Probst}, \& {Weintraub}}]{kwan_molecular_1977}
{Kwan}, J.~H., {Gatley}, I., {Merrill}, K.~M., {Probst}, R., \& {Weintraub}, D.~A. 1977, \apj, 216, 713

\bibitem[{{Lamperti} {et~al.}(2017){Lamperti}, {Koss}, {Trakhtenbrot}, {Schawinski}, {Ricci}, {Oh}, {Landt}, {Riffel}, {Rodr{\'\i}guez-Ardila}, {Gehrels}, {Harrison}, {Masetti}, {Mushotzky}, {Treister}, {Ueda}, \& {Veilleux}}]{lamperti_bat_2017}
{Lamperti}, I., {Koss}, M., {Trakhtenbrot}, B., {et~al.} 2017, \mnras, 467, 540

\bibitem[{{Larkin} {et~al.}(1998){Larkin}, {Armus}, {Knop}, {Soifer}, \& {Matthews}}]{larkin_nearinfrared_1998}
{Larkin}, J.~E., {Armus}, L., {Knop}, R.~A., {Soifer}, B.~T., \& {Matthews}, K. 1998, \apjs, 114, 59

\bibitem[{{Law} {et~al.}(2023){Law}, {E. Morrison}, {Argyriou}, {Patapis}, {{\'A}lvarez-M{\'a}rquez}, {Labiano}, \& {Vandenbussche}}]{law_3d_2023}
{Law}, D.~R., {E. Morrison}, J., {Argyriou}, I., {et~al.} 2023, \aj, 166, 45

\bibitem[{{Lepp} \& {McCray}(1983)}]{lepp_x-ray_1983}
{Lepp}, S. \& {McCray}, R. 1983, \apj, 269, 560

\bibitem[{{L{\'\i}pari} {et~al.}(2003){L{\'\i}pari}, {Terlevich}, {D{\'\i}az}, {Taniguchi}, {Zheng}, {Tsvetanov}, {Carranza}, \& {Dottori}}]{lipari_extreme_2003}
{L{\'\i}pari}, S., {Terlevich}, R., {D{\'\i}az}, R.~J., {et~al.} 2003, \mnras, 340, 289

\bibitem[{{Lonsdale} {et~al.}(2006){Lonsdale}, {Farrah}, \& {Smith}}]{lonsdale_ultraluminous_2006}
{Lonsdale}, C.~J., {Farrah}, D., \& {Smith}, H.~E. 2006, in Astrophysics Update 2, ed. J.~W. {Mason}, 285

\bibitem[{{Lutz} {et~al.}(2003){Lutz}, {Sturm}, {Genzel}, {Spoon}, {Moorwood}, {Netzer}, \& {Sternberg}}]{lutz_iso_2003}
{Lutz}, D., {Sturm}, E., {Genzel}, R., {et~al.} 2003, \aap, 409, 867

\bibitem[{{Lutz} {et~al.}(2020){Lutz}, {Sturm}, {Janssen}, {Veilleux}, {Aalto}, {Cicone}, {Contursi}, {Davies}, {Feruglio}, {Fischer}, {Fluetsch}, {Garcia-Burillo}, {Genzel}, {Gonz{\'a}lez-Alfonso}, {Graci{\'a}-Carpio}, {Herrera-Camus}, {Maiolino}, {Schruba}, {Shimizu}, {Sternberg}, {Tacconi}, \& {Wei{\ss}}}]{lutz_molecular_2020}
{Lutz}, D., {Sturm}, E., {Janssen}, A., {et~al.} 2020, \aap, 633, A134

\bibitem[{{Marasco} {et~al.}(2020){Marasco}, {Cresci}, {Nardini}, {Mannucci}, {Marconi}, {Tozzi}, {Tozzi}, {Amiri}, {Venturi}, {Piconcelli}, {Lanzuisi}, {Tombesi}, {Mingozzi}, {Perna}, {Carniani}, {Brusa}, \& {di Serego Alighieri}}]{marasco_galaxy-scale_2020}
{Marasco}, A., {Cresci}, G., {Nardini}, E., {et~al.} 2020, \aap, 644, A15

\bibitem[{{Maraston} \& {Str{\"o}mb{\"a}ck}(2011)}]{maraston_stellar_2011}
{Maraston}, C. \& {Str{\"o}mb{\"a}ck}, G. 2011, \mnras, 418, 2785

\bibitem[{{Marconcini} {et~al.}(2023){Marconcini}, {Marconi}, {Cresci}, {Venturi}, {Ulivi}, {Mannucci}, {Belfiore}, {Tozzi}, {Ginolfi}, {Marasco}, {Carniani}, {Amiri}, {Di Teodoro}, {Scialpi}, {Tomicic}, {Mingozzi}, {Brazzini}, \& {Moreschini}}]{marconcini_moka3d_2023}
{Marconcini}, C., {Marconi}, A., {Cresci}, G., {et~al.} 2023, \aap, 677, A58

\bibitem[{{Max} {et~al.}(2007){Max}, {Canalizo}, \& {de Vries}}]{max_locating_2007}
{Max}, C.~E., {Canalizo}, G., \& {de Vries}, W.~H. 2007, Science, 316, 1877

\bibitem[{{Max} {et~al.}(2005){Max}, {Canalizo}, {Macintosh}, {Raschke}, {Whysong}, {Antonucci}, \& {Schneider}}]{max_core_2005}
{Max}, C.~E., {Canalizo}, G., {Macintosh}, B.~A., {et~al.} 2005, \apj, 621, 738

\bibitem[{{Medling} {et~al.}(2021){Medling}, {Kewley}, {Calzetti}, {Privon}, {Larson}, {Rich}, {Armus}, {Allen}, {Bicknell}, {D{\'\i}az-Santos}, {Heckman}, {Leitherer}, {Max}, {Rupke}, {Treister}, {Messias}, \& {Wagner}}]{medling_tracing_2021}
{Medling}, A.~M., {Kewley}, L.~J., {Calzetti}, D., {et~al.} 2021, \apj, 923, 160

\bibitem[{{Mingozzi} {et~al.}(2019){Mingozzi}, {Cresci}, {Venturi}, {Marconi}, {Mannucci}, {Perna}, {Belfiore}, {Carniani}, {Balmaverde}, {Brusa}, {Cicone}, {Feruglio}, {Gallazzi}, {Mainieri}, {Maiolino}, {Nagao}, {Nardini}, {Sani}, {Tozzi}, \& {Zibetti}}]{mingozzi_magnum_2019}
{Mingozzi}, M., {Cresci}, G., {Venturi}, G., {et~al.} 2019, \aap, 622, A146

\bibitem[{{Moorwood} {et~al.}(1997){Moorwood}, {Marconi}, {van der Werf}, \& {Oliva}}]{moorwood_origin_1997}
{Moorwood}, A.~F.~M., {Marconi}, A., {van der Werf}, P.~P., \& {Oliva}, E. 1997, \apss, 248, 113

\bibitem[{{Mouri}(1994)}]{mouri_molecular_1994}
{Mouri}, H. 1994, \apj, 427, 777

\bibitem[{{M{\"u}ller-S{\'a}nchez} {et~al.}(2018){M{\"u}ller-S{\'a}nchez}, {Nevin}, {Comerford}, {Davies}, {Privon}, \& {Treister}}]{muller-sanchez_two_2018}
{M{\"u}ller-S{\'a}nchez}, F., {Nevin}, R., {Comerford}, J.~M., {et~al.} 2018, \nat, 556, 345

\bibitem[{{M{\"u}ller-S{\'a}nchez} {et~al.}(2011){M{\"u}ller-S{\'a}nchez}, {Prieto}, {Hicks}, {Vives-Arias}, {Davies}, {Malkan}, {Tacconi}, \& {Genzel}}]{muller-sanchez_outflows_2011}
{M{\"u}ller-S{\'a}nchez}, F., {Prieto}, M.~A., {Hicks}, E.~K.~S., {et~al.} 2011, \apj, 739, 69

\bibitem[{{Nardini} {et~al.}(2009){Nardini}, {Risaliti}, {Salvati}, {Sani}, {Watabe}, {Marconi}, \& {Maiolino}}]{nardini_exploring_2009}
{Nardini}, E., {Risaliti}, G., {Salvati}, M., {et~al.} 2009, \mnras, 399, 1373

\bibitem[{{Nardini} {et~al.}(2013){Nardini}, {Wang}, {Fabbiano}, {Elvis}, {Pellegrini}, {Risaliti}, {Karovska}, \& {Zezas}}]{nardini_exceptional_2013}
{Nardini}, E., {Wang}, J., {Fabbiano}, G., {et~al.} 2013, \apj, 765, 141

\bibitem[{{Ohyama} {et~al.}(2003){Ohyama}, {Yoshida}, \& {Takata}}]{ohyama_superwind-driven_2003}
{Ohyama}, Y., {Yoshida}, M., \& {Takata}, T. 2003, \aj, 126, 2291

\bibitem[{{Oliva} {et~al.}(1994){Oliva}, {Salvati}, {Moorwood}, \& {Marconi}}]{oliva_size_1994}
{Oliva}, E., {Salvati}, M., {Moorwood}, A.~F.~M., \& {Marconi}, A. 1994, \aap, 288, 457

\bibitem[{{Pereira-Santaella} {et~al.}(2016){Pereira-Santaella}, {Colina}, {Garc{\'\i}a-Burillo}, {Alonso-Herrero}, {Arribas}, {Cazzoli}, {Emonts}, {Piqueras L{\'o}pez}, {Planesas}, {Storchi Bergmann}, {Usero}, \& {Villar-Mart{\'\i}n}}]{pereira-santaella_high-velocity_2016}
{Pereira-Santaella}, M., {Colina}, L., {Garc{\'\i}a-Burillo}, S., {et~al.} 2016, \aap, 594, A81

\bibitem[{{Perna} {et~al.}(2024){Perna}, {Arribas}, {Lamperti}, {Pereira-Santaella}, {Ulivi}, {B{\"o}ker}, {Maiolino}, {Bunker}, {Charlot}, {Cresci}, {Rodr{\'\i}guez Del Pino}, {D'Eugenio}, {{\"U}bler}, {Fahrion}, \& {Ceci}}]{perna_no_2024}
{Perna}, M., {Arribas}, S., {Lamperti}, I., {et~al.} 2024, \aap, 690, A171

\bibitem[{{Perna} {et~al.}(2023){Perna}, {Arribas}, {Marshall}, {D'Eugenio}, {{\"U}bler}, {Bunker}, {Charlot}, {Carniani}, {Jakobsen}, {Maiolino}, {Rodr{\'\i}guez Del Pino}, {Willott}, {B{\"o}ker}, {Circosta}, {Cresci}, {Curti}, {Husemann}, {Kumari}, {Lamperti}, {P{\'e}rez-Gonz{\'a}lez}, \& {Scholtz}}]{perna_ultradense_2023}
{Perna}, M., {Arribas}, S., {Marshall}, M., {et~al.} 2023, \aap, 679, A89

\bibitem[{{Pollack} {et~al.}(2007){Pollack}, {Max}, \& {Schneider}}]{pollack_circumnuclear_2007}
{Pollack}, L.~K., {Max}, C.~E., \& {Schneider}, G. 2007, \apj, 660, 288

\bibitem[{{Ramos Almeida} {et~al.}(2009){Ramos Almeida}, {P{\'e}rez Garc{\'\i}a}, \& {Acosta-Pulido}}]{almeida_near-infrared_2009}
{Ramos Almeida}, C., {P{\'e}rez Garc{\'\i}a}, A.~M., \& {Acosta-Pulido}, J.~A. 2009, \apj, 694, 1379

\bibitem[{{Riffel} {et~al.}(2013){Riffel}, {Rodr{\'\i}guez-Ardila}, {Aleman}, {Brotherton}, {Pastoriza}, {Bonatto}, \& {Dors}}]{riffel_molecular_2013}
{Riffel}, R., {Rodr{\'\i}guez-Ardila}, A., {Aleman}, I., {et~al.} 2013, \mnras, 430, 2002

\bibitem[{{Riffel} {et~al.}(2006){Riffel}, {Rodr{\'\i}guez-Ardila}, \& {Pastoriza}}]{riffel_0824_2006}
{Riffel}, R., {Rodr{\'\i}guez-Ardila}, A., \& {Pastoriza}, M.~G. 2006, \aap, 457, 61

\bibitem[{{Riffel} {et~al.}(2021){Riffel}, {Storchi-Bergmann}, {Riffel}, {Bianchin}, {Zakamska}, {Ruschel-Dutra}, {Sch{\"o}nell}, {Rosario}, {Rodriguez-Ardila}, {Fischer}, {Davies}, {Dametto}, {Dahmer-Hahn}, {Crenshaw}, {Burtscher}, \& {Bentz}}]{riffel_agnifs_2021}
{Riffel}, R.~A., {Storchi-Bergmann}, T., {Riffel}, R., {et~al.} 2021, \mnras, 504, 3265

\bibitem[{{Riffel} {et~al.}(2008){Riffel}, {Storchi-Bergmann}, {Winge}, {McGregor}, {Beck}, \& {Schmitt}}]{riffel_mapping_2008}
{Riffel}, R.~A., {Storchi-Bergmann}, T., {Winge}, C., {et~al.} 2008, \mnras, 385, 1129

\bibitem[{{Risaliti} {et~al.}(2006){Risaliti}, {Sani}, {Maiolino}, {Marconi}, {Berta}, {Braito}, {Della Ceca}, {Franceschini}, \& {Salvati}}]{risaliti_double_2006}
{Risaliti}, G., {Sani}, E., {Maiolino}, R., {et~al.} 2006, \apjl, 637, L17

\bibitem[{{Rodr{\'\i}guez-Ardila} {et~al.}(2011){Rodr{\'\i}guez-Ardila}, {Prieto}, {Portilla}, \& {Tejeiro}}]{rodriguez-ardila_near-infrared_2011}
{Rodr{\'\i}guez-Ardila}, A., {Prieto}, M.~A., {Portilla}, J.~G., \& {Tejeiro}, J.~M. 2011, \apj, 743, 100

\bibitem[{{Rodr{\'\i}guez-Ardila} {et~al.}(2005){Rodr{\'\i}guez-Ardila}, {Riffel}, \& {Pastoriza}}]{rodriguez-ardila_molecular_2005}
{Rodr{\'\i}guez-Ardila}, A., {Riffel}, R., \& {Pastoriza}, M.~G. 2005, \mnras, 364, 1041

\bibitem[{{Rupke} \& {Veilleux}(2011)}]{rupke_mrk231_2011}
{Rupke}, D. S.~N. \& {Veilleux}, S. 2011, \apjl, 729, L27

\bibitem[{{Sanders} \& {Mirabel}(1996)}]{sanders_luminous_1996}
{Sanders}, D.~B. \& {Mirabel}, I.~F. 1996, \araa, 34, 749

\bibitem[{{Sanders} {et~al.}(1988){Sanders}, {Soifer}, {Elias}, {Neugebauer}, \& {Matthews}}]{sanders_ultraluminous_1988}
{Sanders}, D.~B., {Soifer}, B.~T., {Elias}, J.~H., {Neugebauer}, G., \& {Matthews}, K. 1988, \apjl, 328, L35

\bibitem[{{Scoville} {et~al.}(2015){Scoville}, {Sheth}, {Walter}, {Manohar}, {Zschaechner}, {Yun}, {Koda}, {Sanders}, {Murchikova}, {Thompson}, {Robertson}, {Genzel}, {Hernquist}, {Tacconi}, {Brown}, {Narayanan}, {Hayward}, {Barnes}, {Kartaltepe}, {Davies}, {van der Werf}, \& {Fomalont}}]{scoville_alma_2015}
{Scoville}, N., {Sheth}, K., {Walter}, F., {et~al.} 2015, \apj, 800, 70

\bibitem[{{Scoville} {et~al.}(1998){Scoville}, {Evans}, {Dinshaw}, {Thompson}, {Rieke}, {Schneider}, {Low}, {Hines}, {Stobie}, {Becklin}, \& {Epps}}]{scoville_arp220_1998}
{Scoville}, N.~Z., {Evans}, A.~S., {Dinshaw}, N., {et~al.} 1998, \apjl, 492, L107

\bibitem[{{Scoville} {et~al.}(1982){Scoville}, {Hall}, {Ridgway}, \& {Kleinmann}}]{scoville_velocity_1982}
{Scoville}, N.~Z., {Hall}, D.~N.~B., {Ridgway}, S.~T., \& {Kleinmann}, S.~G. 1982, \apj, 253, 136

\bibitem[{{Silk}(2013)}]{silk_unleashing_2013}
{Silk}, J. 2013, \apj, 772, 112

\bibitem[{{Souchay} {et~al.}(2013){Souchay}, {Mathis}, \& {Tokieda}}]{souchay_tides_2012}
{Souchay}, J., {Mathis}, S., \& {Tokieda}, T. 2013, {Tides in Astronomy and Astrophysics}, Vol. 861

\bibitem[{{Sternberg} \& {Dalgarno}(1989)}]{sternberg_infrared_1989}
{Sternberg}, A. \& {Dalgarno}, A. 1989, \apj, 338, 197

\bibitem[{{Tacconi} {et~al.}(1999){Tacconi}, {Genzel}, {Tecza}, {Gallimore}, {Downes}, \& {Scoville}}]{tacconi_gas_1999}
{Tacconi}, L.~J., {Genzel}, R., {Tecza}, M., {et~al.} 1999, \apj, 524, 732

\bibitem[{{Tecza} {et~al.}(2000){Tecza}, {Genzel}, {Tacconi}, {Anders}, {Tacconi-Garman}, \& {Thatte}}]{tecza_stellar_2000}
{Tecza}, M., {Genzel}, R., {Tacconi}, L.~J., {et~al.} 2000, \apj, 537, 178

\bibitem[{{Tozzi} {et~al.}(2021){Tozzi}, {Cresci}, {Marasco}, {Nardini}, {Marconi}, {Mannucci}, {Chartas}, {Rizzo}, {Amiri}, {Brusa}, {Comastri}, {Dadina}, {Lanzuisi}, {Mainieri}, {Mingozzi}, {Perna}, {Venturi}, \& {Vignali}}]{tozzi_connecting_2021}
{Tozzi}, G., {Cresci}, G., {Marasco}, A., {et~al.} 2021, \aap, 648, A99

\bibitem[{{Tozzi} {et~al.}(2024){Tozzi}, {Cresci}, {Perna}, {Mainieri}, {Mannucci}, {Marconi}, {Kakkad}, {Marasco}, {Brusa}, {Bertola}, {Bischetti}, {Carniani}, {Cicone}, {Circosta}, {Fiore}, {Feruglio}, {Harrison}, {Lamperti}, {Netzer}, {Piconcelli}, {Puglisi}, {Scholtz}, {Vietri}, {Vignali}, \& {Zamorani}}]{tozzi_super_2024}
{Tozzi}, G., {Cresci}, G., {Perna}, M., {et~al.} 2024, \aap, 690, A141

\bibitem[{{Treister} {et~al.}(2020){Treister}, {Messias}, {Privon}, {Nagar}, {Medling}, {U}, {Bauer}, {Cicone}, {Mu{\~n}oz}, {Evans}, {Muller-Sanchez}, {Comerford}, {Armus}, {Chang}, {Koss}, {Venturi}, {Schawinski}, {Casey}, {Urry}, {Sanders}, {Scoville}, \& {Sheth}}]{treister_molecular_2020}
{Treister}, E., {Messias}, H., {Privon}, G.~C., {et~al.} 2020, \apj, 890, 149

\bibitem[{{Turner} {et~al.}(1977){Turner}, {Kirby-Docken}, \& {Dalgarno}}]{turner_quadrupole_1977}
{Turner}, J., {Kirby-Docken}, K., \& {Dalgarno}, A. 1977, \apjs, 35, 281

\bibitem[{{U} {et~al.}(2022){U}, {Lai}, {Bianchin}, {Remigio}, {Armus}, {Larson}, {D{\'\i}az-Santos}, {Evans}, {Stierwalt}, {Law}, {Malkan}, {Linden}, {Song}, {van der Werf}, {Gao}, {Privon}, {Medling}, {Barcos-Mu{\~n}oz}, {Hayward}, {Inami}, {Rich}, {Aalto}, {Appleton}, {Bohn}, {B{\"o}ker}, {Brown}, {Charmandaris}, {Finnerty}, {Howell}, {Iwasawa}, {Kemper}, {Marshall}, {Mazzarella}, {McKinney}, {Muller-Sanchez}, {Murphy}, {Sanders}, \& {Surace}}]{U_resolving_2022}
{U}, V., {Lai}, T., {Bianchin}, M., {et~al.} 2022, \apjl, 940, L5

\bibitem[{{U} {et~al.}(2012){U}, {Sanders}, {Mazzarella}, {Evans}, {Howell}, {Surace}, {Armus}, {Iwasawa}, {Kim}, {Casey}, {Vavilkin}, {Dufault}, {Larson}, {Barnes}, {Chan}, {Frayer}, {Haan}, {Inami}, {Ishida}, {Kartaltepe}, {Melbourne}, \& {Petric}}]{u_spectral_2012}
{U}, V., {Sanders}, D.~B., {Mazzarella}, J.~M., {et~al.} 2012, \apjs, 203, 9

\bibitem[{{Ulivi} {et~al.}(2025){Ulivi}, {Perna}, {Lamperti}, {Arribas}, {Cresci}, {Marconcini}, {Rodr{\'\i}guez Del Pino}, {B{\"o}ker}, {Bunker}, {Ceci}, {Charlot}, {D'Eugenio}, {Fahrion}, {Maiolino}, {Marconi}, \& {Pereira-Santaella}}]{ulivi_arp220_2024}
{Ulivi}, L., {Perna}, M., {Lamperti}, I., {et~al.} 2025, \aap, 693, A36

\bibitem[{{van der Werf} {et~al.}(1993){van der Werf}, {Genzel}, {Krabbe}, {Blietz}, {Lutz}, {Drapatz}, {Ward}, \& {Forbes}}]{van_der_werf_near-infrared_1993}
{van der Werf}, P.~P., {Genzel}, R., {Krabbe}, A., {et~al.} 1993, \apj, 405, 522

\bibitem[{{Veilleux} {et~al.}(2013){Veilleux}, {Mel{\'e}ndez}, {Sturm}, {Gracia-Carpio}, {Fischer}, {Gonz{\'a}lez-Alfonso}, {Contursi}, {Lutz}, {Poglitsch}, {Davies}, {Genzel}, {Tacconi}, {de Jong}, {Sternberg}, {Netzer}, {Hailey-Dunsheath}, {Verma}, {Rupke}, {Maiolino}, {Teng}, \& {Polisensky}}]{veilleux_fast_2013}
{Veilleux}, S., {Mel{\'e}ndez}, M., {Sturm}, E., {et~al.} 2013, \apj, 776, 27

\bibitem[{{Veilleux} {et~al.}(2003){Veilleux}, {Shopbell}, {Rupke}, {Bland-Hawthorn}, \& {Cecil}}]{veilleux_search_2003}
{Veilleux}, S., {Shopbell}, P.~L., {Rupke}, D.~S., {Bland-Hawthorn}, J., \& {Cecil}, G. 2003, \aj, 126, 2185

\bibitem[{{Wang} {et~al.}(2014){Wang}, {Nardini}, {Fabbiano}, {Karovska}, {Elvis}, {Pellegrini}, {Max}, {Risaliti}, {U}, \& {Zezas}}]{wang_fast_2014}
{Wang}, J., {Nardini}, E., {Fabbiano}, G., {et~al.} 2014, \apj, 781, 55

\bibitem[{{Yamada} {et~al.}(2021){Yamada}, {Ueda}, {Tanimoto}, {Imanishi}, {Toba}, {Ricci}, \& {Privon}}]{yamada_comprehensive_2021}
{Yamada}, S., {Ueda}, Y., {Tanimoto}, A., {et~al.} 2021, \apjs, 257, 61

\end{thebibliography}

\begin{appendix}
\onecolumn

\section{Cube1 stellar subtraction} \label{sec: ppxf cube1}
\begin{figure*}[!h]
    \centering
    \includegraphics[width=1.\linewidth]{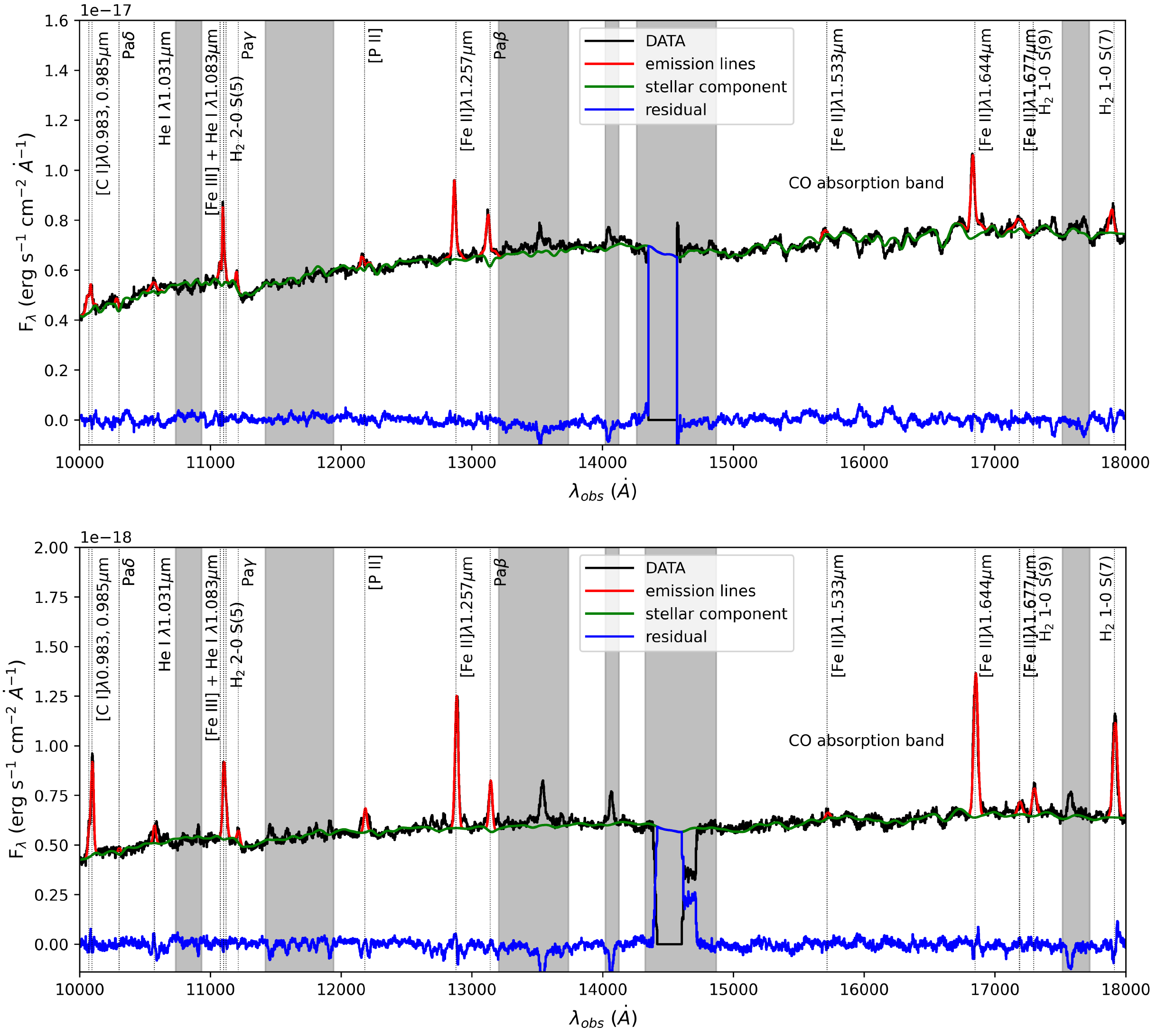}
    \caption{Examples of continuum and emission lines fitting in a spaxel from the southern nucleus (top panel) and one from the region between nuclei (bottom panel), with three Gaussian components for each line. (See Fig. \ref{fig: ppxf example} for details.)}
    \label{fig: ppxf cube1}
\end{figure*}

In this appendix, we present the analog examples of Fig. \ref{fig: ppxf example} for cube1. In Fig. \ref{fig: ppxf cube1} we plot the fit cube1 spectra from the same spaxel of Fig. \ref{fig: ppxf example}, one centered on the southern nucleus (upper panel) and the other located in the region between the two nuclei (lower panel). In both, we show data (in black), the total model of emission lines and stellar component (in red), the model of only stars (in green), and the residuals between data and total model (in blue). The vertical lines mark all features modeled with pPXF procedure, while the gray bands show the spectral regions we masked during the fitting procedure, corresponding to the gap between the detectors and covering a few emission lines not relevant to this work.

\FloatBarrier 
\FloatBarrier 

\section{Moment maps of broad and narrow components}\label{app moment maps}
\begin{figure*}[!]
    \centering
    \includegraphics[width=1\linewidth]{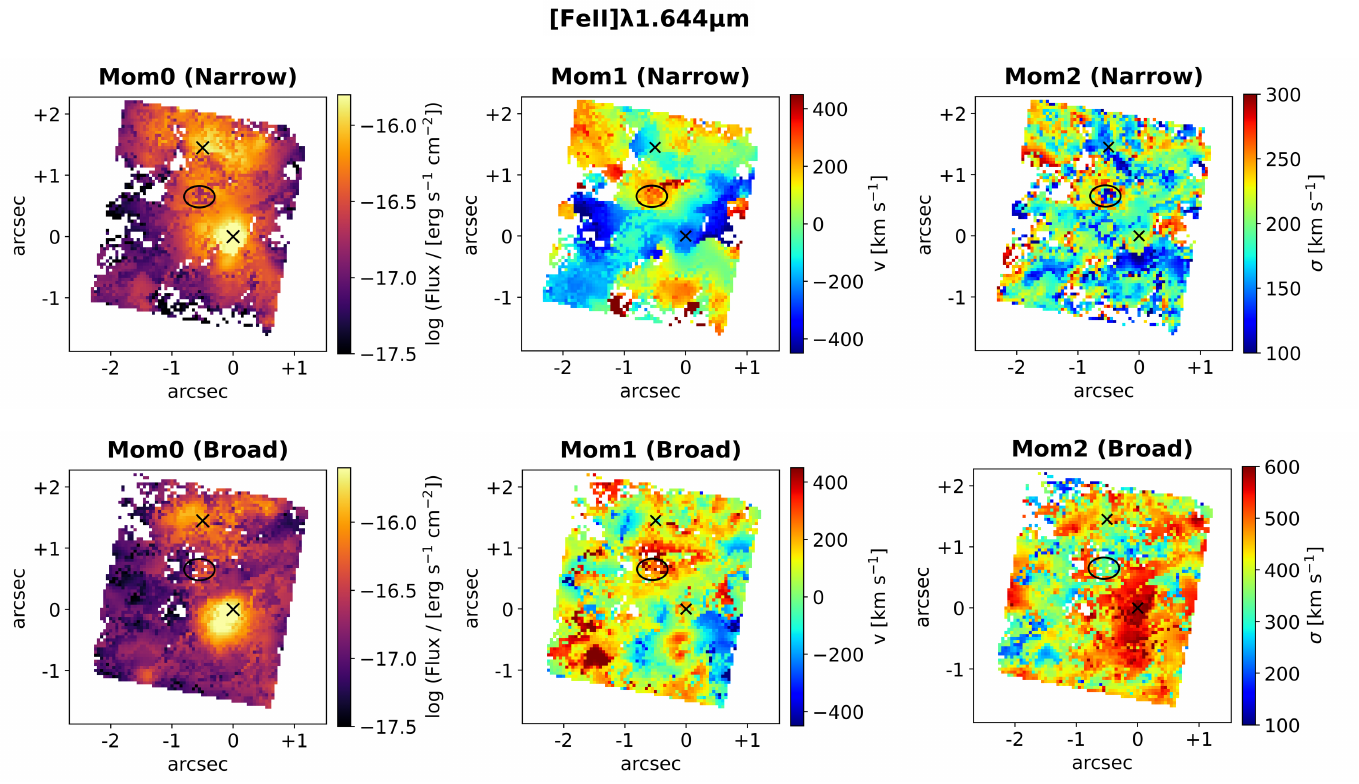}
    \caption{Moment maps of [Fe II]$\lambda$1.644$\mu$m narrow and broad components. (See Fig. \ref{fig: H2 NvsB} for details.)}
    
    \label{fig: FeII NvsB}
\end{figure*}

\begin{figure*}[!] 
    \centering
    \includegraphics[width=1\linewidth]{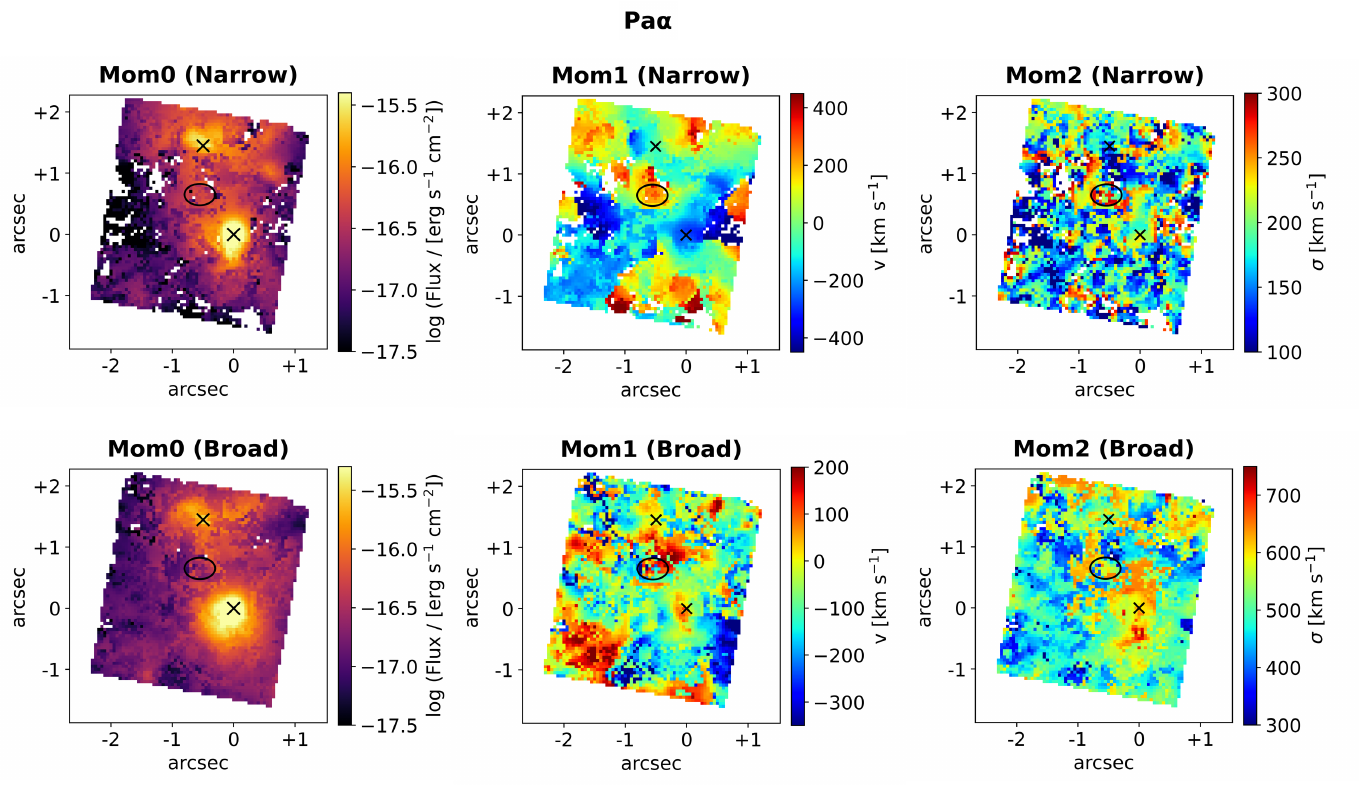}
    \caption{Moment maps of Pa$\alpha$ narrow and broad components. (See Fig. \ref{fig: H2 NvsB} for details.)
    \label{fig: PaA NvsB}}
\end{figure*}

\FloatBarrier 
In Figs. \ref{fig: FeII NvsB} and \ref{fig: PaA NvsB} we plot the moment maps of [Fe II]$\lambda$1.644$\mu$m and Pa$\alpha$ emission lines. In the upper (lower) panels are shown the maps of the narrow (broad) components.

\FloatBarrier 

\end{appendix}

\label{lastpage}
\end{document}